\documentclass[a4paper,fleqn,usenatbib,useAMS]{mnras}

\usepackage{graphicx}	% Including figure files
\usepackage{amsmath}	% Advanced maths commands
\usepackage{multicol}   % Multi-column entries in tables
\usepackage{pdflscape}	% Landscape pages
\usepackage[colorinlistoftodos]{todonotes}

\def\las{\mathrel{\hbox{\rlap{\hbox{\lower3pt\hbox{$\sim$}}}\hbox{\raise2pt\hbox{$<$}}}}}
\def\gas{\mathrel{\hbox{\rlap{\hbox{\lower3pt\hbox{$\sim$}}}\hbox{\raise2pt\hbox{$>$}}}}}
\usepackage[T1]{fontenc}
\usepackage{ae,aecompl}

\usepackage{newtxtext,newtxmath}
\title[MNRAS Exomoons from TTVs]{Exomoon Candidates from Transit Timing Variations\newline \textit{Eight Kepler systems with TTVs explainable by photometrically unseen exomoons} }

\author[Fox, Wiegert]{Chris Fox$^{1,2}$\thanks{Contact e-mail: \href{mailto:cfox53@uwo.ca}{cfox53@uwo.ca}}, Paul Wiegert$^{1,2}$
\\
$^{1}$Department of Physics \& Astronomy, The University of Western Ontario, London, Ontario, Canada\\
$^{2}$Institute for Earth and Space Exploration (IESX), The University of Western Ontario, London, Ontario, Canada}
\date{Accepted for publication November 26 2020 by the Monthly Notices of the Royal Astronomical Society}
\pubyear{2020}

\begin{document}
\label{firstpage}
\pagerange{\pageref{firstpage}--\pageref{lastpage}}
\maketitle

% Abstract of the paper
\begin{abstract}
If a transiting exoplanet has a moon, that moon could be detected directly from the transit it produces itself, or indirectly via the transit timing variations it produces in its parent planet. There is a range of parameter space where the Kepler Space Telescope is sensitive to the TTVs exomoons might produce, though the moons themselves would be too small to detect photometrically via their own transits. The Earth's Moon, for example, produces TTVs of 2.6 minutes amplitude by causing our planet to move around their mutual centre of mass. This is more than Kepler's short-cadence interval of 1 minute and so nominally detectable (if transit timings can be measured with comparable accuracy), even though the Moon's transit signature is only 7\% that of Earth's, well below Kepler's nominal photometric threshold.

Here we examine several Kepler systems, exploring the hypothesis that an exomoon could be detected solely from the TTVs it induces on its host planet.  We compare this with the alternate hypothesis that the TTVs are caused by an non-transiting planet in the system.  We examine 13 Kepler systems and find 8 where both hypotheses explain the observed TTVs equally well.  Though no definitive exomoon detection can be claimed on this basis, the observations are nevertheless completely consistent with a dynamically stable moon small enough to fall below Kepler's photometric threshold for transit detection, and these systems warrant further observation and analysis.
\end{abstract}

% Select between one and six entries from the list of approved keywords.
% Don't make up new ones.

%%%%%%%%%%%%%%%%%%%%%%%%%%%%%%%%%%%%%%%%%%%%%%%%%%
%%%%%%%%%%%%%%%%% BODY OF PAPER %%%%%%%%%%%%%%%%%%

\section{Introduction}
Most of the planets found by the Kepler Space Telescope have been via the transit method \citep{bor2010}.  However, additional non-transiting planets have been discovered by examining the variability of transit timings.  Gravitational perturbations between planets can result in deviations from perfectly Keplerian orbits, seen as transit timing variations (TTVs) \citep{agol2005, holmur2005} which can reveal the presence of otherwise undetected planets.  

Here we consider the role of a companion in orbit of a planet, a companion which we term an {\it exomoon}, in producing TTVs.  In particular, we consider the parameter space where an exomoon could produce TTVs while {\it not} producing a photometrically detectable transit.  We then examine the Kepler data set for such signals, and discuss several planetary systems that exhibit TTVs consistent with exomoons, and compare the hypothesis that these TTVs are caused by an exomoon with the hypothesis that they are caused by another planet in the system.

Exomoons have been studied in depth from a number of perspectives.  The role of exomoons as habitable worlds has been explored by \citet{hinkelkane2013,hillkane2018,martinez2019}.  The possible formation mechanisms of exomoons have been examined by \cite{barrmoon2017} and \cite{malamudmoon2020}.  

That exomoons may induce TTVs upon their host planet has been examined by other authors.  The properties of TTVs generated by hypothetical exomoons has been explored by \cite{sarsch1999, kipp2009, heller2016}. These papers were theoretical in nature and did not examine observed light curves.  \citet{szabo2013} searched for exomoons in the Kepler data set by means of Fourier Transforms of the entire transit timing data set. No definitive exomoon detections were made.  More recently, \citet{rodenbeck2020} examined exomoon indicators from TTV, TDV and TRV (transit timing, duration and radius variations respectively), modelling under realistic conditions, but did not apply these techniques to the Kepler data set.

The most prominent search for exomoons has been the HEK (Hunt for Exomoons with Kepler \citep{hek3, hek4, hek5, hek6} project, which  uses a photodynamical approach, modelling the expected photometric signal of an exoplanet-exomoon combination from transit to transit within a Bayesian framework.  However, no search to date has made a positive identification of an exomoon.  Arguably the best moon transit candidate to date comes from the HEK project: the Kepler-1625 system \citep{hek1625}. However, alternative explanations for the signal \citep{heller2019, kreid2019} have also been proposed, and Kepler-1625 remains an unconfirmed and controversial exomoon candidate.

In this work, we examine two questions. The first is, could an exomoon too small to generate a photometrically-detectable transit still create detectable TTVs? The answer is yes and we examine the circumstances in detail in Section \ref{section:theory}. The second is, does the Kepler data set contain TTVs signals consistent with those produced by an exomoon below Kepler's photometric detection threshold? The answer is again yes, though the statistical validity of these signals must be addressed with care. We discuss this further in sections \ref{section:targetselection}, \ref{section:methods} and \ref{section:results}.

While other searches have been primarily focused on finding an exomoon transit signal (such as those of HEK e.g. \citet{hek6}), what makes this study unique is that we specifically consider only exomoons that would be too small to create detectable photometric (transit) signals. We will actually exclude from our sample systems where the TTV signals are too large to be consistent with photometrically unseen moons, concentrating solely on moons lying in the niche of phase space where their TTVs are detectable by Kepler but their transits are not. As a result, this work does not examine Kepler's photometric data at all, and we will exclude from consideration any exomoon candidates large enough to be easily seen from their transit signals, tacitly assuming that these would have been seen (if present) by earlier dedicated studies.  

We examine two classes of models where the TTVs are created by 1) another planet in the system and 2) a moon in orbit around the planet, to determine which might provide a better fit to the TTV signals seen in the Kepler data set. One constraint we impose on our exomoon model is that the moon's contribution to the transit signal is small enough to remain undetected.   To first order, Kepler is sensitive to transiting bodies the size of the Earth orbiting a Sun-sized star \citep{gill2011}.  We will consider a body, whether planet or moon, with a transit confidence level equivalent to an Earth-sized body orbiting a Sun-sized star to be above Kepler's {\it photometric} sensitivity limit.  Similarly, Kepler can detect TTVs of order the interval between its short-cadence observations, or about one minute \citep{bor2010}, which sets its {\it TTV} sensitivity limit.  In practice, the TTV sensitivity limit will be set by the accuracy to which the transit timings can be determined and we will consider realistic timing errors here.  However, the net result is that for many Kepler systems there is a region of parameter space where an exomoon could create TTVs that are above Kepler's TTV sensitivity limit while having a cross-section that puts it below Kepler's photometric sensitivity. This is the scenario that we examine in this paper.

Initially we also examined a third model with two moons orbiting the planet in circular coplanar orbits.  However, as will be discussed later, this hypothesis resulted in systems that were highly unstable and we did not find any viable two moon systems that could explain the TTV patterns better than the other two models.

\section{Theoretical Basis for Moon-Induced TTVs} \label{section:theory}
\begin{figure}
 \includegraphics[width=\columnwidth]{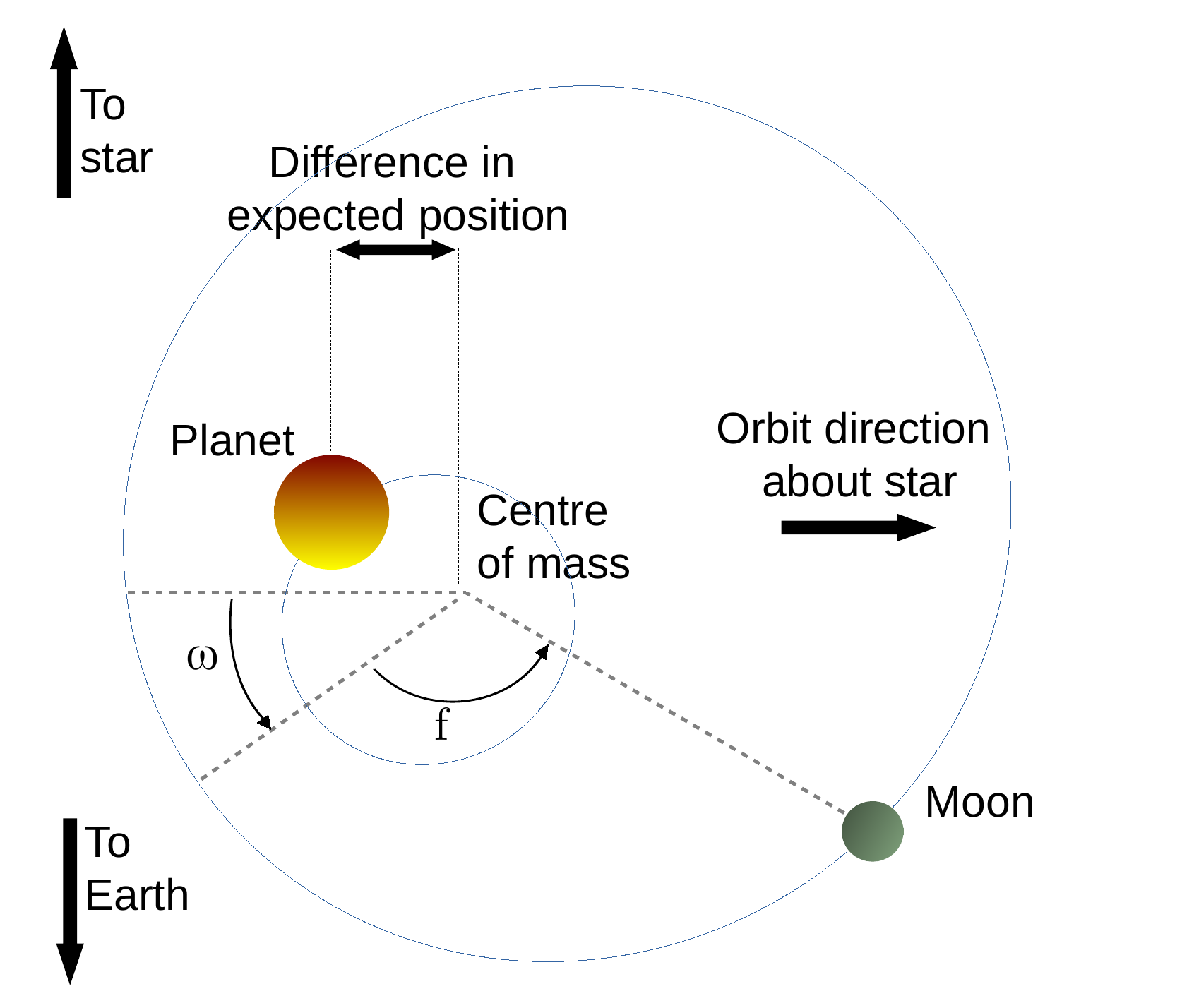}
 \caption{Simple model of planet-moon system.  \label{fig:planetmoonsimple} }
\end{figure}

In this section we discuss the conditions under which an exomoon could be detectable from its Kepler-derived TTVs even if it were below that spacecraft's photometric detection threshold. We use a simplified model of a planet-moon system to model planetary TTVs resulting from the moon, as illustrated in Figure \ref{fig:planetmoonsimple}. 
The planet and moon orbit their mutual centre of mass.  In the absence of any other influences, the centre of mass of the planet-moon system will orbit the parent star with a fixed period.  The transit timing of the planet is then offset by an amount that depends on the orientation of the planet-moon system during the transit.  The TTV for single transit for this simple model is expressed as:
\begin{equation}
  \label{eq:ttv}
    TTV = \Bigg(\frac{P_{p}}{2\pi G M_{*}}\Bigg)^{1/3}  \frac{M_{m}a_{pm}}{M_{p}+M_{m} } \frac{(1-e^{2})}{1+e\cos(f) } \sin\Big(\omega+f-\frac{\pi}{2}\Big) 
\end{equation}
where $a_{pm}$ is the distance of the moon to the planet (not the barycentre), $f$ the true anomaly, and $\omega$ the argument of periastron of the moon.  These definitions are consistent with those of \citet{kipp2009}.

\label{text:periodduration}The moon necessarily orbits the planet with a period much shorter than the period of the planet about the star because it must orbit within the planet's Hill sphere \citep{kipp2009}.  Notwithstanding this, our model will assume that the period of the moon is significantly greater than the transit duration; that is that there is no motion of the planet with respect to the moon-planet centre of mass during the transit. A moon on too small an orbit could produce substantial reflex motion of the planet during the transit, shortening or lengthening the transit depending on the moon's phase. This can have an impact upon the measured transit centre and timing measurement, thereby making the modelling more complex. However, we will see that in all the cases we examine here, the hypothetical moon's period is long enough that such effects can be safely ignored.

\begin{figure}
 \includegraphics[width=\columnwidth]{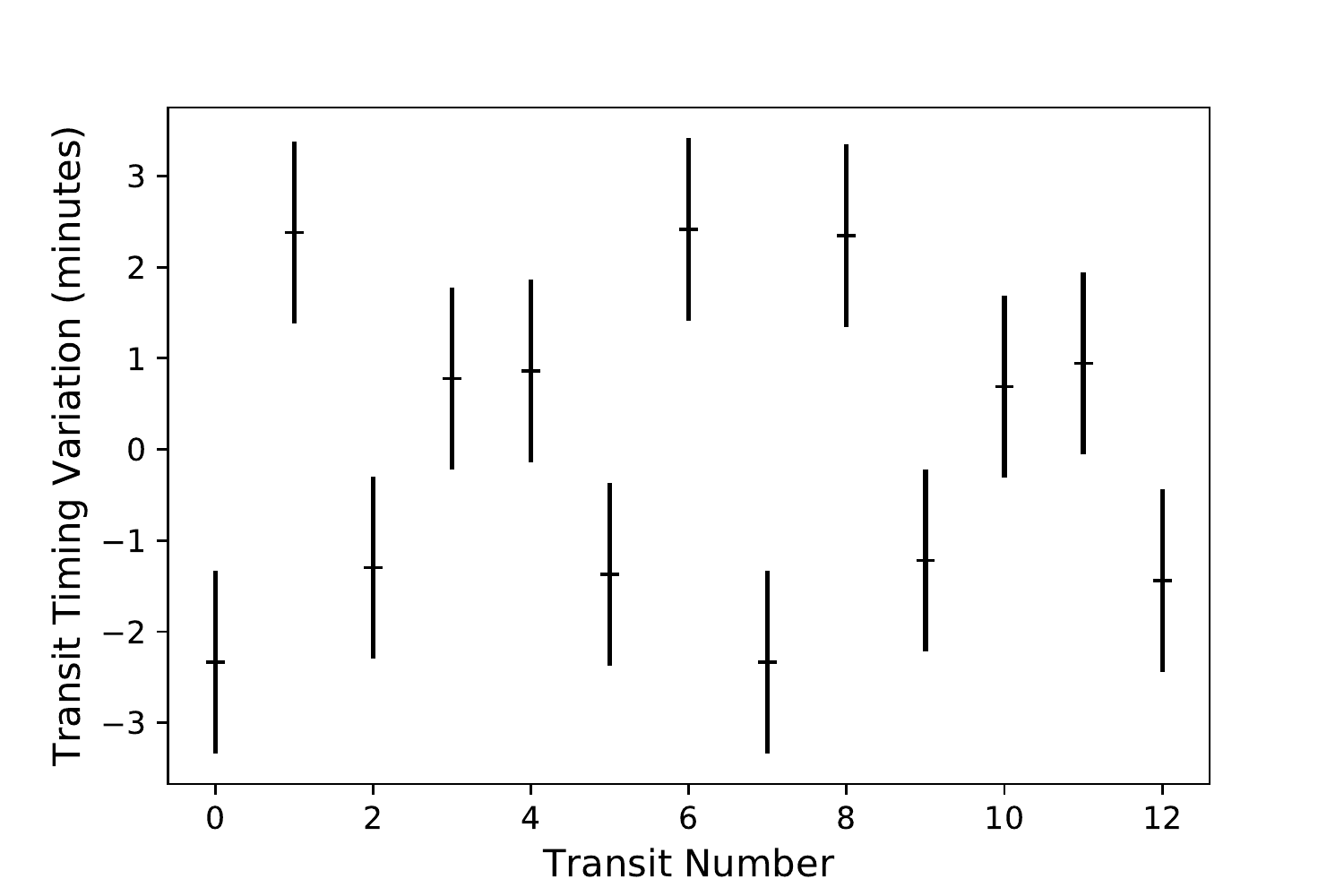}
 \caption{Simulated ideal TTV pattern of an Earth-Moon analog, with 1 minute error bars.  \label{fig:earthmoonExact} }
\end{figure}

As an example of the type of system we are examining here, applying Equation \ref{eq:ttv} to the Earth-Moon system yields a TTV amplitude of 2.58 minutes. This exceeds Kepler's short-cadence interval of one minute so is nominally detectable, though we do note that transit timing uncertainties well below the cadence interval are possible as demonstrated in \citet{hm2016}.   The Moon's cross-section of only 7\% of Earth's puts it below Kepler's photometric detection threshold.  We note that Earth orbits with a period of 365.26 days, which would only produce four transits in the four-year Kepler data set.  For a confident detection of an exomoon, more transits would be needed. In particular, our attempts to compare models by determining the best-fit parameters require, at a minimum, one transit per parameter and ideally many more.  TDVs, transit duration variations, could potentially double the number of data points, but as will be seen later, these have relatively high errors compared to their amplitude and thus have limited utility. Though our own Earth-Moon system may or may not be recoverable from the Kepler data set, it illustrates the principle that, in some cases, Kepler is more sensitive to TTV variations from exomoons than it is to their photometric signals.  Figure \ref{fig:earthmoonExact} shows the expected TTV pattern of an Earth-Moon system with an error bar of one minute added,  providing an illustration of the possible signal. 

While TTV signals of this magnitude and with low error are relatively rare in the Kepler data set, hundreds do exist and can provide valuable insights.  We also note that as TTV strength is linearly proportional to the moon's mass and its semi-major axis (Equation \ref{eq:ttv}), modest increases in either could produce a significantly stronger TTV signal than what is produced by the Earth-Moon system.

\subsection{Detectability from TTVs versus Transits} \label{ttvs-vs-transits}
\begin{figure}
 \includegraphics[width=\columnwidth]{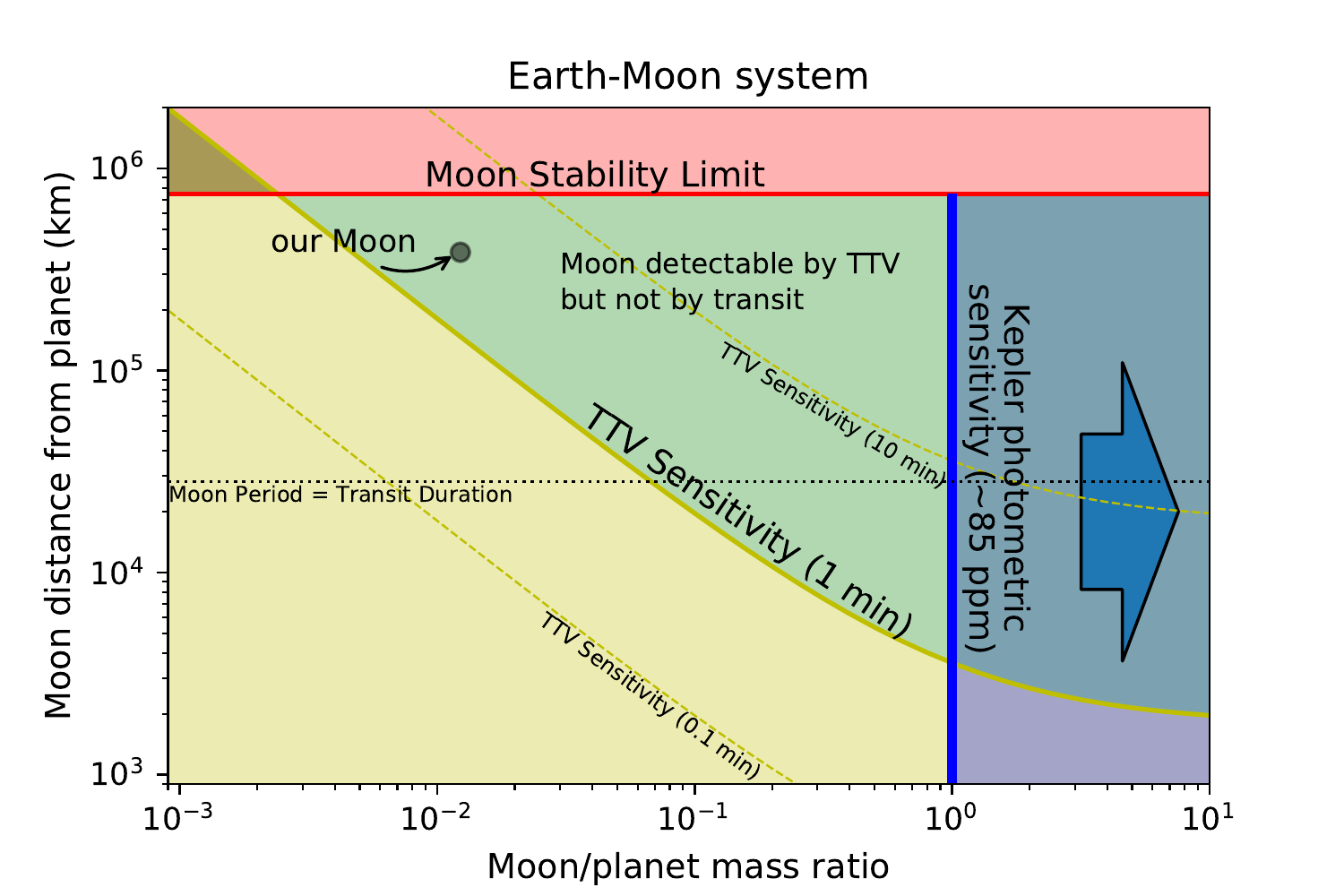}
 \caption{Parameter space of a Moon-Earth Analog.  The diagonal yellow line represents a TTV sensitivity of 1 minute.  Alternate sensitivity lines of 10 and 0.1 minutes are also shown for comparison.  The horizontal red line represents the orbital stability of the moon at 0.5 of the Earth's Hill radius.  The vertical blue line is the Kepler photometric detection limit, using an Earth mass as proxy for an Earth radius. A dotted horizontal line delimits where the moon orbital period is equal to the planet transit duration: moons well above this line can be considered stationary with respect to the planet during the transit. The grey dot indicates Earth's Moon.  Being inside the green region, our Moon would be nominally detectable by Kepler from its TTVs, but its transit would be below Kepler's photometric sensitivity. \label{fig:earthmoonSens} }
\end{figure}

Before proceeding to a search of the Kepler data set, it will be useful to construct a illustrative diagram of the parameter space which can be used to assess whether candidate systems are broadly consistent with our criteria. 

 We define our region of interest or the 'green zone' to be the region where the transit signature of an exomoon would be below Kepler's photometric threshold, but the planetary TTVs induced by such a moon would larger than the uncertainty in those TTVs.  This picture provides an informative first look at the parameter space.  Figure \ref{fig:earthmoonSens} illustrates the region of interest (in green) using an Earth-Moon analog, where we have assumed a TTV sensitivity of one minute, equal to Kepler's short cadence interval. 

We will construct a similar but more realistic diagram for each of our candidate systems based on its individual stellar and planetary parameters. The lines in the diagram, which are described below, represent the approximate location of various thresholds related to our search. The green zone is the parameter space in which an exomoon could produce TTVs while being too small to be observed photometrically, and is where we will focus our attention.  Moon parameters that fall well outside of the green zone will not be considered here. The lines in Figure \ref{fig:earthmoonSens} and following diagrams are:

\begin{enumerate}
\item Equation \ref{eq:ttvlimline} represents Kepler's sensitivity to exomoon-generated TTVs, expressed in terms of the moon's parameters.  This is a restatement of equation \ref{eq:ttv}, assuming low eccentricity moon orbit, a fixed TTV sensitivity on the part of Kepler, and reorganized to write the moon's distance from the planet a function of the moon-planet mass ratio. 
\begin{equation}
    a_{pm} = TTV \Bigg(\frac{2\pi G M_{*}}{P_{p}}\Bigg)^{1/3} \Bigg(1+\Big({\frac{M_{m}}{M_{p}}}\Big)^{-1}\Bigg)
    \label{eq:ttvlimline}
\end{equation}
Equation \ref{eq:ttvlimline} is shown by the yellow line in Figure \ref{fig:earthmoonSens} where a TTV sensitivity is 1 minute is assumed for that case. When constructing this diagram for our target systems, we take the TTV sensitivity to be the timing precision appropriate for each star; the average TTV error from \cite{hm2016}. This is typically on the order of 3 or 4 minutes in the systems we examine, and pushes this line upwards, making the green zone smaller.    To demonstrate the effect of TTV sensitivity, alternative TTV sensitivity lines are shown on Figure \ref{fig:earthmoonSens} as well, corresponding to 0.1 and 10 minutes.

\item The red horizontal line in Figure \ref{fig:earthmoonSens}, expressed by Equation \ref{eq:hilllim}, represents one-half of the Hill radius of the planet. This serves as our outer limit for the stability of exomoons.
\begin{equation}
    a\big|_{0.5 Hill} = 0.5 a_{p}\Bigg(\frac{M_{p}}{3M_{*}}\Bigg)^{1/3} = 0.5  \Bigg(\frac{G (M_{*}+M_{p}) {P_{p}}^{2}M_{p} }{12\pi^{2}M_{*}}\Bigg) ^{1/3} \label{eq:hilllim}
\end{equation}
 Numerical studies have shown that prograde moons are not stable beyond about 0.3 $R_{Hill}$ \citep{holwie99} though retrograde moons can survive out as far as $0.5~R_{Hill}$ \citep{niccukshe08}. As a result, any modelled fit to the TTVs that requires an exomoon above the red line would be unstable. In practice we restrict our searches to less than $0.3~R_{Hill}$ to ensure that not only do the moons remain bound to the planet, but their orbits do not vary strongly with time (due to stellar gravitational perturbations), so that our assumption of a fixed elliptical moon orbit is valid. The use of the Hill sphere becomes questionable as the moon/planet mass approaches unity, but it provides us with a useful zeroth-order limit:  any moon with an orbital radius of more than half the Hill radius is unlikely to be stable.

\item The blue vertical line in Figure \ref{fig:earthmoonSens}, is expressed by Equation \ref{eq:translim} and corresponds to an Earth-sized body transiting a Sun-sized star with 29 ppm noise.  We adopt this as representative of Kepler's photometric detection threshold for our initial survey of the parameter space.  
\begin{equation}
    \frac{M_{m}}{M_{p}} \Bigg|_{threshold} = \Bigg( \frac{M_{\earth}}{M_{p}}\Bigg) \Bigg( \frac{R_{*}}{R_{\sun}} \Bigg)^{3} \Bigg( \frac{CDPP_{*}}{29 ppm} \Bigg)^{3/2}\label{eq:translim}
\end{equation}
Here, the CDPP is interpolated to the specific transit duration of the planet, using the Combined Differential Photometric Precision values in the Kepler Stellar tables \citep{kep8cat2018}.  This provides a measure of the actual photometric uncertainty for the transit of each planet.  Each star's CDPP value is compared to the typical CDPP that Kepler achieved in its sample, 29 ppm \citep{gill2011}.  A moon with a mass (and hence cross-section greater than an Earth-Sun equivalent) would appear to the right of this line, and we will consider it photometrically detectable in Kepler data. This detection limit is expressed as the moon/planet mass ratio instead of just in terms of the moon's mass for consistency with the previous equations.  We assume a terrestrial planet density for simplicity though some targets may be better described using a Neptune-like density, and that case will also be explored.  

We note that while the transit detection limit line expressed by Equation~\ref{eq:translim} may be placed at different values of $M_{m}$/$M_{p}$ for different systems, it always represents the same transit detection threshold, adjusted for stellar radius and noise levels.  We also note that successful transit detections are subject to more factors than we have included here; we use this limit as our first order guideline only; it is not a hard limit.  

\item The dashed horizontal line in Fig.~\ref{fig:earthmoonSens} is the distance from the planet where the moon's period is equal to the transit duration of the planet. A moon near or below this line moves significantly during the transit; however our simplified model assumes little or no motion of the moon relative to the planet during the transit.  A moon near or below this line may require more advanced modelling for reasons discussed in section \ref{text:periodduration}, and we will only consider systems which lie well above this line.
\end{enumerate}

The four lines described above divide the parameter space in ways which will help illustrate the properties of the different modelled moon systems, and similar diagrams based on the appropriate stellar and planetary properties of various Kepler systems will be discussed in more detail later.

\subsection{Transit Duration Variations}
Though valuable sources of information, TTVs from exomoons are subject to a degeneracy between the mass and semi-major axis of the exomoon.  Transit Duration Variations (TDVs) can be brought to bear to resolve this degeneracy \citep{kipp2009, heller2016} and we incorporated the TDVs provided by \citet{hm2016} as part of our exomoon analysis.  

All TDVs in our sample were found to be comparable in magnitude to their errors.  The ratio of standard deviation of the TDVs to the average error of the TDVs is typically just above 1 (see Table \ref{tab:tdvs} later in Section~\ref{section:targetselection}).  While small, these TDVs may still be useful by providing constraints; any proposed model that would create a large TDV could be ruled out.  The TDV signal is given as a fractional value, and can be described by:
\begin{equation}
    \frac{TDV}{\langle D \rangle} =  \frac{P_{p} M_{m} }{2\pi a_{p} } \Bigg[ \Bigg(\frac{G} {(M_{p}+M_{m})}\Bigg)\frac{ 1 + e^{2} + 2e \cos f }{a_{pm}(1-e^{2})}\Bigg]^{1/2}  \cos\theta \label{eq:tdv}
\end{equation}
where $\theta$ = $\omega + f - \phi$, and $\phi = \arctan \left( \frac{1+e \cos f}{e \sin f} \right)$,  $a_{pm}$ is the semi-major axis of the planet-moon (not moon-barycentre) orbit, $a_{p}$ is the semi-major axis of the planet around its parent star, $f$ is the true anomaly of the moon about the planet, and $\omega$ the argument of periastron of the moon's orbit.  This equation is consistent with the derivation by \citet{kipp2009}.  For comparison, our Moon produces a fractional duration variation upon Earth of 0.000418.

\section{Target Selection} \label{section:targetselection}
Having determined that there is a region of phase space where exomoons could produce TTVs without appearing above the noise level in the system's Kepler light curve, we proceed to ask whether there are any signals consistent with such exomoons in the Kepler data.  We make use of the DR25 data\footnote{Kepler DR25 release notes: https://archive.stsci.edu/missions-and-data/kepler/documents/data-release-notes} \citep{kep8cat2018}, retrieved from the NASA Exoplanet Archive \citep{nep2013}.

To find a list of targets for analysis, we start with the 2599 systems with TTVs reported by \cite{hm2016}, but restrict our analysis to 779 systems for which TDVs were also calculated. We also require at least 10 measured transits.  This is determined from the need for sufficient data to fit the parameters of our models. The planet hypothesis has the most free parameters (10). As a result, we require a minimum of 10 Kepler-observed transits so as to provide sufficient constraints to that model.  Given Kepler's primary mission lifespan, $\ge 10$ transits corresponds to a period of approximately 160 days or less (assuming no missing transits in the data). This condition effectively restricts our candidates to hotter planets orbiting relatively close to their star. This last criterion reduces our sample to 618 systems.

We will also require that the planet 1) either have a status of "confirmed" or a disposition score of 1 from NASA's Exoplanet Archive \citep{nep2013, kep8cat2018}; and 2) have no known siblings (that is, there is only one known planet in the system). Though there is no physical reason not to expect moons simply because there is another planet present, that planet may generate TTVs in and of itself, and so this second criterion simplifies our analysis, though undetected planets may still be present. With these two conditions, our sample is reduced to 272 systems. 

Here we define the signal-to-noise of the TTVs to be the ratio of standard deviation of the TTVs to their average uncertainty. These quantities are calculated  with all outliers identified by \cite{hm2016} removed. In particular the reported uncertainties in the transit timings are used, not any theoretical value derived from the short-cadence timing interval. However, we note that if there are unmodelled timing errors in the catalog, the uncertainties could be larger than reported.  To extract the strongest signals, we required the signal-to-noise to be at least 1.5, with additional tests of the TTVs statistical significance to be undertaken in a later step. This leaves us with 40 systems.

The next step is to exclude targets whose TTV signals require a moon too large to have plausibly remained undetected photometrically by Kepler.  Recall that we are assuming in this work that such large moons would have been revealed by other studies. In particular we note that \citet{hek6}, using TTVs to create phase-folded light curves to search for evidence of exomoons among the Kepler systems, did not report any findings, so any existing moons must be below Kepler's photometric threshold.

To exclude transit-visible moons, we consider the minimum moon mass required to induce the moon's TTV amplitude (by using a coplanar and circular version of equation \ref{eq:ttv}, solving for $M_{m}$) and comparing that to the photometric detectability limit of the star (using equation \ref{eq:translim}).  This comparison allows us to remove systems with TTVs too large to be consistent with undetected exomoons.  This leaves us with 15 targets.

The remaining systems are now examined more carefully for their statistical significance. Weak TTV signals could be produced spuriously by noise and these should be excluded. We run a simple Monte Carlo test. Consider a Kepler system that has $N$ transits, a typical transit timing uncertainty $\sigma$  and a signal-to-noise of $S$. Sets of $N$ random deviates were chosen from a Gaussian distribution with a standard deviation of $\sigma$, and from these the probability of achieving a signal-to-noise of $S$ by chance determined. Any systems whose TTV signals have a chance greater than 1 in 5000 of being generated spuriously are excluded (recall our initial sample size is 2599).  We find that 2 of these systems fail this test.  Of the remaining 13 targets. we can have high confidence that their TTV signal is not spurious.

The 13 systems remaining are:  KOI-63, KOI-268, KOI-303, KOI-318, KOI-1302, KOI-1472, KOI-1848, KOI-1876, KOI-1888, KOI-1925, KOI-2469, KOI-2728, KOI-3220.  These systems are summarized in Table~\ref{tab:sysprops}.  At this point, these targets have TTVs roughly consistent with those expected from unseen exomoons, but whether the TTV / TDV pattern is reproducible in detail is to be determined by our subsequent simulations.

\begin{table}
 \caption {Measured Properties of Target Systems} 
 \label{tab:sysprops}
 \begin{tabular}{llllll}
  KOI & Kepler & Spec & Star Mass & Star Radius & CDPP  \\
  ID  & ID & Type & ($M_{\sun}$) & ($R_{\sun}$) & (ppm)  \\
  \hline
  63.01 & 63b & G5 & $0.943\substack{+0.051\\-0.069}$  &  $0.886\substack{+0.120\\-0.051}$  & 63.5    \\ \\  
  268.01 & & F7$^{*}$ & $1.175\substack{+0.058\\-0.065}$  &  $1.359\substack{+0.062\\-0.068}$  &  25.6  \\ \\
  303.01 & 517b & G6V & $0.871\substack{+0.071\\-0.043}$  &  $1.023\substack{+0.142\\-0.142}$  &  38.1 \\ \\
  318.01 & 522b & F7$^{*}$ & $1.486\substack{+0.126\\-0.154}$  &  $1.927\substack{+0.353\\-0.431}$  & 90.8   \\ \\
  1302.01 & 809b &G0$^{*}$ & $1.050\substack{+0.124\\-0.138}$  &  $0.962\substack{+0.297\\-0.099}$  & 95.7 \\ \\
  1472.01 & 857b & G5$^{*}$& $0.966\substack{+0.050\\-0.055}$  &  $0.938\substack{+0.127\\-0.054}$  & 117.7  \\ \\
  1848.01 & 978b & F6IV & $1.097\substack{+0.073\\-0.066}$  &  $1.184\substack{+0.193\\-0.123}$  & 61.9  \\ \\
  1876.01 & 991b & K5$^{*}$ & $0.584\substack{+0.031\\-0.027}$  &  $0.580\substack{+0.026\\-0.029}$  & 168.9  \\ \\
  1888.01 & 1000b & F6IV &  $1.406\substack{+0.086\\-0.086}$  &  $1.467\substack{+0.24\\-0.111}$  & 97.4 \\ \\
  1925.01 & 409b & K0 &  $0.902\substack{+0.050\\-0.055}$  &  $0.888\substack{+0.036\\-0.036}$  & 109.7 \\ \\  
  2469.01 &  & K2$^{*}$ & $0.774\substack{+0.048\\-0.028}$  &  $0.803\substack{+0.028\\-0.060}$  & 143.1  \\ \\
  2728.01 & 1326b & F4$^{*}$ & $1.535\substack{+0.219\\-0.267}$  &  $2.632\substack{+0.471\\-0.875}$  &  100.7 \\ \\
  3220.01 & 1442b & F7$^{*}$ & $1.323\substack{+0.098\\-0.088}$  &  $1.401\substack{+0.263\\-0.132}$  &  73.7 \\ \\
  \hline
 \end{tabular} \\
  All values are from \citet{math2017}, except Spectral Types are from Simbad \citep{simbad2000}.  Spectral Types indicated with an * are estimates based on effective temperature. CDPP values are interpolated from Kepler Stellar Tables to the specific transit duration of the planet.
\end{table}

\begin{table}
 \caption{Planet Properties Estimates}
 \label{tab:radiusMassEst}
 \begin{tabular}{lllll}
   & & & & Avg \\
   & & & & TTV \\
   &  Radius & Mass & Average Period & Error \\
  KOI & ($R_{\earth}$) & ($M_{\earth}$) & (days) & (min) \\
  \hline
  63.01 & $5.89\substack{+0.57\\-0.52}$ & $28.84\substack{+23.64\\-12.62}$ & 9.434$\pm$0.00004 & 9.38\\ \\
  268.01 & $3.02\substack{+0.14\\-0.14}$ & $9.33\substack{+7.65\\-4.08}$ & 110.37838$\pm$0.00069 & 3.10\\ \\
  303.01 & $2.57\substack{+0.42\\-0.23}$ & $7.59\substack{+6.21\\-3.42}$ & 60.92833$\pm$0.00018 & 3.11\\ \\
  318.01 & $6.17\substack{+1.42\\-0.92}$ & $32.36\substack{+36.82\\-15.38}$ & 38.584780$\pm$0.000086 & 2.39\\ \\
  1302.01 & $3.24\substack{+0.75\\-0.48}$ & $10.96\substack{+9.93\\-5.08}$ & 55.639286$\pm$0.000589 & 10.63\\ \\
  1472.01 & $6.76\substack{+0.65\\-0.59}$ & $38.02\substack{+34.42\\-17.13}$ & 85.351419$\pm$0.000191 & 2.34\\ \\  
  1848.01 & $2.69\substack{+0.47\\-0.29}$ & $8.13\substack{+7.01\\-3.66}$ & 49.622065$\pm$0.000426 & 9.38\\ \\    
  1876.01 & $2.39\substack{+0.17\\-0.11}$ & $6.61\substack{+4.87\\-2.81}$ & 82.534607$\pm$0.000595 & 6.61\\ \\  
  1888.01 & $4.68\substack{+0.57\\-0.51}$ & $19.95\substack{+16.36\\-8.99}$ & 120.01918$\pm$0.000650 & 5.21\\ \\
  1925.01 & $1.0\substack{+0.05\\-0.05}$ & $1.00\substack{+0.78\\-0.34}$ & 68.95832$\pm$0.00045 & 5.01\\ \\
  2469.01 & $2.40\substack{+0.17\\-0.16}$ & $6.61\substack{+5.14\\-2.81}$ & 131.187139$\pm$0.002623 & 15.50\\ \\  
  2728.01 & $5.25\substack{+1.51\\-0.98}$ & $24.55\substack{+26.74\\-11.96}$ & 42.35120$\pm$0.00035 & 7.26\\ \\  
  3220.01 & $3.80\substack{+0.57\\-0.41}$ & $14.13\substack{+12.17\\-6.37}$ & 81.41635$\pm$0.00042 & 4.68\\ \\
  \hline
 \end{tabular} \\
  Periods and average TTV errors computed using data from \citet{hm2016}.  Radius and Mass estimates (including 1$\sigma$ errors) from \citet{ck2018} 
\end{table}

\begin{table}
 \caption {Signal to Noise of Target Systems TTVs and TDVs} 
 \label{tab:tdvs}
 \begin{tabular}{lllllll}
   & TTV & TTV & & TDV & TDV &   \\
   & Std Dev & Avg Err & TTV & Std Dev & Avg Err & TDV \\ 
   KOI & (min) & (min) & SNR & (min) & (min) & SNR \\
  \hline
  63.01 & 23.10 & 9.38 & 2.46 & 0.080 & 0.058 & 1.38\\ 
  268.01 & 7.33 & 3.10 & 2.37 & 0.019 & 0.009 & 2.16\\
  303.01 & 4.85 & 3.11 & 1.56 & 0.020 & 0.018 & 1.08 \\
  318.01 & 9.11 & 2.39 & 3.81 & 0.015 & 0.008 & 1.87 \\
  1302.01 & 19.02 & 10.63 & 1.79 & 0.055 & 0.056 & 0.97 \\
  1472.01 & 3.88 & 2.34 & 1.66 & 0.023 & 0.013 & 1.74 \\
  1848.01 & 23.10 & 9.38 & 2.46 & 0.080 & 0.058 & 1.38 \\
  1876.01 & 20.70 & 6.61 & 3.13 & 0.145 & 0.054 & 2.66 \\
  1888.01 & 9.56 & 5.21 & 1.84 & 0.019 & 0.016 & 1.13 \\
  1925.01 & 7.87 & 5.01 & 1.57 & 0.061 & 0.064 & 0.96 \\
  2469.01 & 37.45 & 15.50 & 2.42 & 0.139 & 0.099 & 1.40 \\
  2728.01 & 12.38 & 7.26 & 1.71 & 0.045 & 0.035 & 1.31 \\
  3220.01 & 7.82 & 4.68 &1.67 & 0.021 & 0.014 & 1.48 \\
  \hline
 \end{tabular} \\
  All values computed from \citet{hm2016} data. 
\end{table}

\begin{figure*}
 \caption{Transit Timing Variations (TTVs) and Transit Duration Variations (TDVs) of Target Systems. } 
 \label{fig:basettvs}
 \begin{tabular}{lll}
 \includegraphics[scale=0.37]{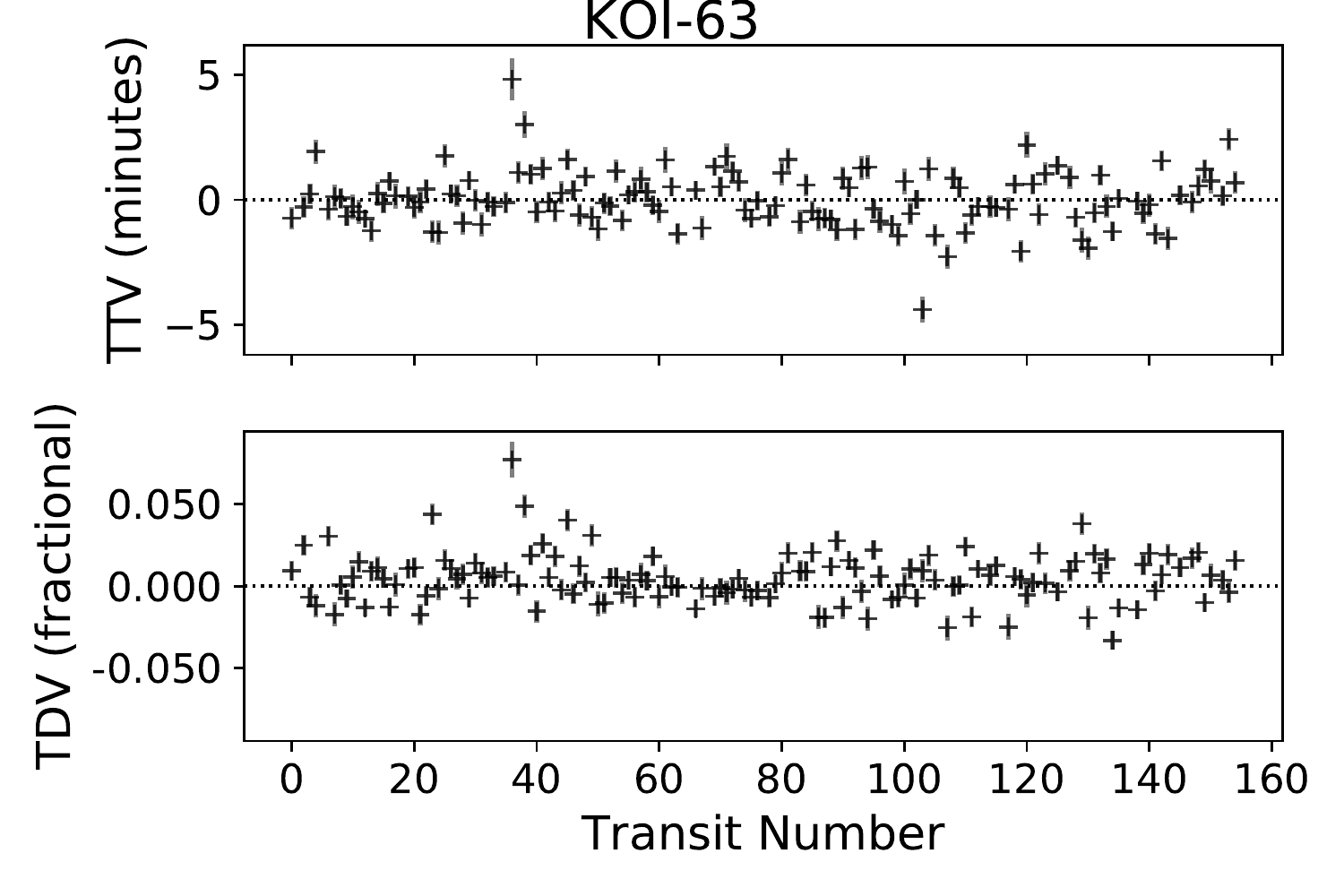} & \includegraphics[scale=0.37]{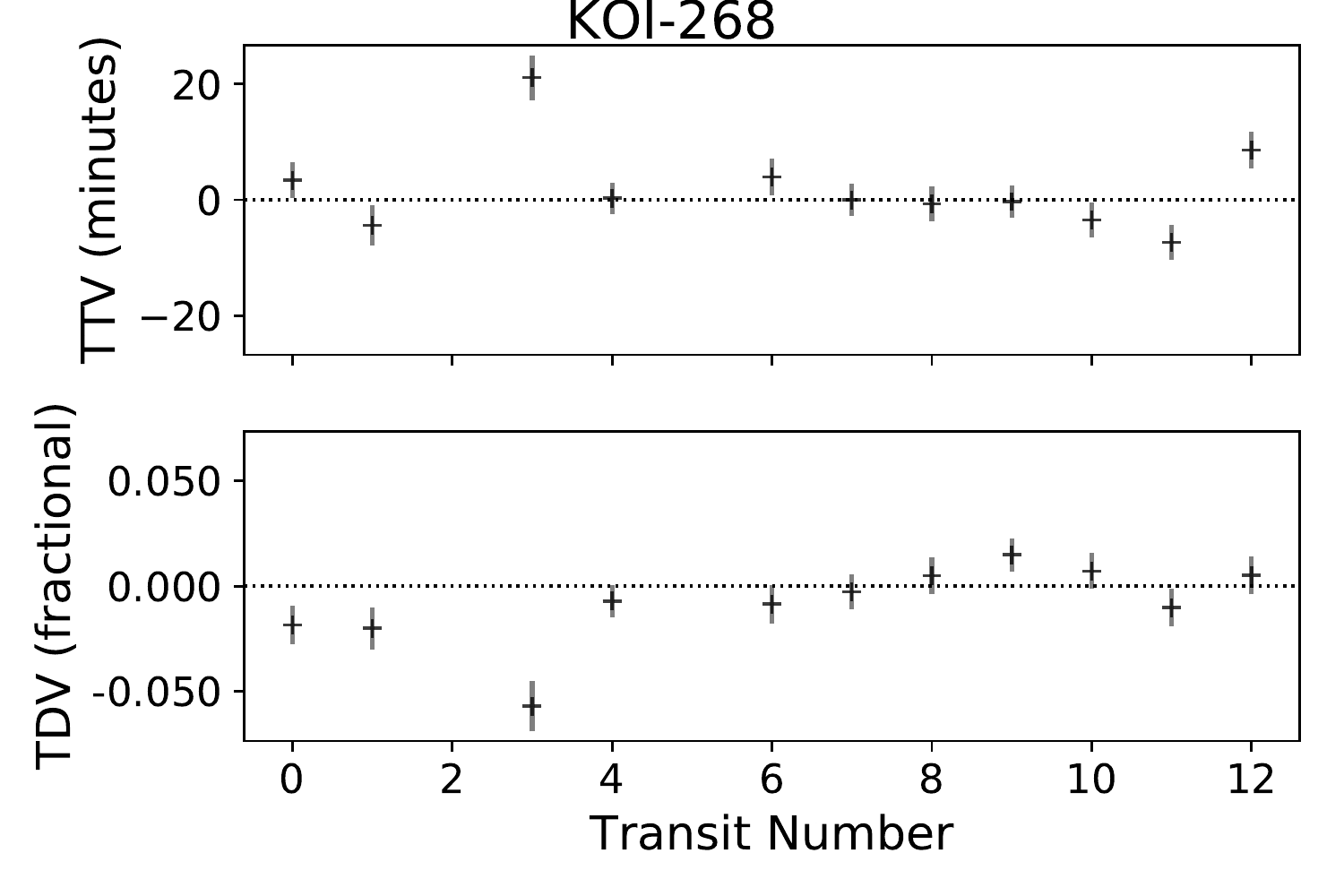} & \includegraphics[scale=0.37]{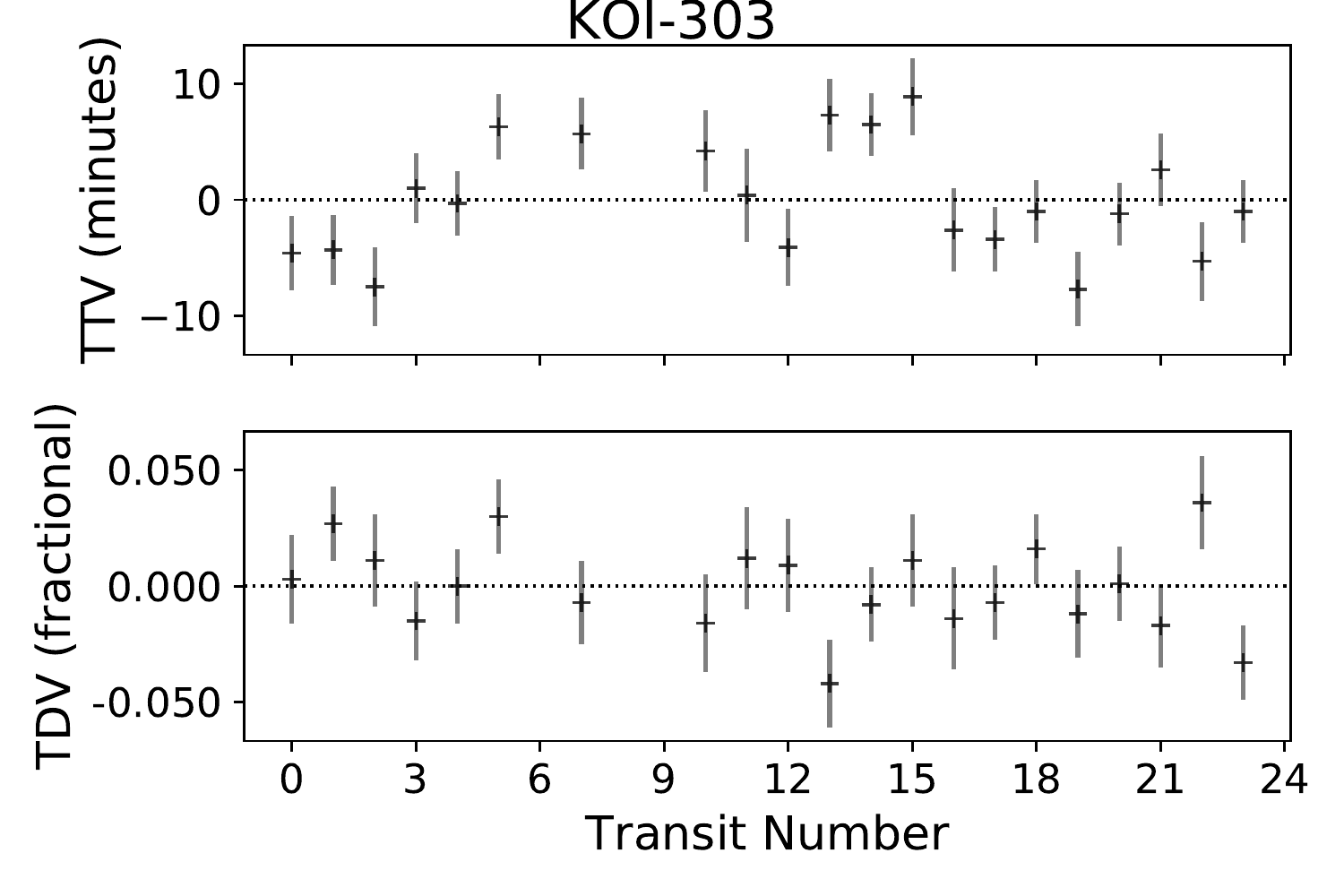} \\ 
 \includegraphics[scale=0.37]{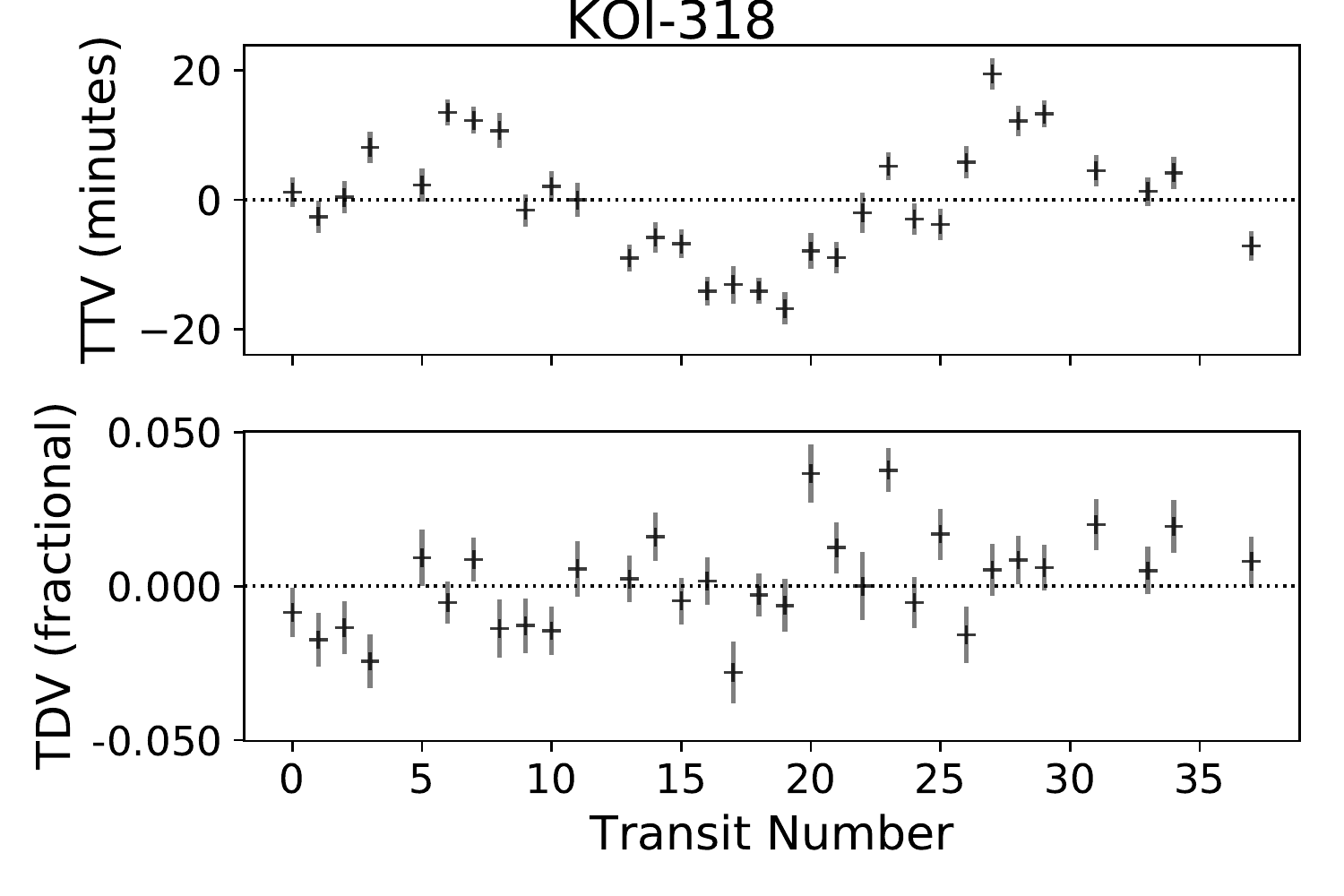} & \includegraphics[scale=0.37]{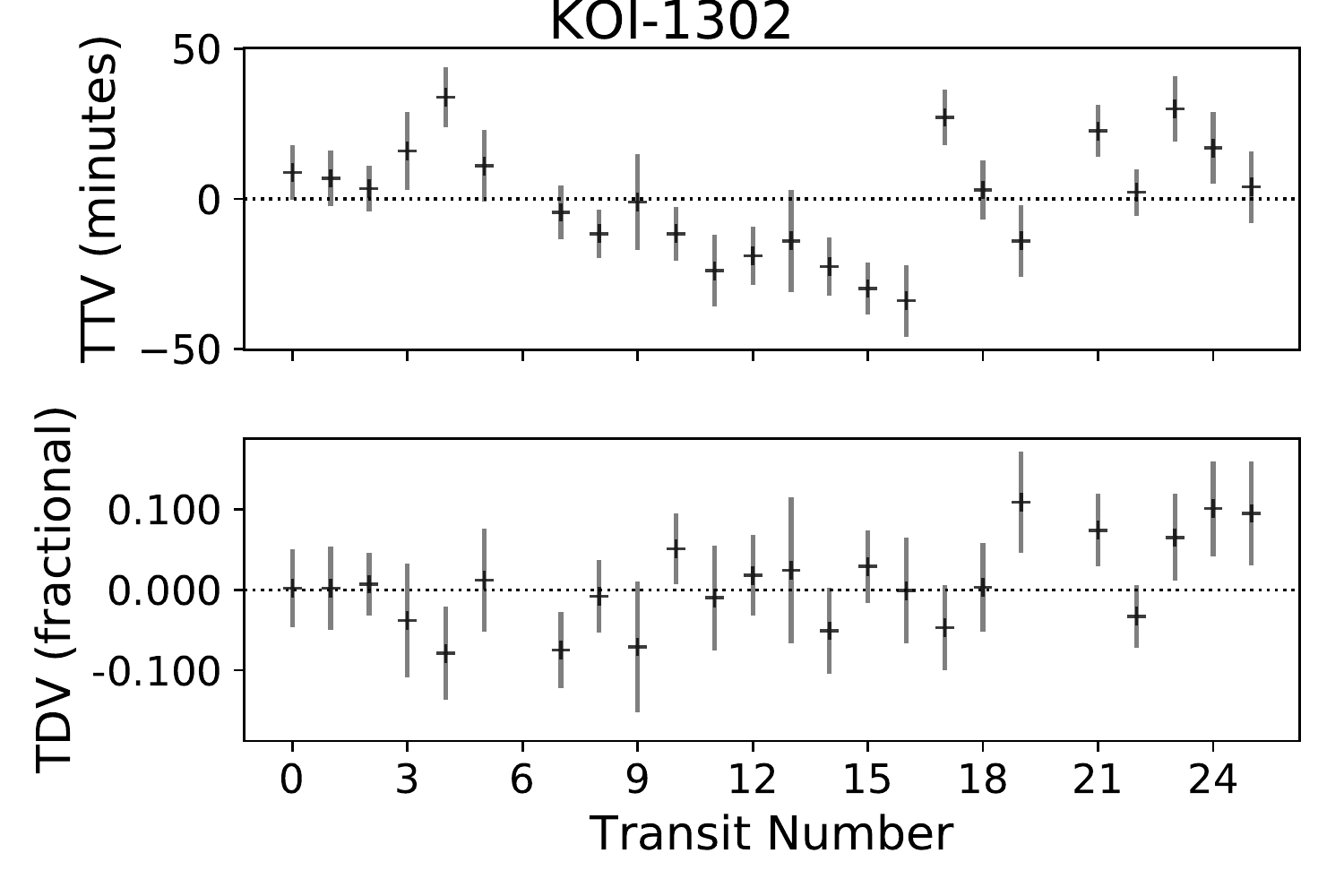} & \includegraphics[scale=0.37]{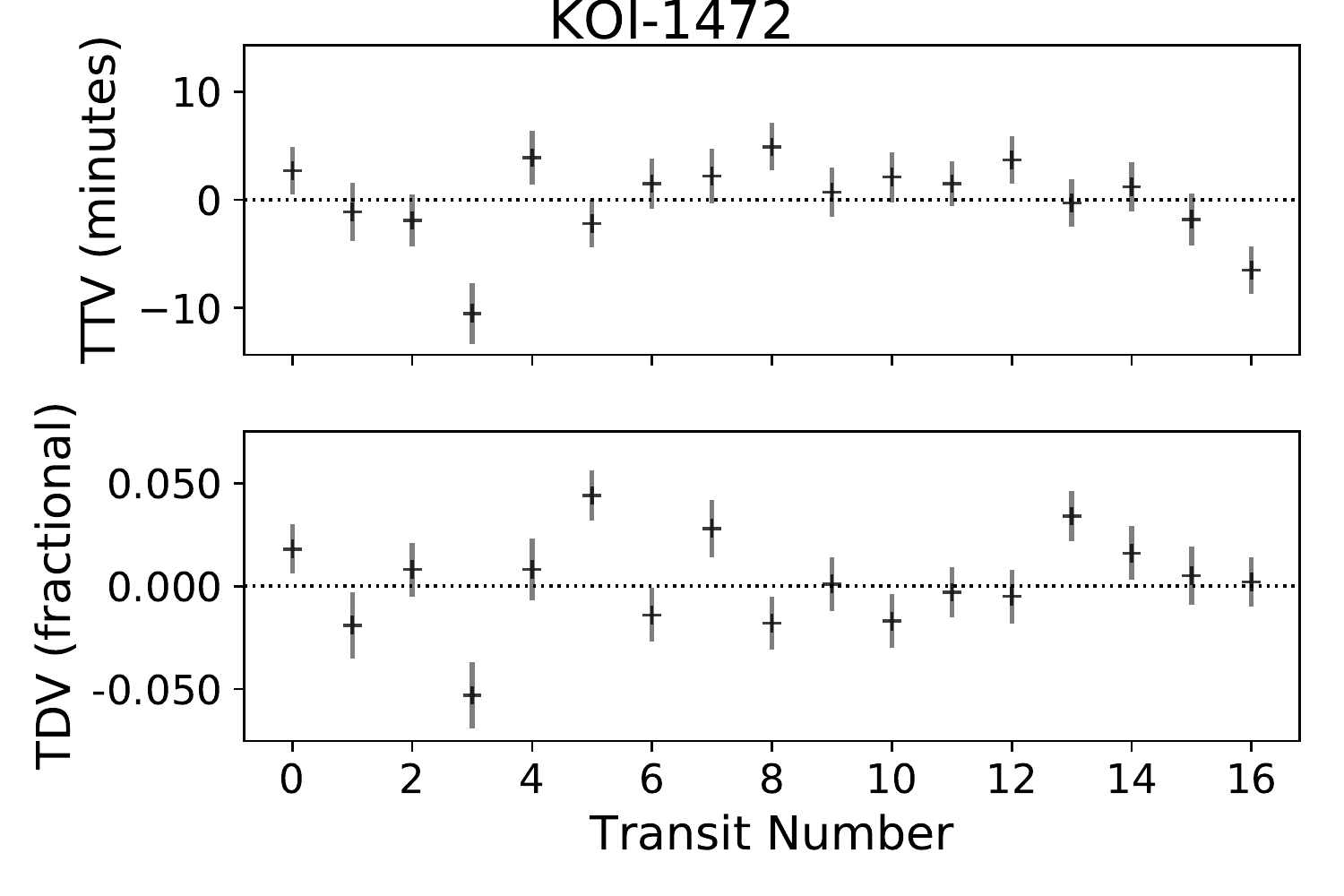} \\
 \includegraphics[scale=0.37]{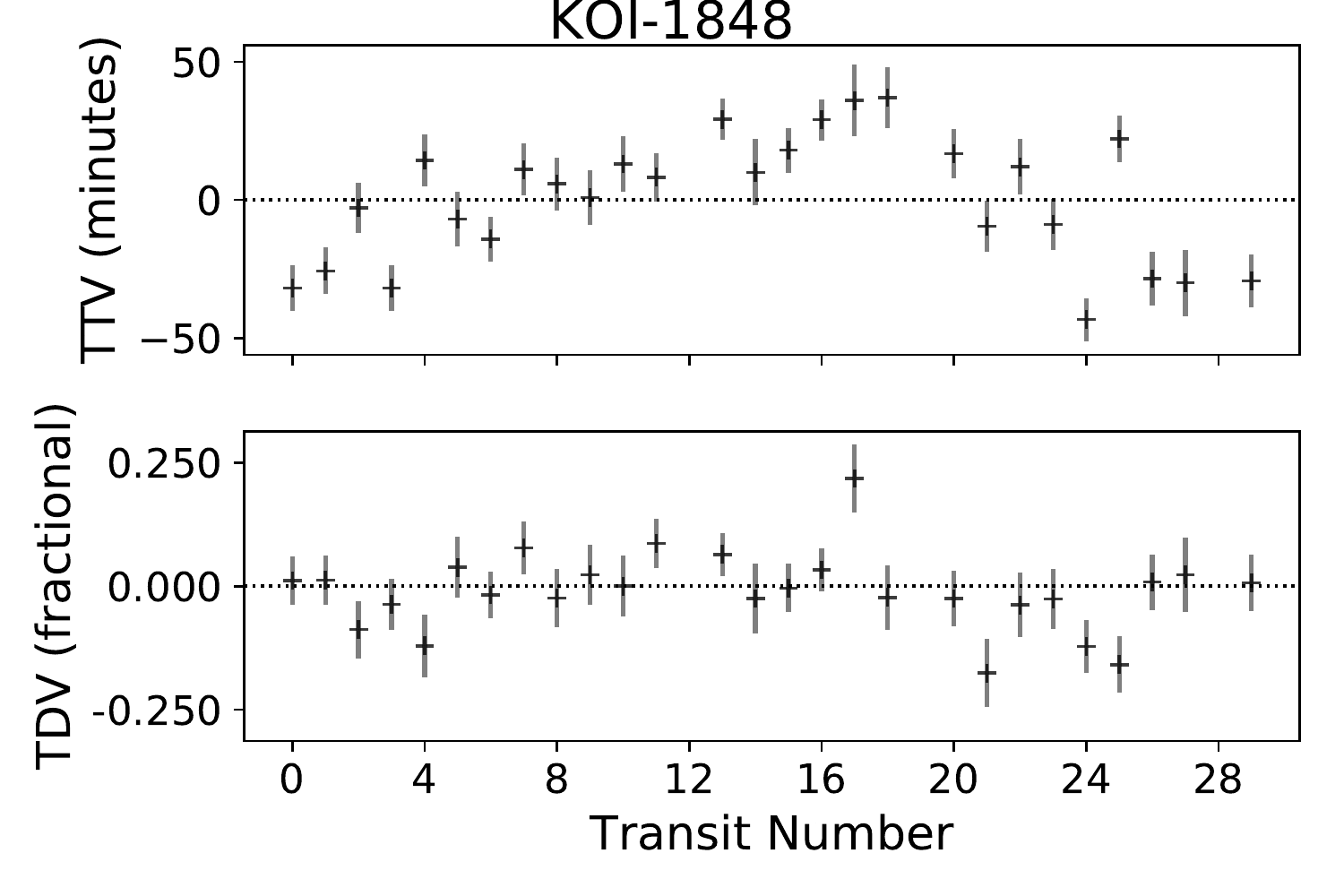} & \includegraphics[scale=0.37]{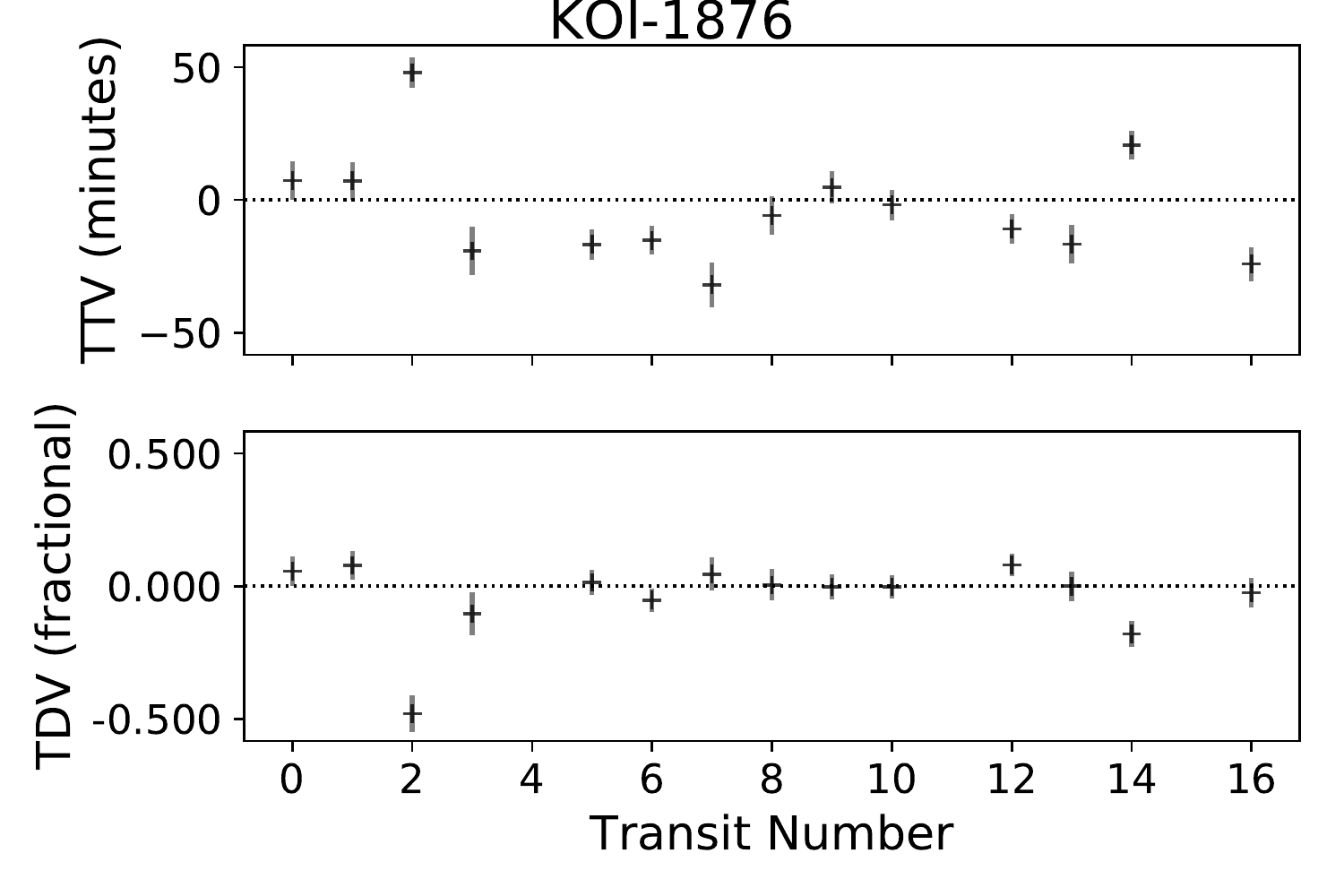} & \includegraphics[scale=0.37]{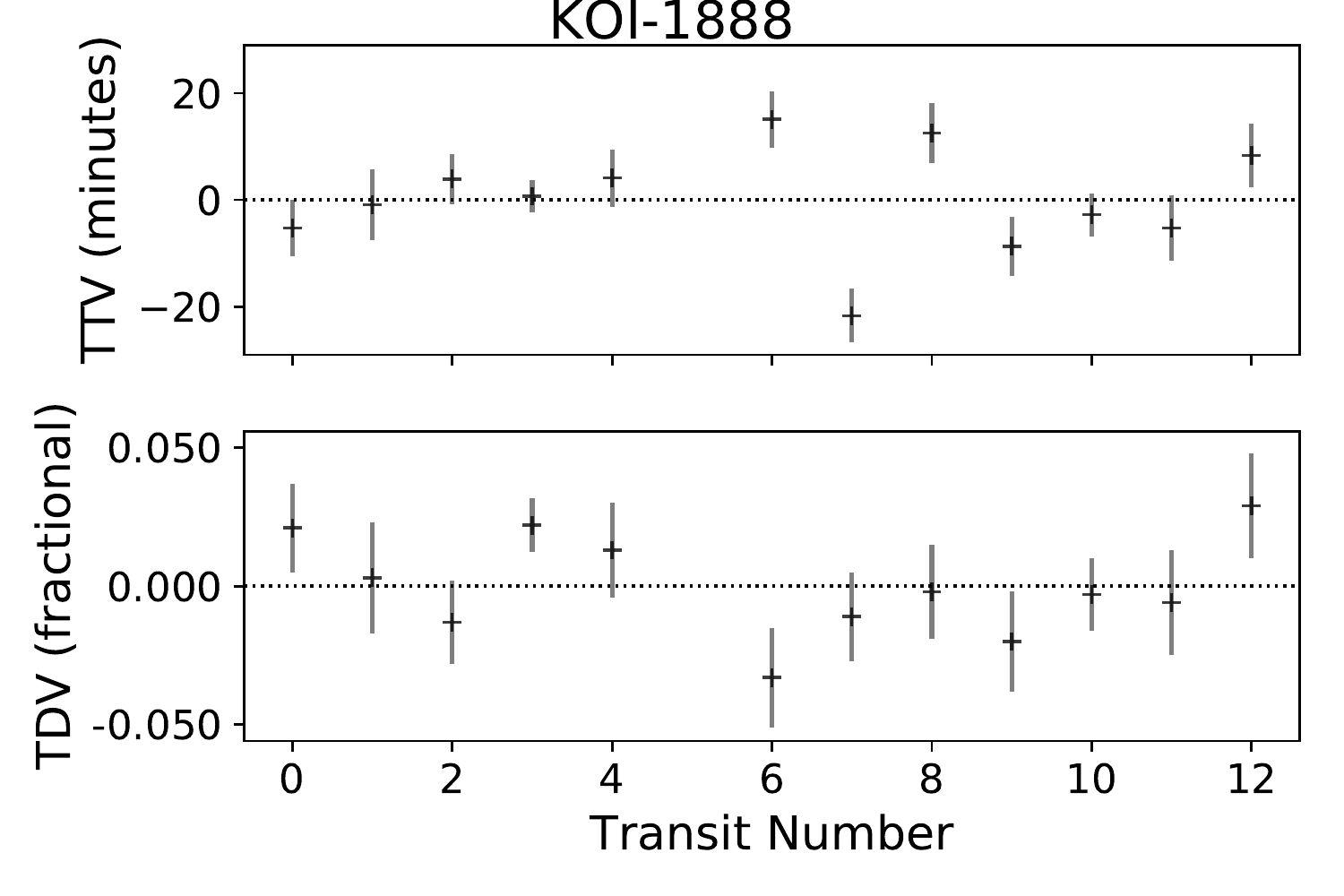} \\
 \includegraphics[scale=0.37]{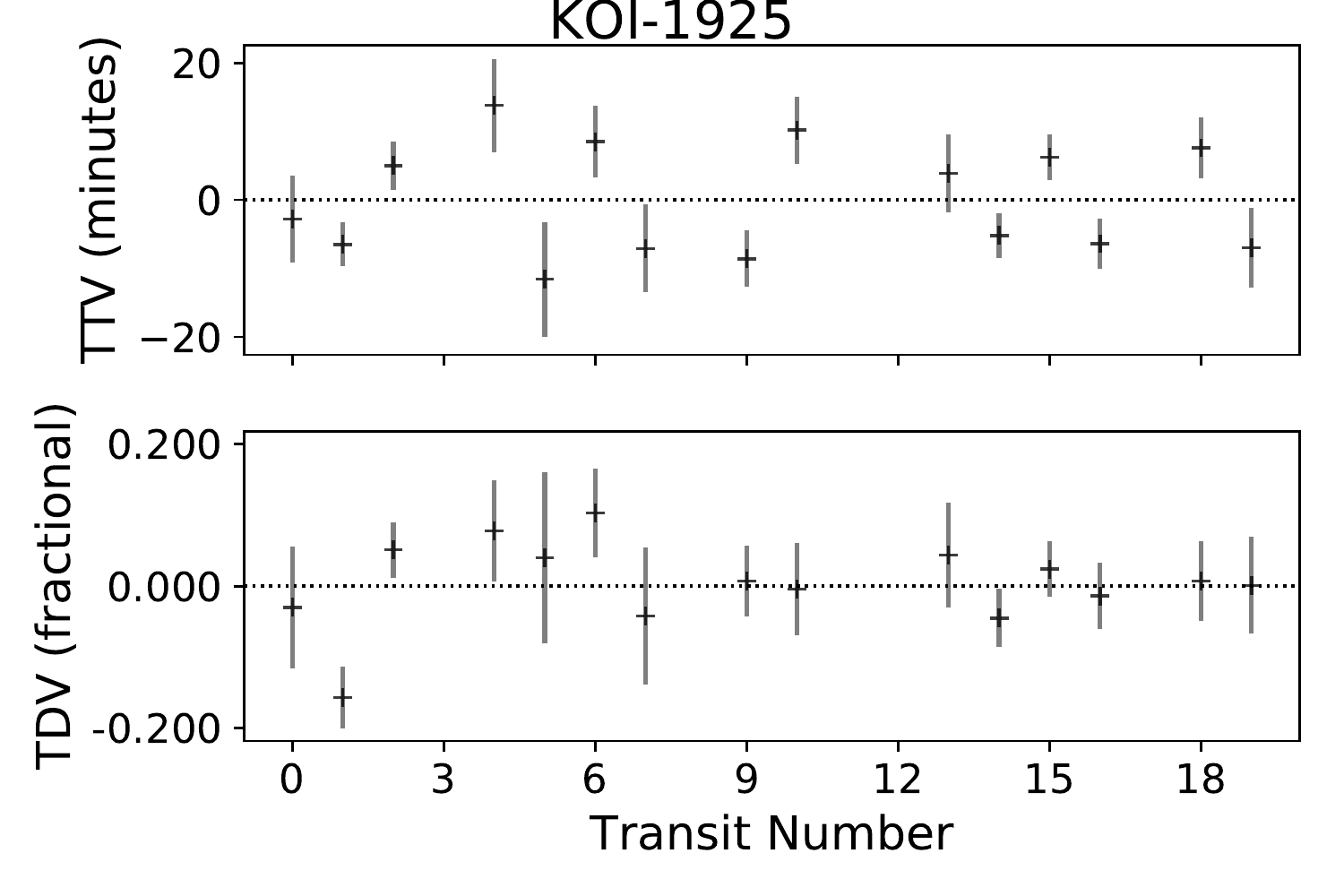} & \includegraphics[scale=0.37]{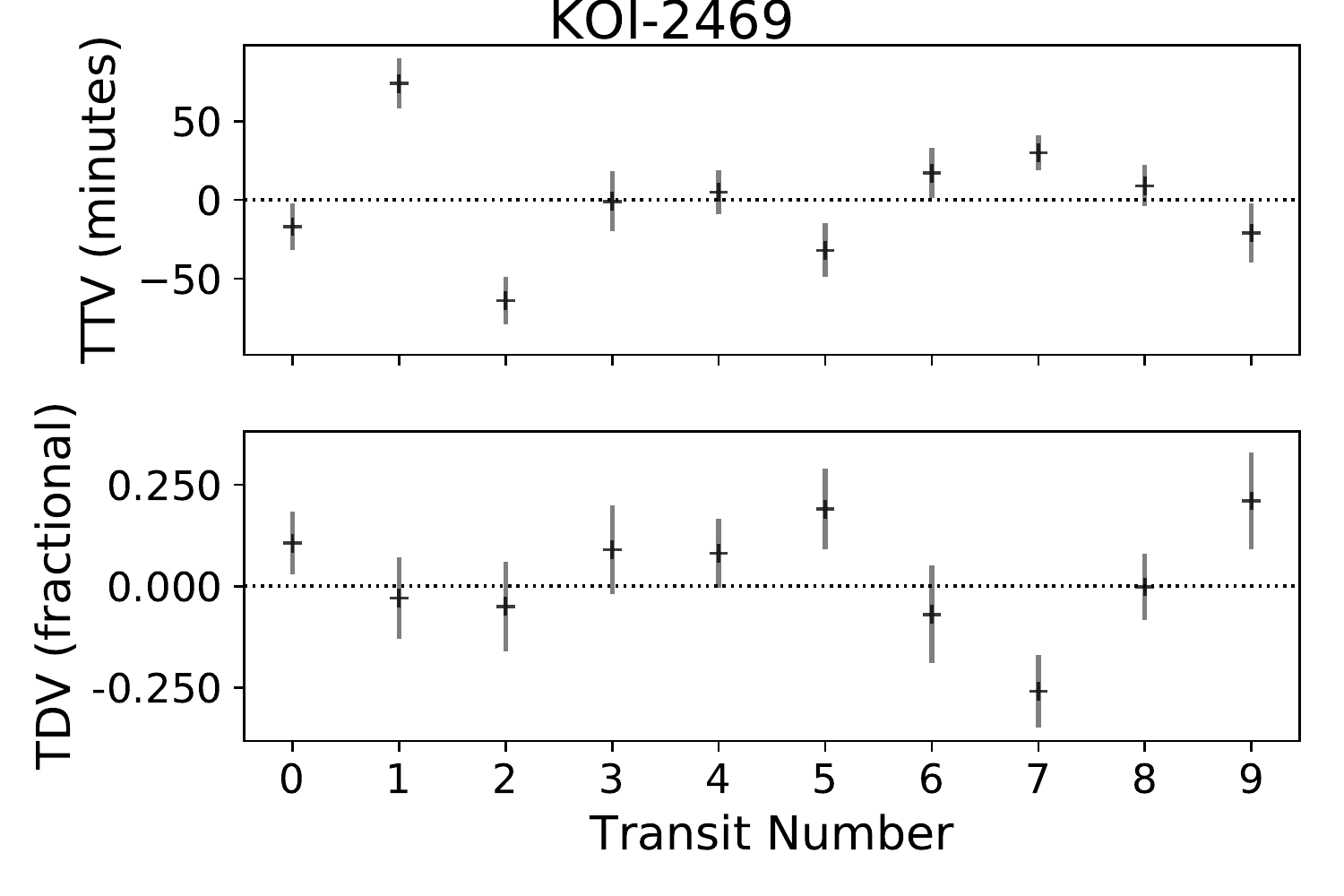} & \includegraphics[scale=0.37]{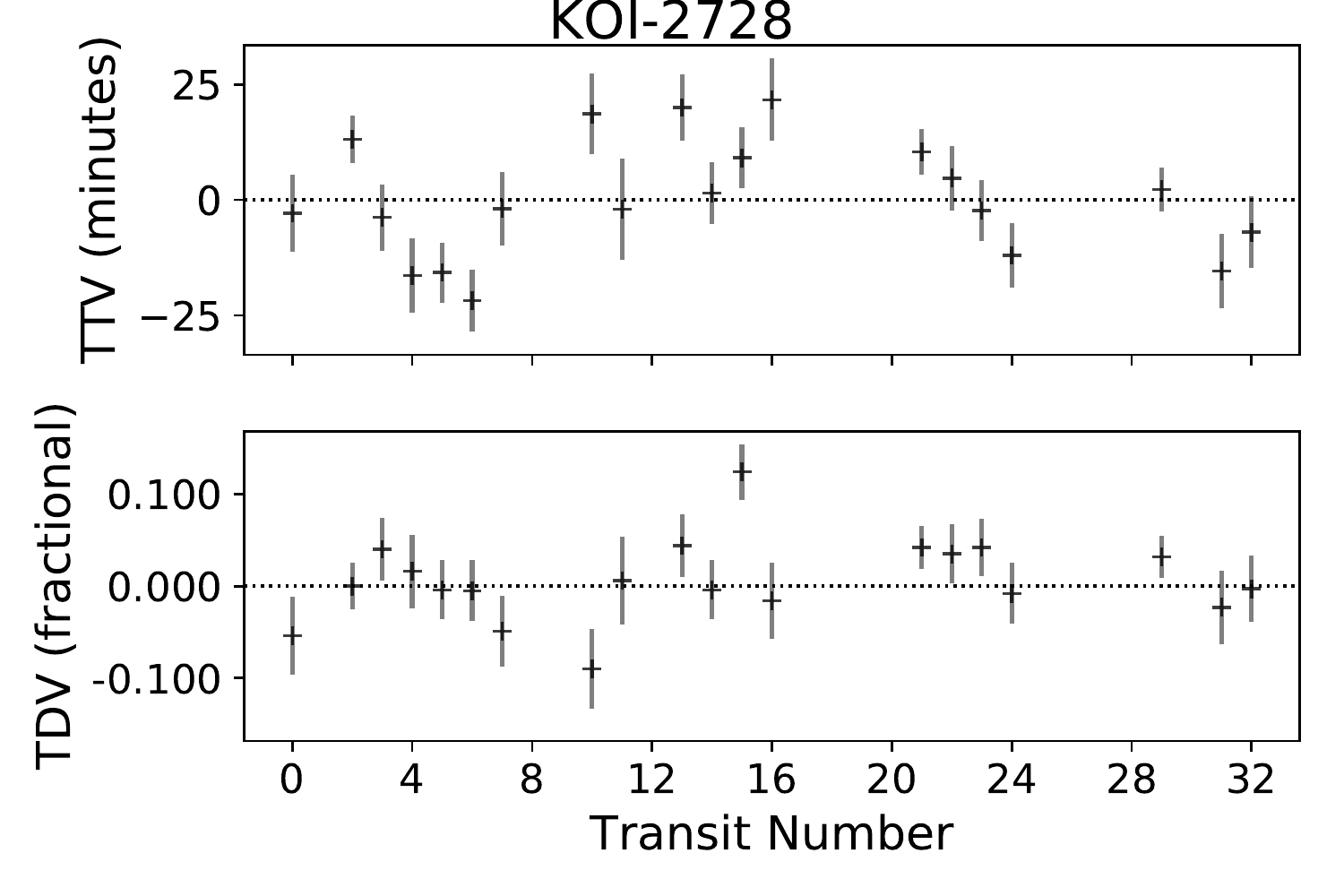} \\
 \includegraphics[scale=0.37]{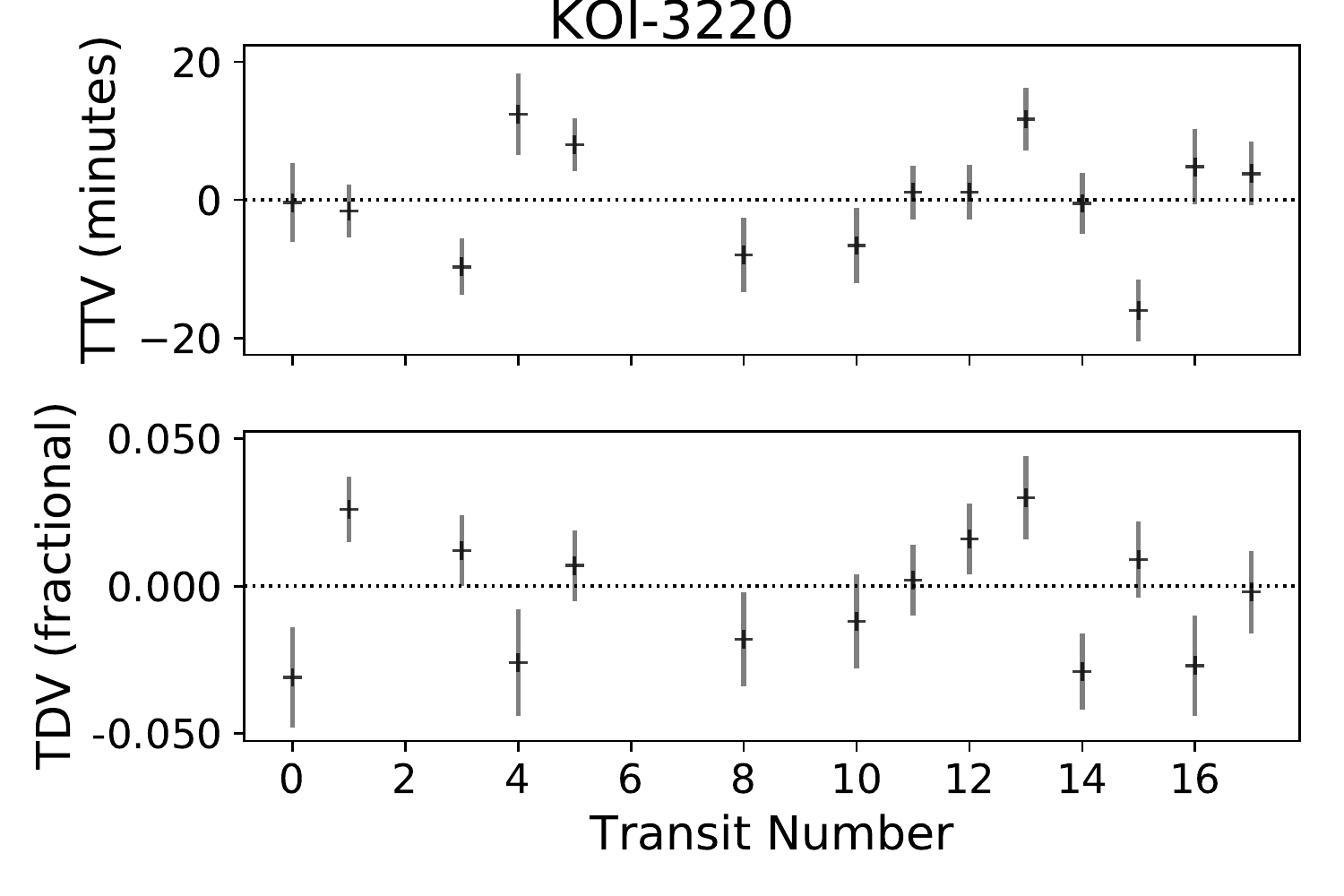} 
 \end{tabular}
\end{figure*}

\section{Methods and Setup} \label{section:methods}
\subsection{Simulating Systems and Finding Parameters}
To examine the hypothesis that the TTVs and TDVS observed by Kepler were produced by an exomoon, we model the TTVs and TDVs induced by either one or two exomoons in orbit around the planet. Our analysis employed two publicly available software packages.  The first was TTVFast \citep{da2014} which simulates the orbits of the planets around a star and  calculates the TTVs resulting from planetary gravitational interactions. This code was used to assess the competing hypothesis that the observed TTVs were induced by a non-transiting planet.  The second package was MultiNest \citep{feroz2009}, (which we used via its Python interface, PyMultiNest \citep{buch2014}), a Bayesian Inference tool which we used to search the parameter space for possible solutions for both the exomoon and exoplanet hypotheses.  

For each system, the observed transit times and durations come from \citet{hm2016}. Those points (and only those points) they flag as outliers are removed. The quality of fit for each simulation is based on the usual $\chi^2$ value, which is converted to a log-likelihood value for MultiNest.  For the exomoon hypothesis, both the TTVs and TDV were fitted. For the exoplanet hypothesis only the TTVs were fitted, as these were sufficient to demonstrate the plausibility of an additional planet as a competing hypothesis. 

\subsection{System Stability}
\label{sec:stability}
As part of our analysis, systems that showed reasonable TTV and/or TDV fits for either the exomoon or additional planet models were further tested for long-term stability.  The stability simulation codes used different algorithms for the case where only planets were included, and ones that included moons. The code used for the purely planetary case is a symplectic one based on the Wisdom-Holman algorithm \citep{wishol1999}.  This code uses a timestep less than 1/20th of the period of the innermost planet in all cases, and includes post-Newtonian general relativistic effects. This is the same code as used in \cite{foxwie19}. In cases where the stability of moons is examined, the RADAU15 \citep{eve85} algorithm is used, with a tolerance of $10^{-12}$.

These stability studies provide an additional check on our results, as some planet and/or moon parameter values which provide good matches to the TTVs over the course of Kepler's lifetime may be unstable on longer times, and are thus unlikely to represent the real configuration of these systems. All planet hypothesis results discussed in this work were found to be in stable configurations over 10 million years, so we cannot exclude the additional planet hypothesis on the basis of system instability.  The single-moon hypothesis results discussed later were all found to be stable for at least 100 (Earth) years, which corresponds to 300,000 to 5 million moon orbits, depending on the system.  

Stability is of particular concern with regards to the two-moon models which we ultimately did not pursue. The analyses of \cite{gla93} and \cite{chawetbos96} on the stability of multi-planet systems are likely approximately applicable here. Even though the stability of moons is really quite a different problem, our restriction to moons orbiting inside 0.3 $R_{Hill}$ means that stability results for planetary systems are likely to provide a useful guide. Those authors find that stability (more precisely Hill stability, that is the absence of close encounters, but in practice these encounters result in the ejection of one or both of the moons) of a two moon system is only expected where the moons are more than 2$\sqrt{3}$ mutual Hill radii apart.  Combining the planetary 0.3 $R_{Hill}$ condition with the lunar 2$\sqrt{3}$ mutual Hill radii  results in a significant restriction to our model. The need for the moon to generate significant TTVs tends to favour models with a large moon near 0.3 $R_{Hill}$, and the resulting large mutual Hill radius forces the second moon to be very near the planet. The TTVs then are primarily driven by the outer moon, not dissimilar to the single moon scenario and providing little improvement to the fit.  In addition, this configuration always proved to be rapidly unstable. It is conceivable that configurations of moons in mean-motion resonance with each other could stabilize themselves, we do not examine resonant configurations here.  While multiple exomoons could certainly exist around exoplanets, the size required to produce the TTVs of our sample systems preclude the existence of multiple {\it massive} moons, and we do not examine the multiple moon scenario further.

\subsection{Parameters and Priors}
The two models (exomoon vs additional planet) have a different set of priors and allowed parameter ranges.  The mass of the known transiting planet is taken to be fixed in both cases, with the nominal mass taken from \citet{ck2018}.

\subsubsection{Exomoon hypothesis priors}
When considering the exomoon model, the planet is taken to have a circular orbit around the star.  The moon is assumed to orbit the planet in the same plane that the planet does the star; any difference between these planes results in a mass-inclination degeneracy. Thus, our derived mass results can be considered as minimum masses. The moon is also taken to orbit in the same (prograde) direction as the planet. Similar TTVs and TDVs could be created by a retrograde moon and such moons could be stable out to larger radii (see Section~\ref{ttvs-vs-transits}). Nevertheless, we choose prograde moons as the more likely and more conservative assumption, since we cannot distinguish the two cases from our data.

The other parameters are the mass of the moon, its semi-major axis, mean anomaly, eccentricity, and argument of periastron, for a total of 5 parameters. The moon is allowed to have a non-circular orbit, but stellar gravitational perturbations are ignored; its orbit is considered fixed. The stability simulations of exomoon candidates (discussed in Section \ref{sec:stability} and \ref{section:results}) showed only small changes to the moon orbits during the time examined, so this assumption is valid.   

For the moon hypotheses, the moon mass prior was uniform from zero though to a maximum value equal to the planet's mass.  While this choice runs against some of our actual prior knowledge about the system, that is, that moons have not been detected photometrically within them despite extensive searches, it ensures we cover the full range of possible masses.  Because of the degeneracy between the moon's mass and semi-major axis, we represent the greater likelihood of a smaller and farther-out moon through a triangular prior on the semi-major axis.  Such a prior also assists in keeping the moon above the duration-period limit where our model would break down.  This triangular semi-major axis prior has a probability density of zero at the planet, and a linearly increasing probability density up to a maximum at 0.3 Hill (this latter limit is chosen for reasons of stability as discussed in Section~\ref{ttvs-vs-transits}). Note that this choice of prior does not affect the quality (that is, the $\chi^2$) of any particular fit, though it does influence MultiNest's choices and the resulting posteriors towards larger $a_{pm}$. 

The prior distributions for the remaining moon orbital elements were all uniform. Eccentricity was allowed to go as high as 0.5, and the angular elements could run from 0 to 360 degrees.

\subsubsection{Exoplanet hypothesis priors}
\label{sec:planethyppri}
When examining the additional planet model, there are a total of 10 parameters.  Each planet has 7 parameters: 6 orbital elements plus its mass.  The known transiting planet has 3 parameters known to high precision: the period, inclination and longitude of the ascending node, and we use a fixed mass, the nominal value from \cite{ck2018}, to remove one additional parameter.  The inclination (with respect to the planet of the sky) must be near 90\textdegree\ or else a transit would not be observed.  Slight deviations in inclination have minimal effect on the observed TTVs \citep{agol2005} so we set the inclination to 90\textdegree\ for the known planet.  Finally, the longitude of the ascending node, while not known in a true sense, can be set as our reference orientation of 0\textdegree, leaving 3 orbital elements.  The second hypothesized new planet has nothing known about it, so it has 7 parameters to be fit: 6 orbital parameters plus its mass.  This means a total of 10 parameters to fit the additional planet hypothesis.  In all cases, the proposed new planet had a period prior ranging from 1 d to beyond the 4:1 resonance outside of the known planet.  The mass prior upper limit was 1500 $M_{\earth}$ (approximately 5 Jupiter masses).  The mean anomaly prior ran from 0\textdegree\ to 360\textdegree.  The eccentricity prior ran from 0 to 0.5.  The ascending node prior was allowed to vary uniformly from -45\textdegree\ to +45\textdegree\.  The inclination prior was allowed to vary from 45\textdegree\ to 135\textdegree\ and was uniform in $i$ (not $\cos i$).  All other priors were uniform.

\section{Results} \label{section:results}
Multinest parameter-fitting simulations were performed for each model (exomoon and additional planet) at least three times for each system, to help ensure we found the best solution and not a local minimum.  Here we report both the best-fit results as well as the Bayesian posteriors for the runs that resulted in the lowest $\chi^{2}$. The model $\chi^2$ values are occasionally less than (but always of order) unity, suggesting 'over-fitting' (either too many parameters or underestimated errors) in some cases. 

Each set of model parameters (for both the additional planet and exomoon hypotheses) also had to be stable in long term dynamical simulations (see Section~\ref{sec:stability}).  Configurations that were not stable were to be disregarded, but all exomoon and additional planet models reported on here proved dynamically stable over the time scales tested. 

Eight systems are assessed to be consistent with exomoon-generated TTVs, with five others being excluded for various reasons to be discussed below. The observed TTVs, TDVs and associated errors for all thirteen systems examined are shown in Figure \ref{fig:basettvs}.  Comparison of the modelled TTVs with the observations are shown separately for each candidate (in order of KOI number) below. 

Full results for both best fit parameters and Bayesian posteriors are included in the appendix.
\subsection{Best exomoon candidate systems}
\subsubsection{KOI-268}
\begin{tabular}{rlrl}
    Spectral Type & F7 & Planet Period & 110.38 d \\
    Star Radius & 1.36 $R_{\sun}$ & Planet Radius & 3.0 $R_{\earth}$ \\
    Star Mass & 1.18 $M_{\sun}$ & Planet Mass & 9.3 $M_{\earth}$ \\
    CDPP (12.2 hr) & 25.6 ppm  & Avg TTV Err & 3.1 min \\ 
\end{tabular}
\\ \\
KOI-268 is an unconfirmed target (and hence has no Kepler designation), but has a disposition score of 1 from NASA's Exoplanet Archive \citep{nep2013, kep8cat2018}, indicating there is high confidence that this is an actual planet.  It has one of the highest SNR (standard deviation / average error) in both its TTVs and TDVs.  It also orbits the least noisy star in our sample, with a CDPP of less than 26 ppm.

\begin{figure}
 \includegraphics[scale=0.6]{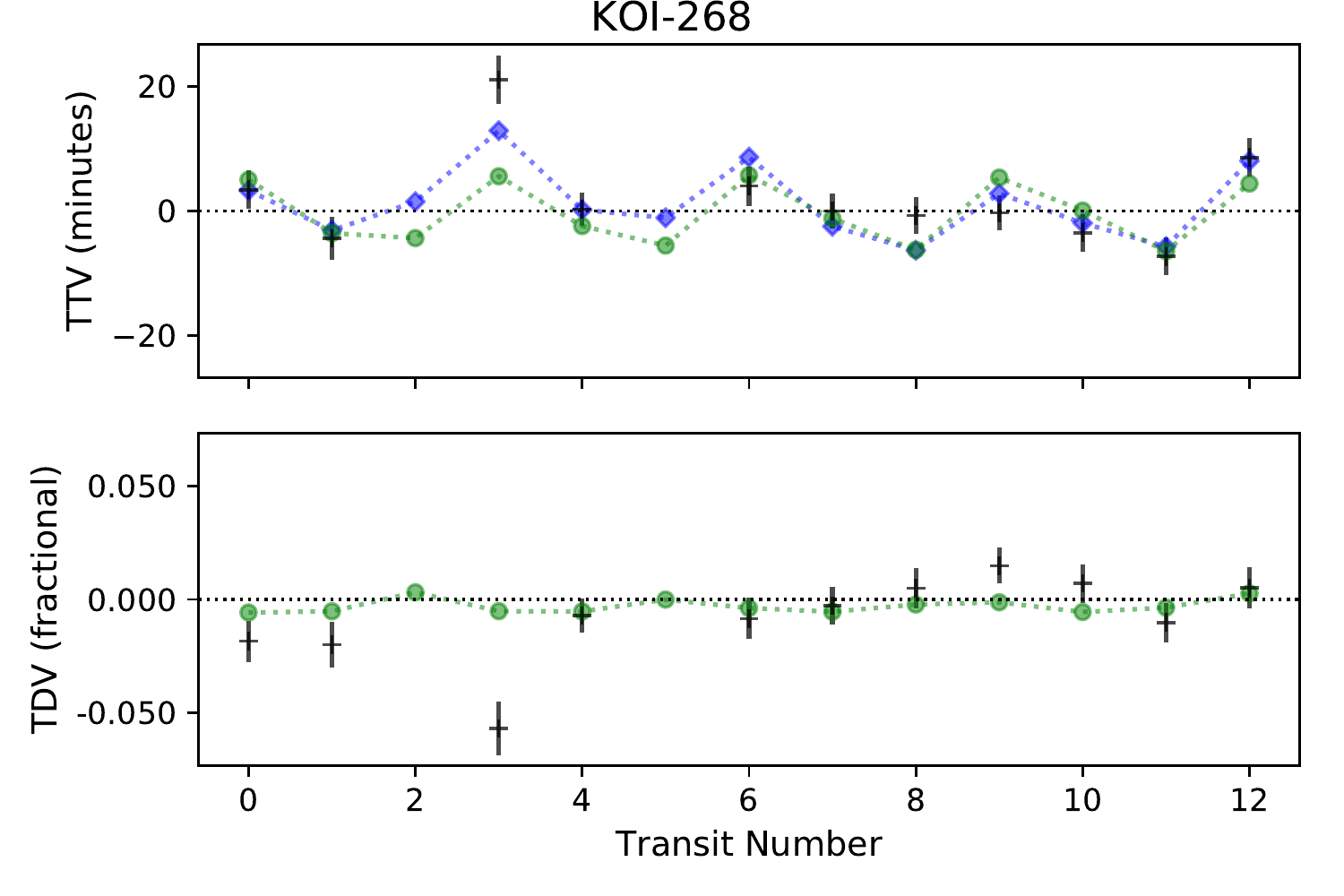}
 \includegraphics[scale=0.6]{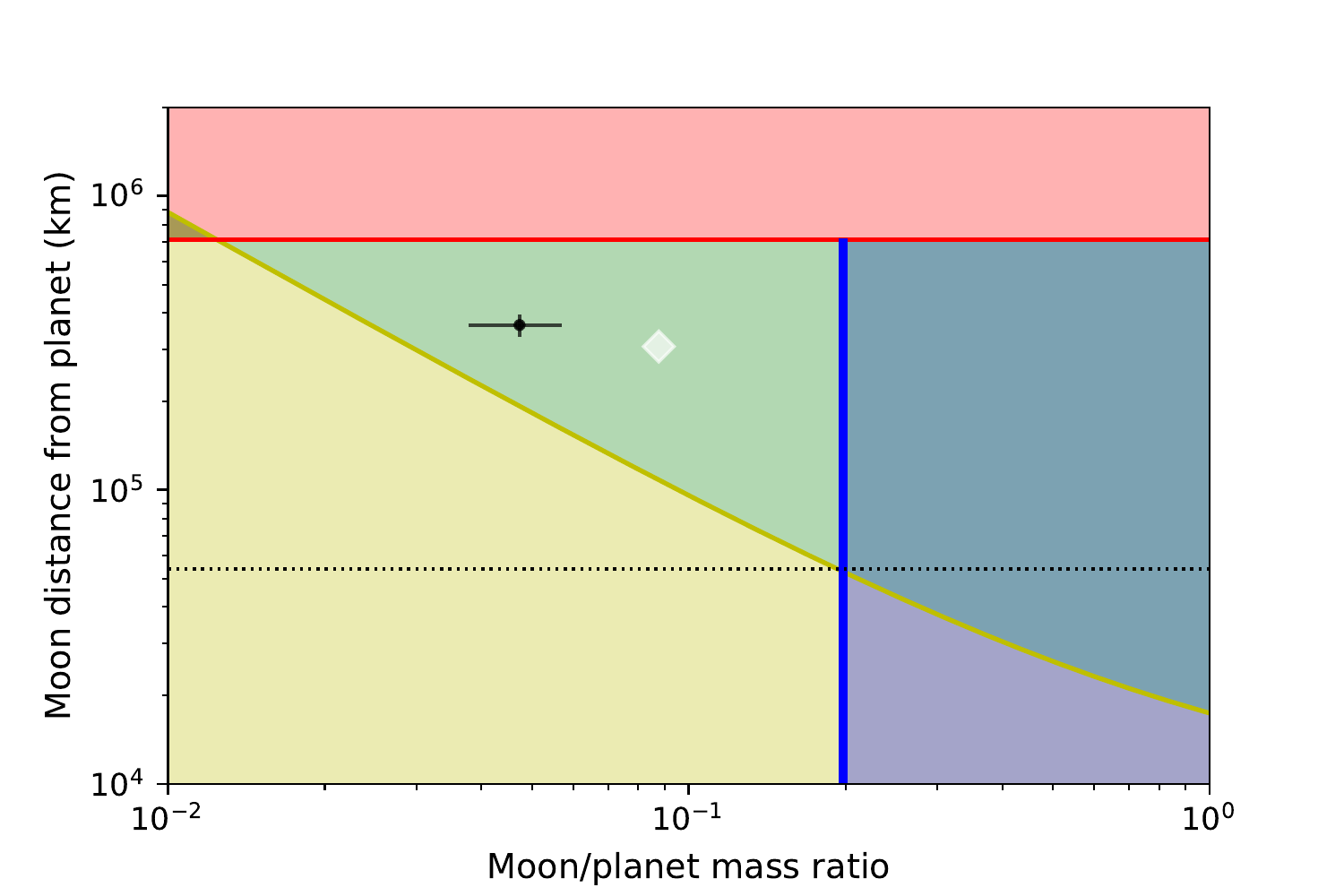}
 \caption{\label{fig:koi268bests}Quality of fit and sensitivity plot for KOI-268.01\newline
  In the TTV and TDV plots, the black points are the observed TTVs from Kepler (including error), the blue diamonds show the model results of the planet hypothesis and green dots indicate those for the moon hypothesis.  In the sensitivity plot, the white diamond is the best-fit solution and the black dot is the peak of the posterior distribution with the extended lines indicating the 1$\sigma$ uncertainty to either side. }
\end{figure}

The additional planet hypothesis produces a better TTV fit than the exomoon hypothesis, with a reduced $\chi^2$ value of nearly 0.6 compared to 1.5.  However, much of the difference in these values is attributable to a single data point, transit 3, which shows a particularly large TTV value more than double any other.  This transit also produces an abnormally low TDV.  Neither hypothesis can reproduce this transit, but the planet simulation gets closer to the TTV than the moon simulation.  The best fit moon is nearly 1 $M_{\earth}$ in size, but due to this star's large size, the moon is well below Kepler's photometric sensitivity and in the green zone of the sensitivity plot (Figure \ref{fig:koi268bests}).  Given that both hypotheses give reduced $\chi^2 \sim 1$, neither hypothesis is statistically favoured over the other.
\subsubsection{KOI-303}
\begin{tabular}{rlrl}
    Spectral Type & G6V & Planet Period & 60.93 d \\
    Star Radius & 1.02 $R_{\sun}$ & Planet Radius & 2.6 $R_{\earth}$ \\
    Star Mass & 0.87 $M_{\sun}$ & Planet Mass & 7.6 $M_{\earth}$ \\
    CDPP (6.3 hr)& 38.1 ppm & Avg TTV Err & 3.1 min \\ 
\end{tabular}
\\ \\
KOI-303.01 (Kepler-517b) has the second smallest TTV amplitude of our candidates, with no TTV larger than 10 minutes.  Even though it has one of the lowest average errors in the TTV data, at only 3.1 minutes, the TTV SNR is the lowest of our sample at 1.56.  

\begin{figure}
 \includegraphics[scale=0.6]{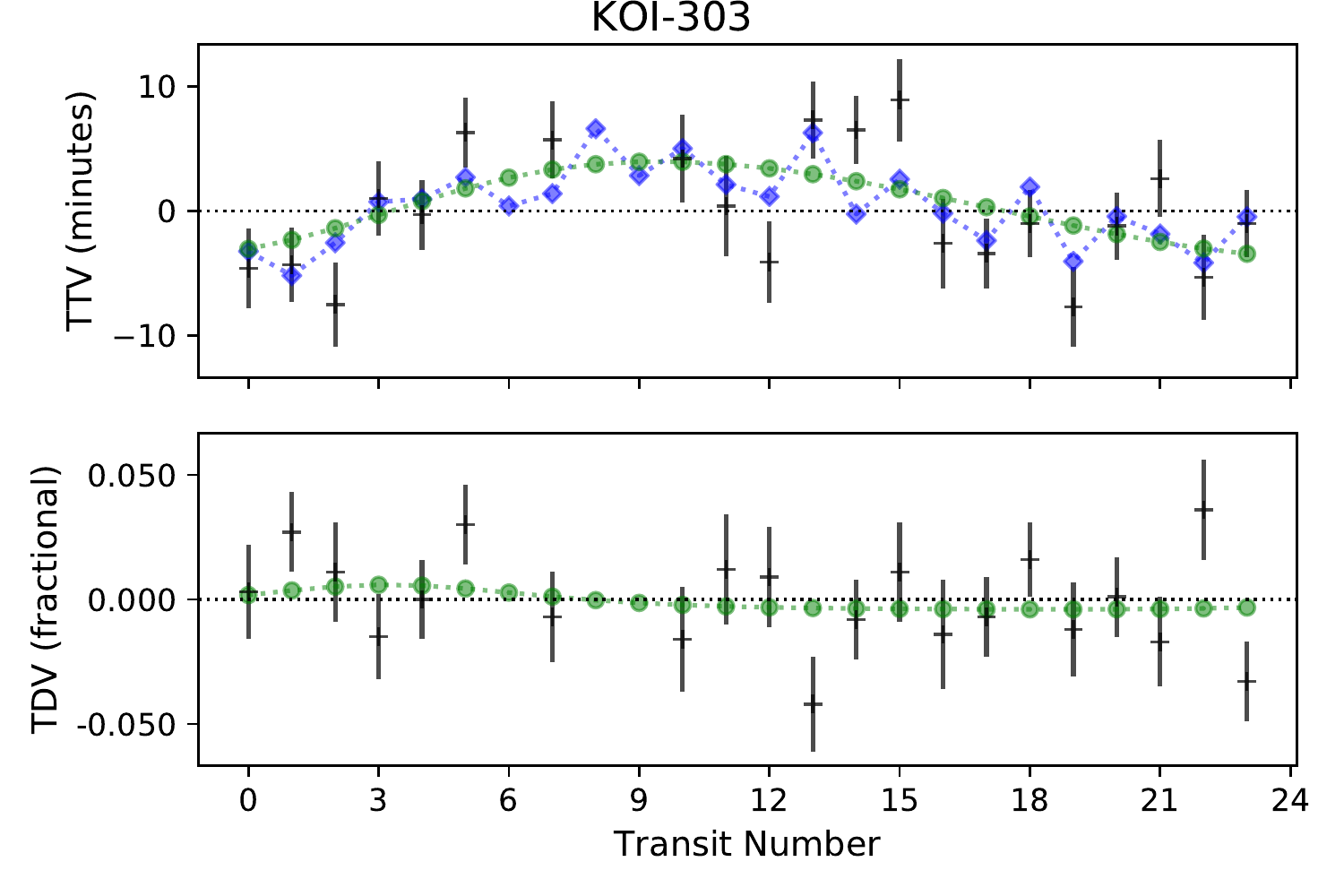}
 \includegraphics[scale=0.6]{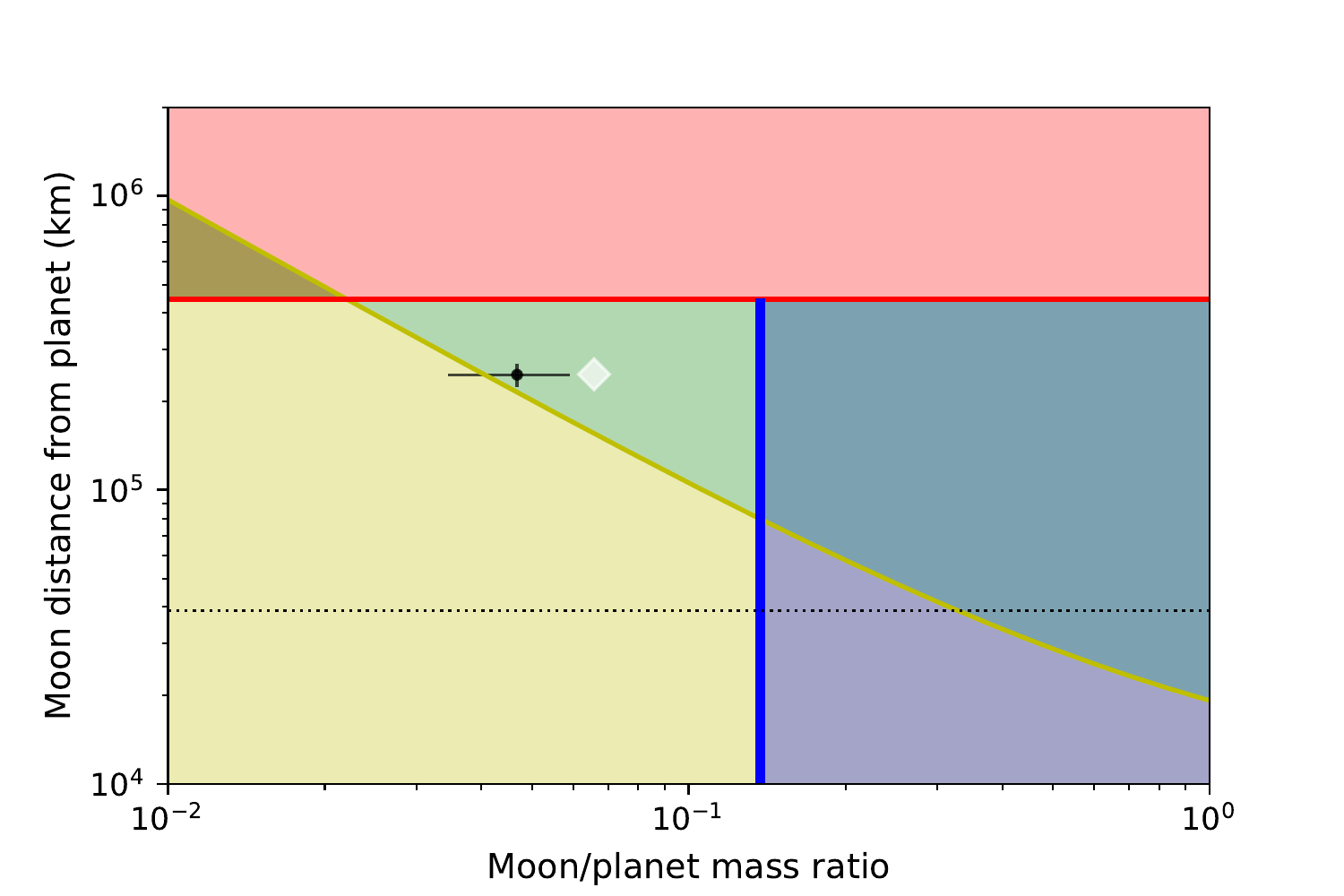}
 \caption{\label{fig:koi303bests}Quality of fit and sensitivity plot for KOI-303.01\newline
  The symbols used are the same as in Figure \ref{fig:koi268bests}.  }
\end{figure}

We find that both hypotheses can provide reasonable fits, with reduced $\chi^2$ values less than 1.  The moon hypothesis requires a moon mass of approximately 0.36 $M_{\earth}$ at an orbital distance of 0.28 $R_{Hill}$.  At that mass, assuming a bulk density equal to Earth, the expected radius of this moon would be $\approx 0.65 R_{\earth}$, putting it in the green zone as shown on the sensitivity plot (Figure \ref{fig:koi303bests}).  Like KOI-268, the posterior suggests an even lower mass value.  We conclude that the TTVs of KOI-303.01 are equally well explainable by a moon as a sibling planet.

\subsubsection{KOI-1302}
\begin{tabular}{rlrl}
    Spectral Type & G0 & Planet Period & 60.93 d \\
    Star Radius & 0.96 $R_{\sun}$ & Planet Radius & 3.2 $R_{\earth}$ \\
    Star Mass & 1.05 $M_{\sun}$ & Planet Mass & 11.0 $M_{\earth}$ \\
    CDPP (7.3 hr) & 95.7 ppm & Avg TTV Err & 10.6 min \\ 
\end{tabular}
\\ \\
KOI-1302.01 (Kepler-809b) has the second highest average TTV error, but has one of our larger amplitudes resulting in a moderate SNR value.  

\begin{figure}
 \includegraphics[scale=0.6]{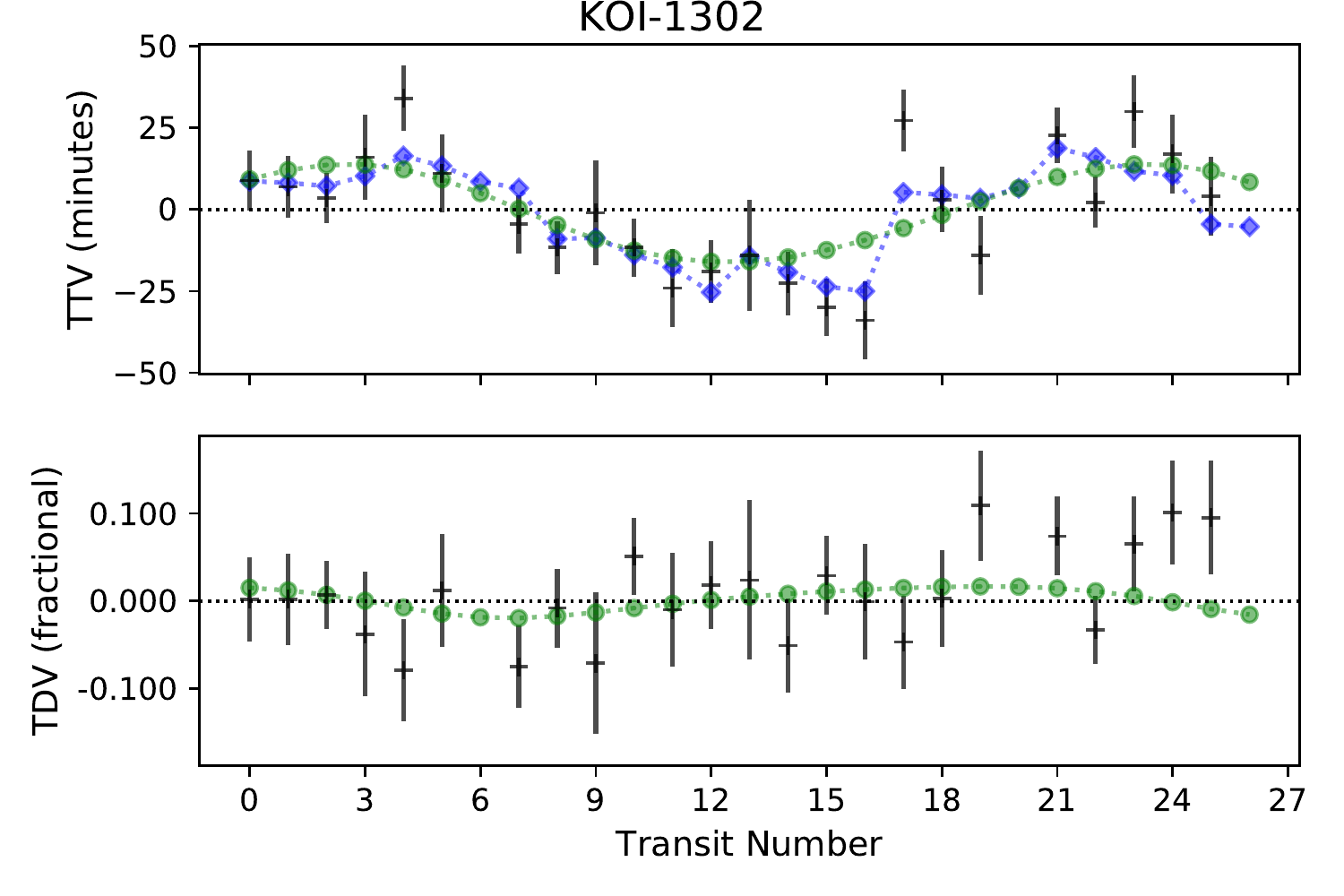}
 \includegraphics[scale=0.6]{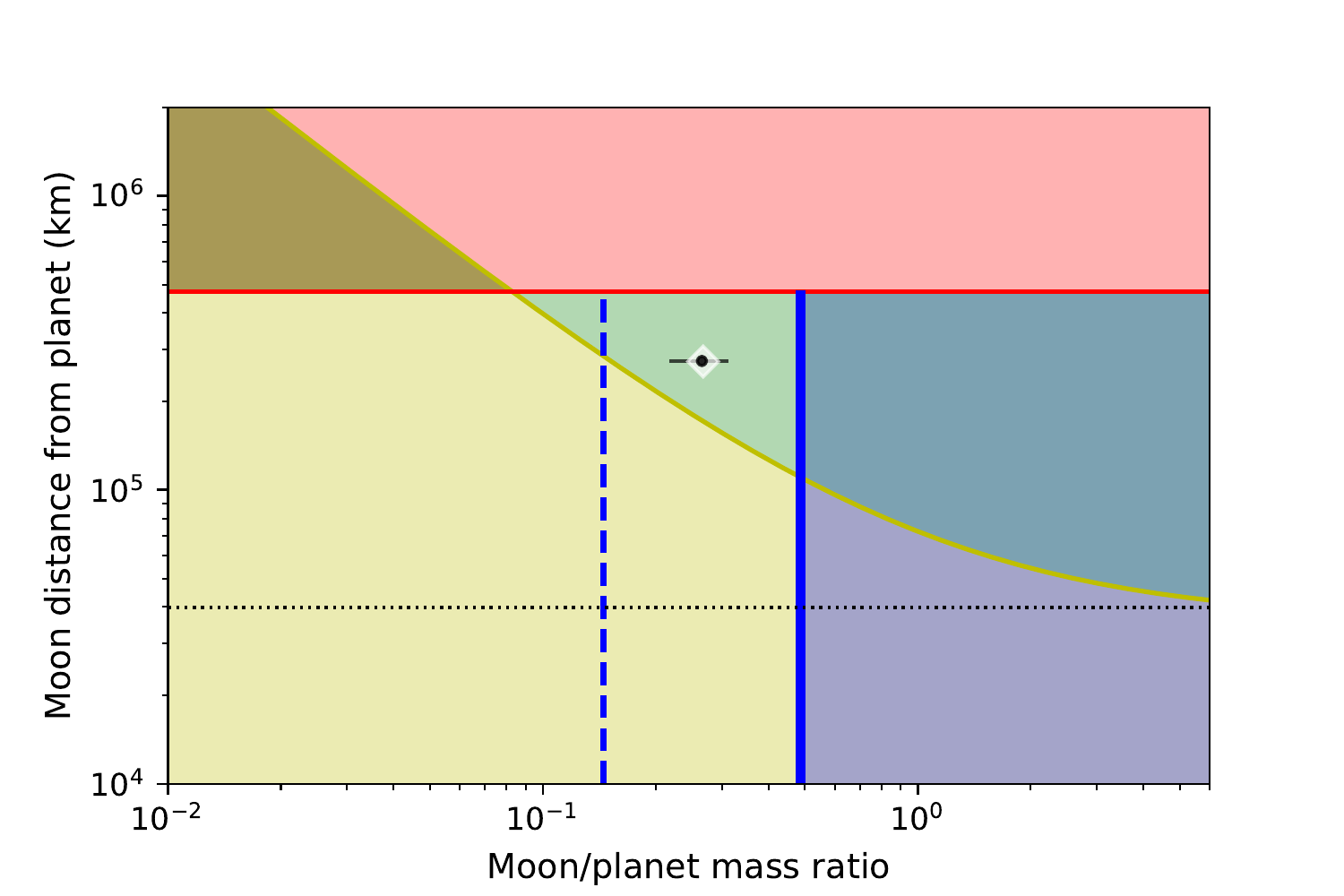}
 \caption{\label{fig:koi1302bests}Quality of fit and sensitivity plot for KOI-1302.01\newline
  The symbols used are the same as in Figure \ref{fig:koi268bests}.  The dashed blue line represents the detection threshold when assuming a Neptunian density rather than terrestrial.}
\end{figure}

We find that both planet and moon hypotheses can provide good fits. The planet's $\chi^2$ value is nominally lower at 0.5 compared to 0.8, though this may just be the result of the additional parameters available in the planet model.  The moon hypothesis requires a moon mass of approximately 2.9 $M_{\earth}$ at an orbital distance of 0.28 $R_{Hill}$ (Figure \ref{fig:koi1302bests}).  Such a mass straddles the boundary of Super-Earths and Mini-Neptunes, so its density becomes problematic.  The solid blue line on Figure \ref{fig:koi1302bests} assumes a terrestrial density, while the dashed blue assumes a Neptune-like density.  A lower-density moon has a reduced green zone, as its cross-section is larger for a given mass. We would expect a low-density moon massive enough to generate the TTVs to be visible in transit, whereas a terrestrial moon of the same mass would be below the photometric limit.  We conclude that the TTVs of KOI-1302.01 are equally well explainable by a terrestrial moon as a sibling planet, but are not the result of a moon of Neptune-like density.

\subsubsection{KOI-1472}
\begin{tabular}{rlrl}
    Spectral Type & G5 & Planet Period & 85.35 d \\
    Star Radius & 0.94 $R_{\sun}$ & Planet Radius & 6.8 $R_{\earth}$ \\
    Star Mass & 0.97 $M_{\sun}$ & Planet Mass & 38.0 $M_{\earth}$ \\
    CDPP (6.8 hr)& 118 ppm & Avg TTV Err & 2.3 min \\ 
\end{tabular}
\\ \\
KOI-1472.01 (Kepler-857b) is our largest plnaetary target, with a nominal mass of 38 $M_{\earth}$, more than double Neptune.  It also has the lowest average TTV error, at only 2.3 minutes, despite orbiting one of the noisier stars.

We find that both hypotheses can provide good fits, though the planet's reduced $\chi^2$ value is lower, 0.3 versus 0.9.  There is significant difference between the best fit values and the nominal posterior values of the mass (1.6 and 0.7 $M_{\earth}$ respectively), but both are well inside of the green zone.  This discrepancy results from a difference in the eccentricity.  The best fit is found with an eccentricity of nearly 0.5, easily our most eccentric model fit.  However the posteriors centre around a lower eccentricity and lower mass solution.  In terms of mass ratios, this moon would be the smallest in our sample.  The best-fit moon-planet mass ratio is 0.043, nearly a factor of 4 greater than the Earth-Moon system.  However, the peak of the mass posterior gives a moon-planet mass ratio of only 0.018. We conclude that the TTVs of KOI-1472.01 are explainable by either a moon or a sibling planet.
\begin{figure}
 \includegraphics[scale=0.6]{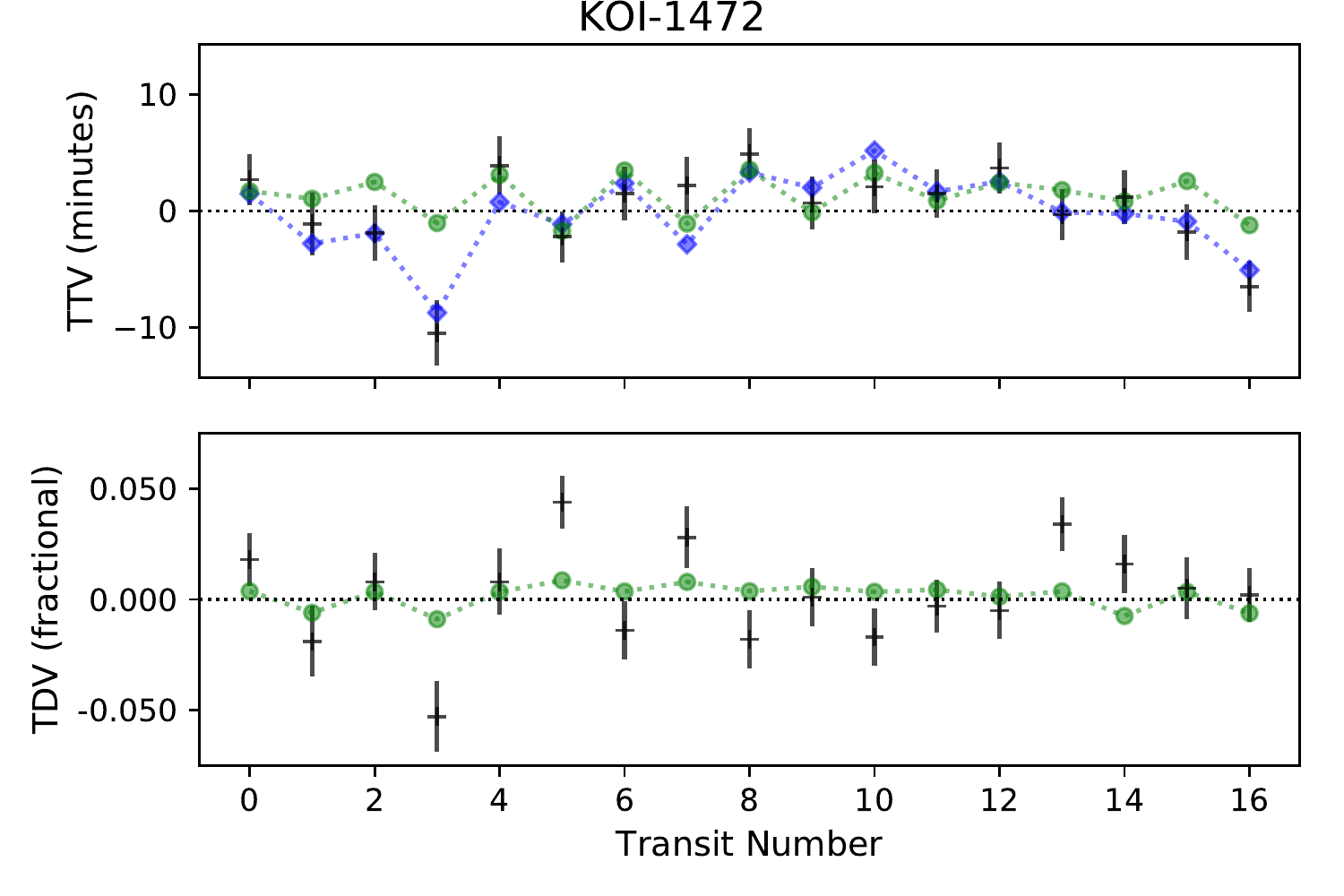}
 \includegraphics[scale=0.6]{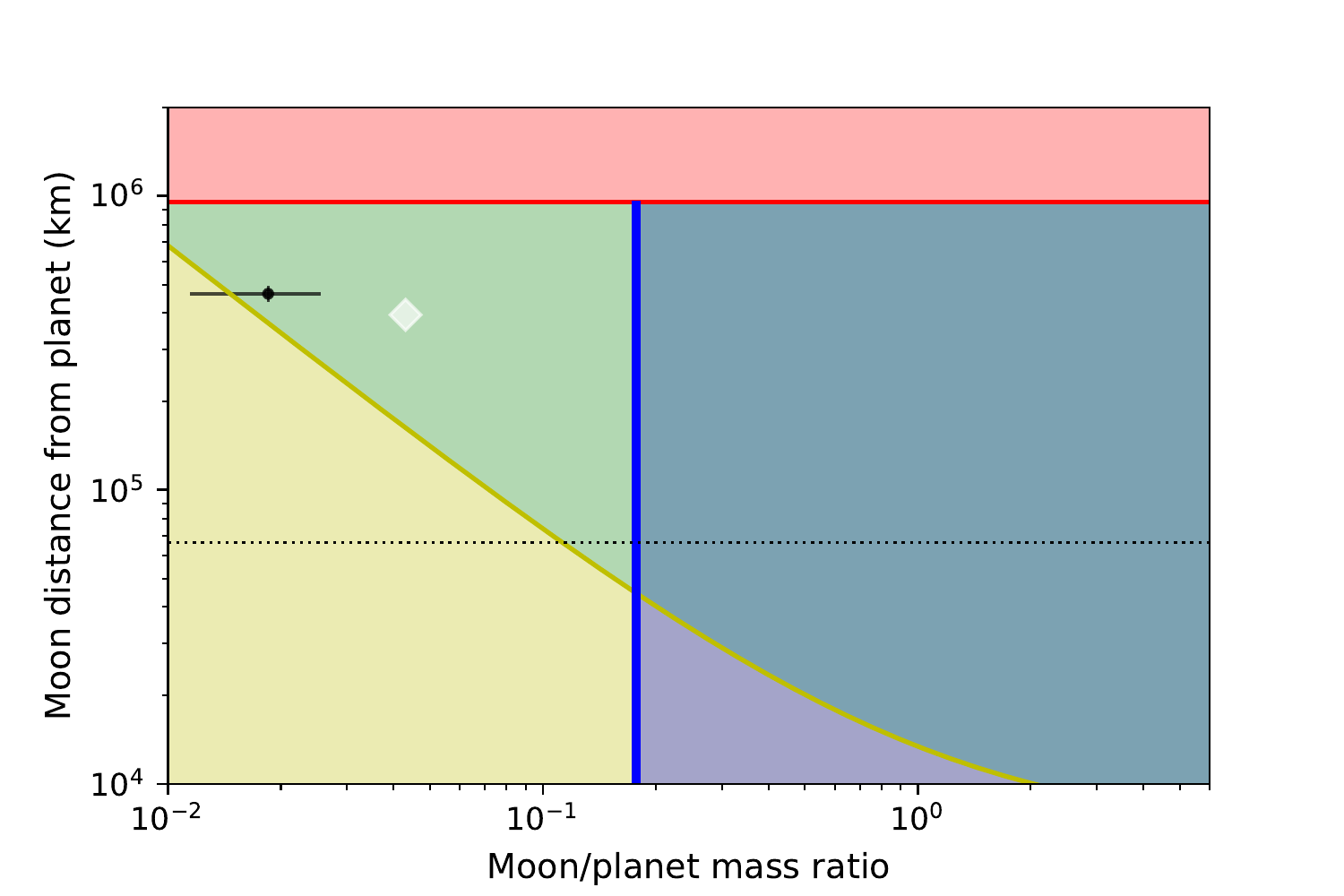}
 \caption{\label{fig:koi1472bests}Quality of fit and sensitivity plot for KOI-1472.01\newline
  The symbols used are the same as in Figure \ref{fig:koi268bests}.  }
\end{figure}

\subsubsection{KOI-1888}
\begin{figure}
 \includegraphics[scale=0.6]{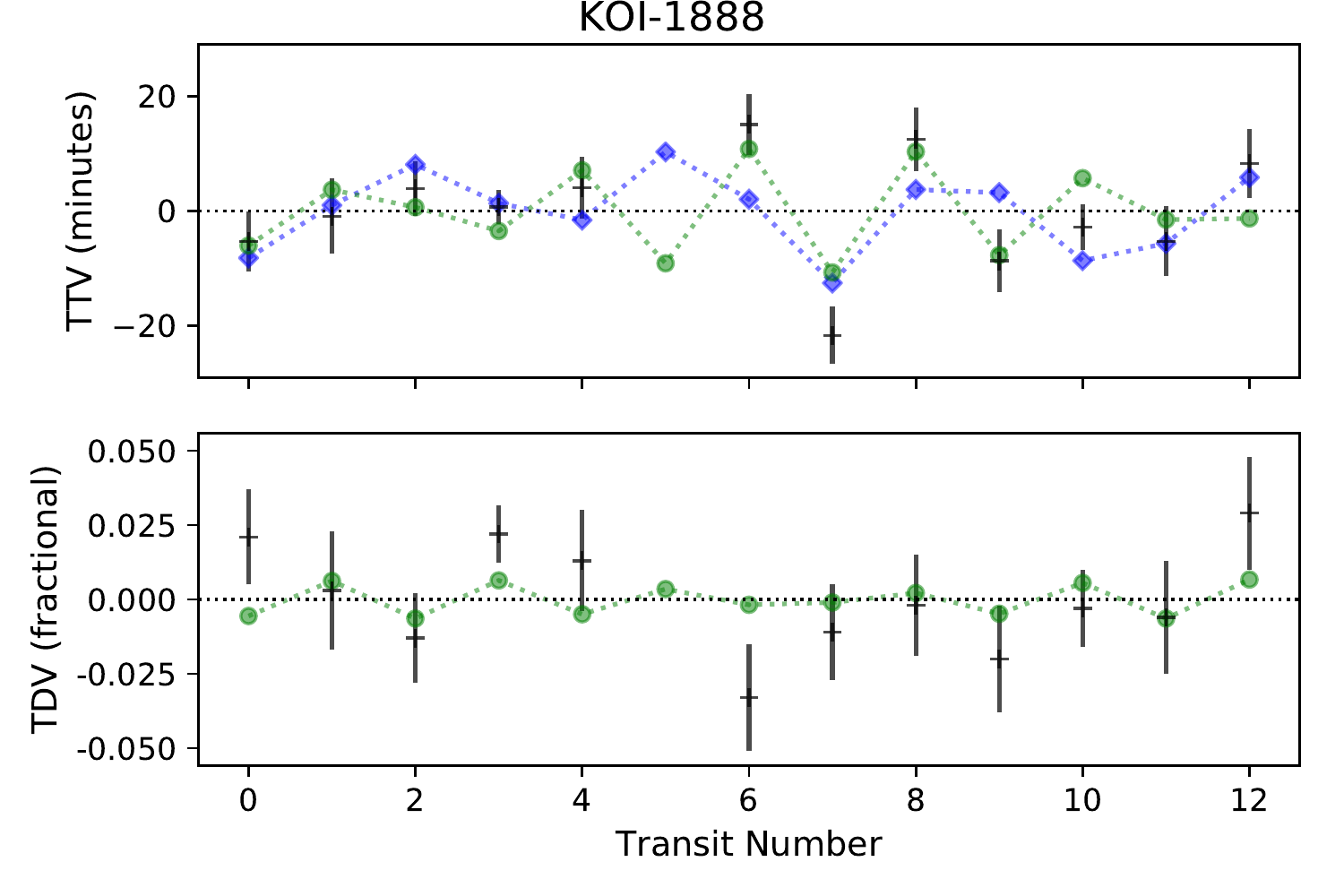}
 \includegraphics[scale=0.6]{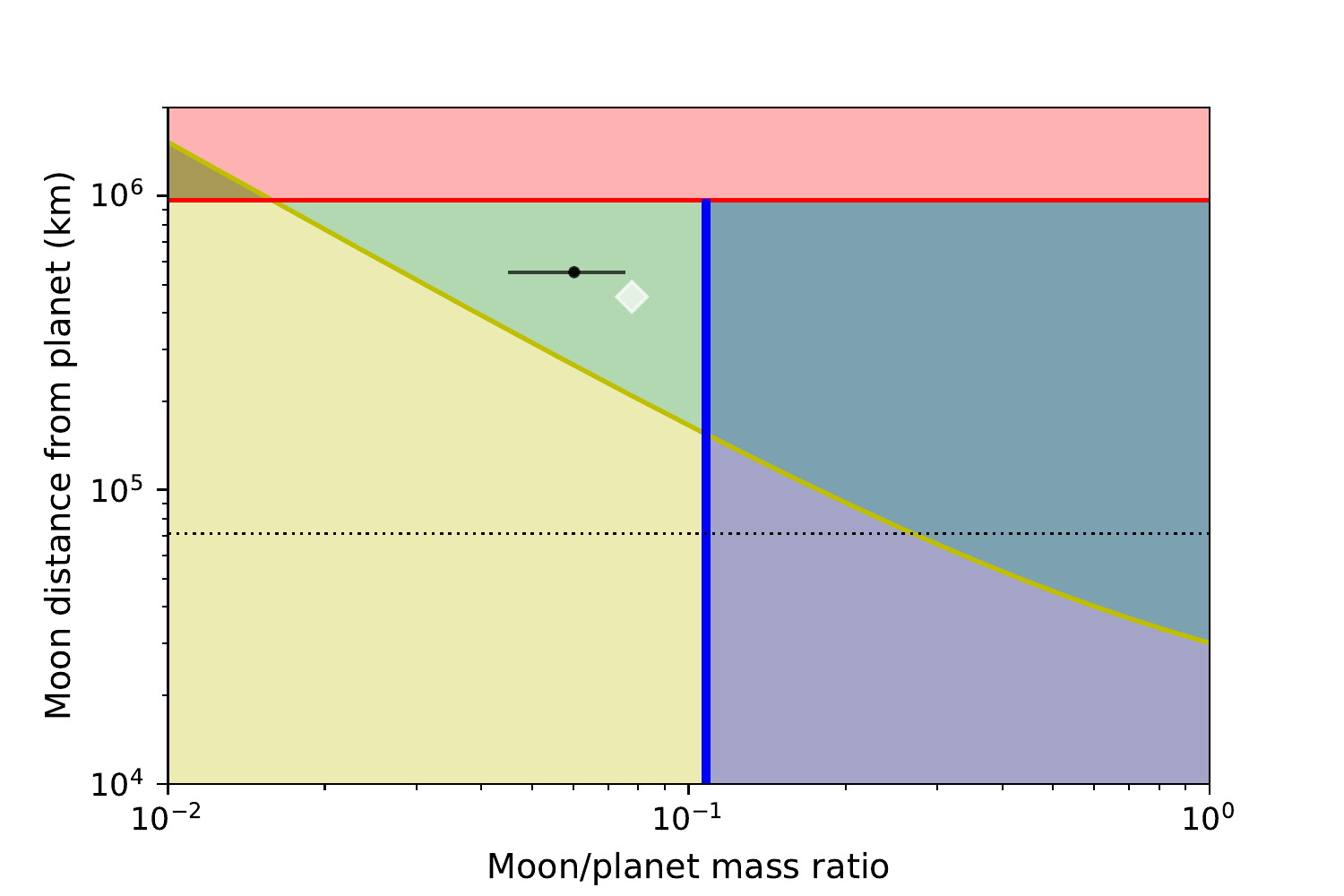}
 \caption{\label{fig:koi1888bests}Quality of fit and sensitivity plot for KOI-1888.01\newline
  The symbols used are the same as in Figure \ref{fig:koi268bests}.}
\end{figure}

\begin{tabular}{rlrl}
    Spectral Type & F6IV & Planet Period & 120.02 d \\
    Star Radius & 1.47 $R_{\sun}$ & Planet Radius & 4.7 $R_{\earth}$ \\
    Star Mass & 1.41 $M_{\sun}$ & Planet Mass & 20.0 $M_{\earth}$ \\
    CDPP  (11.6 hr) & 97.4 ppm & Avg TTV Err & 5.2 min \\ 
\end{tabular}
\\ \\
KOI-1888.01 (Kepler-1000b) is a confirmed planet orbiting a sub-giant F star, and the TTV pattern has our second highest SNR.   We obtained excellent fits from both model hypotheses.  The planet's reduced $\chi^2$ is a bit higher than the moon's (0.88 vs 0.68), but both are less than 1.  The best-fit moon is 1.5 Earth in mass, and in conjunction with the sub-giant star, the best-fit moon is inside the green zone of the sensitivity plot (Figure \ref{fig:koi1888bests}).  The posterior places the moon at slightly smaller mass, farther inside the green zone.  Thus we conclude this TTVs are as well fit by a moon as by an additional planet.

\subsubsection{KOI-1925}
\begin{tabular}{rlrl}
    Spectral Type & K0 & Planet Period & 68.96 d \\
    Star Radius & 0.89 $R_{\sun}$ & Planet Radius & 1.0 $R_{\earth}$ \\
    Star Mass & 0.90 $M_{\sun}$ & Planet Mass & 1.0 $M_{\earth}$ \\
    CDPP (3.1 hr) & 110 ppm & Avg TTV Err & 5.0 min \\ 
\end{tabular}
\\ \\
KOI-1925.01 (Kepler-409b) is our candidate most comparable to Earth in size and mass, with nominal values of 1.0 Earth in both values \citep{ck2018}.  Because this planet is easily the smallest of our candidates the transit depth of this system is also the least, at 0.012\% (120 ppm).  

The reduced $\chi^2$ values are less than 1 for both planet and moon hypotheses (0.66 and 0.62 respectively).  Our algorithm found the best-fit moon mass of about 0.3 $M_{\earth}$ in a close orbit just over 0.2 Hill.  This is physically the smallest moon of our set, but as the planet is only 1 $M_{\earth}$, it constitutes are rather high moon-planet mass ratio.   However, as the posteriors show there is a wide range of possible masses below this value.  If this moon's mass was near the lower end of the posterior, then it would be comparable in its mass ratio to Earth as Charon is to Pluto (0.13 versus 0.12).  This is our smallest potential moon, though would still be significantly more massive than our own moon by a factor of 10.  We conclude that a moon is a legitimate hypothesis, but the planet hypothesis is just as compelling.
\begin{figure}
 \includegraphics[scale=0.6]{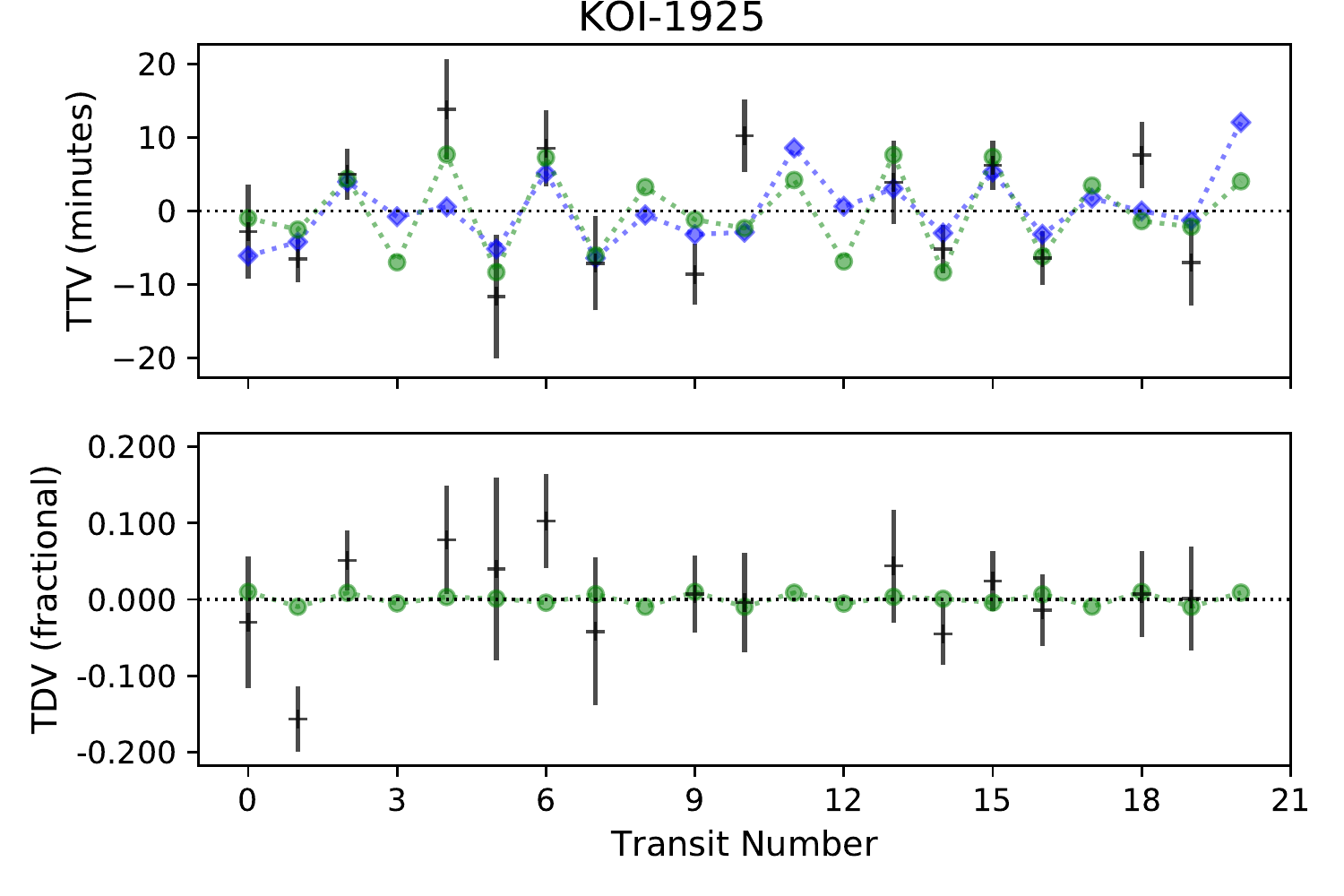}
 \includegraphics[scale=0.6]{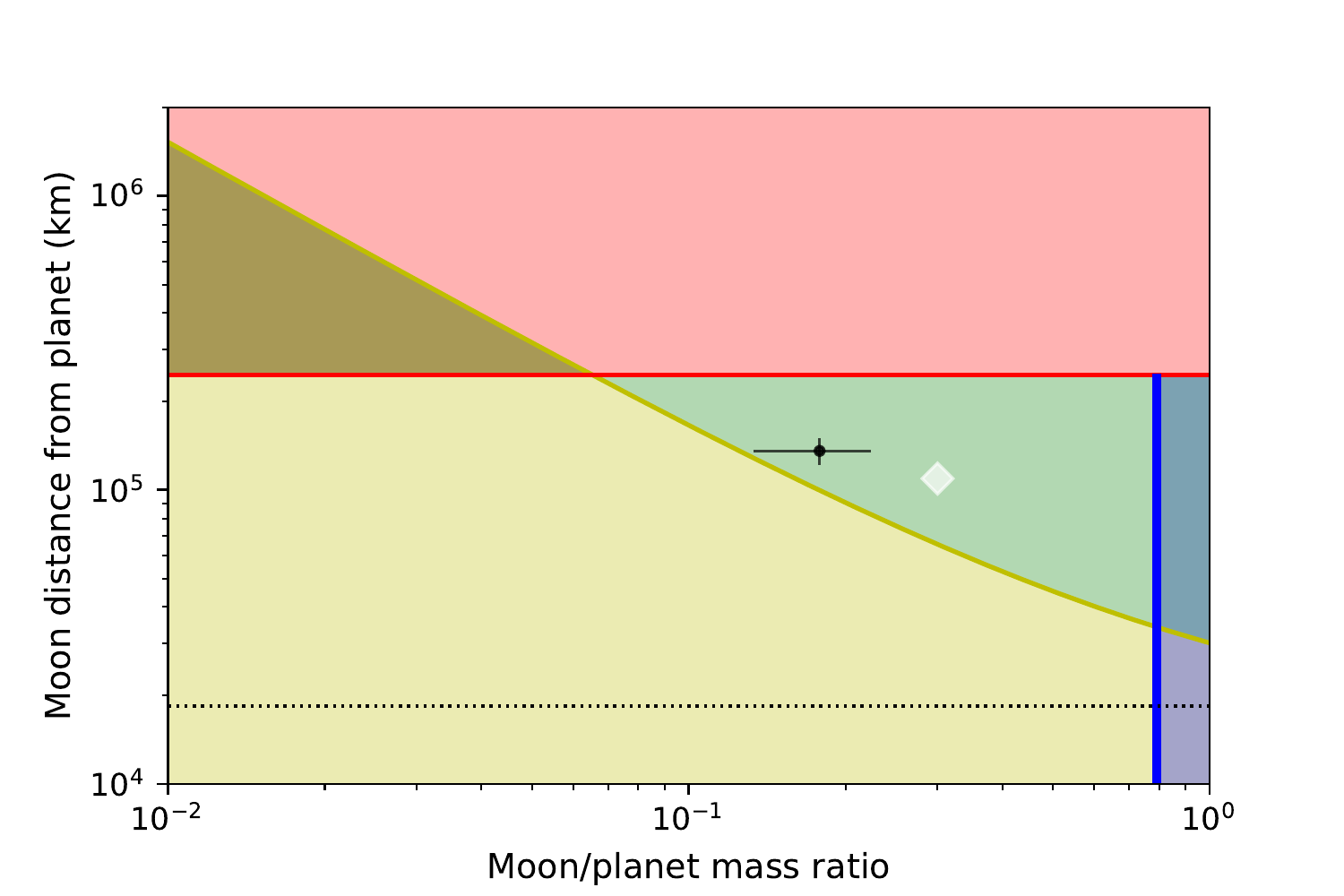} 
 \caption{\label{fig:koi1925bests}Quality of fit and sensitivity plot for KOI-1925.01\newline
  The symbols used are the same as in Figure \ref{fig:koi268bests}.}
\end{figure}

\subsubsection{KOI-2728}
\begin{tabular}{rlrl}
    Spectral Type & F4 & Planet Period & 42.35 d \\
    Star Radius & 2.63 $R_{\sun}$ & Planet Radius & 5.3 $R_{\earth}$ \\
    Star Mass & 1.54 $M_{\sun}$ & Planet Mass & 24.6 $M_{\earth}$ \\
    CDPP (7.8 hr) & 101 ppm & Avg TTV Err & 7.3 min \\ 
\end{tabular}
\\ \\
KOI-2728.01 (Kepler-1326b) is the extreme of our sample in several categories.  The host star is the hottest and most massive of our candidates and has a significantly larger radius than any other.  The planet is the largest in estimated radius and mass of our candidates.  Because of the star's size, this massive planet only gives the third-lowest transit depth of our sample.

\begin{figure}
 \includegraphics[scale=0.6]{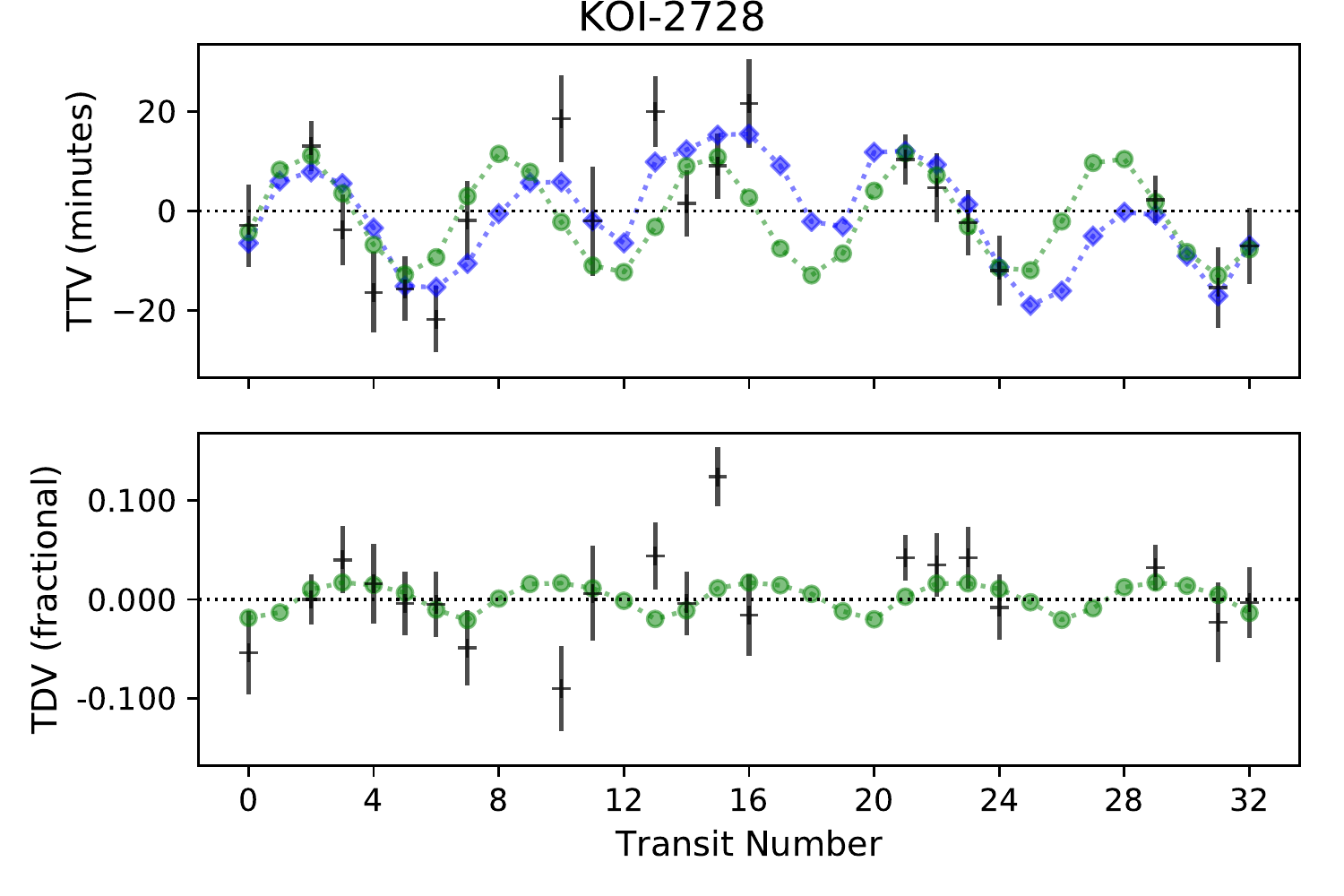}
 \includegraphics[scale=0.6]{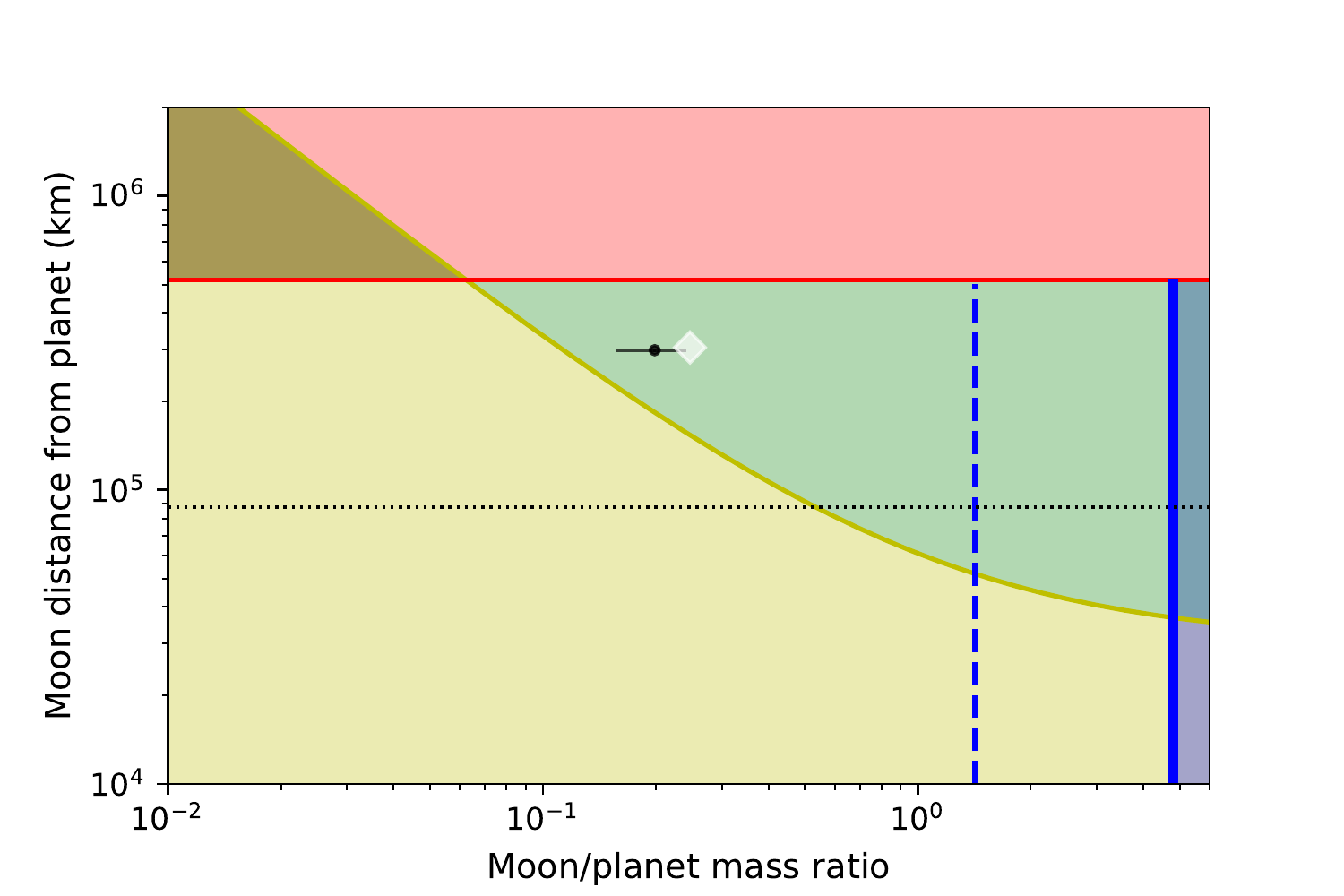} 
 \caption{\label{fig:koi2728bests}Quality of fit and sensitivity plot for KOI-2728.01\newline
  The symbols used are the same as in Figure \ref{fig:koi268bests}.  The dashed blue line represents the detection threshold when assuming a Neptunian density rather than terrestrial.}
\end{figure}

The reduced $\chi^2$ values for both planet and moon hypotheses are well below 1 (0.43 and 0.75 respsectively).  The best-fit moon is super-Earth sized, at 6 $M_{\earth}$, the largest of any of our exomoon fits and suggesting a lower density.  But the sub-giant nature of the star, with a radius more than 2.6 times that of our sun, combined with one of the highest CDPP values in our sample, makes this a more difficult transit detection than its mass would otherwise suggest. Figure \ref{fig:koi2728bests} shows our standard detection threshold assuming a terrestrial density, but even with a Neptunian density (the dashed blue line), this body would still be well below the detection limit.  We conclude these TTVs could be induced either by a massive moon or by a sibling planet.

\subsubsection{KOI-3220}
\begin{tabular}{rlrl}
    Spectral Type & F7 &  Planet Period & 81.42 d \\
    Star Radius & 1.4 $R_{\sun}$ &  Planet Radius & 3.8 $R_{\earth}$ \\
    Star Mass & 1.3 $M_{\sun}$ &  Planet Mass & 14.1 $M_{\earth}$ \\
    CDPP (14.2 hr)& 73.7 ppm & Avg TTV Err & 4.7 min \\ 
\end{tabular}
\\ \\
KOI-3220.01 (Kepler-1442b) is another planet around a large hot star.  The planet is our second largest candidate, and may resemble Uranus (3.8 $R_{\earth}$ and 14 $M_{\earth}$).  Its orbital period of 81 days is right in the middle of our sample.  This planet shows a TTV pattern similar to KOI-1925.01, but with slightly lower error (4.6 vs 5.0 minutes).  Unlike KOI-1925, the TDV pattern of KOI-3220 is one of the strongest, showing significant scatter across the entire range.

\begin{figure}
 \includegraphics[scale=0.6]{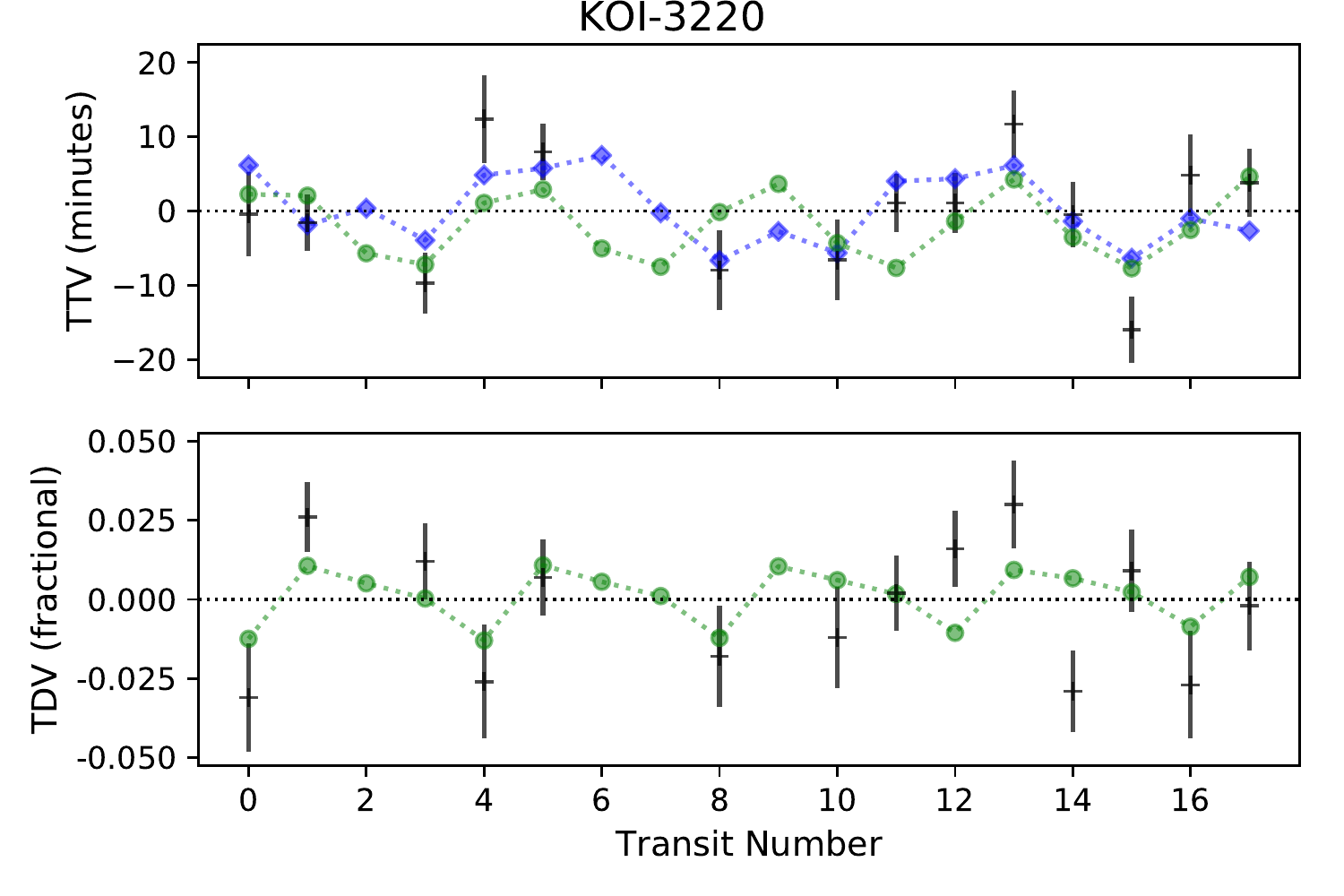}
 \includegraphics[scale=0.6]{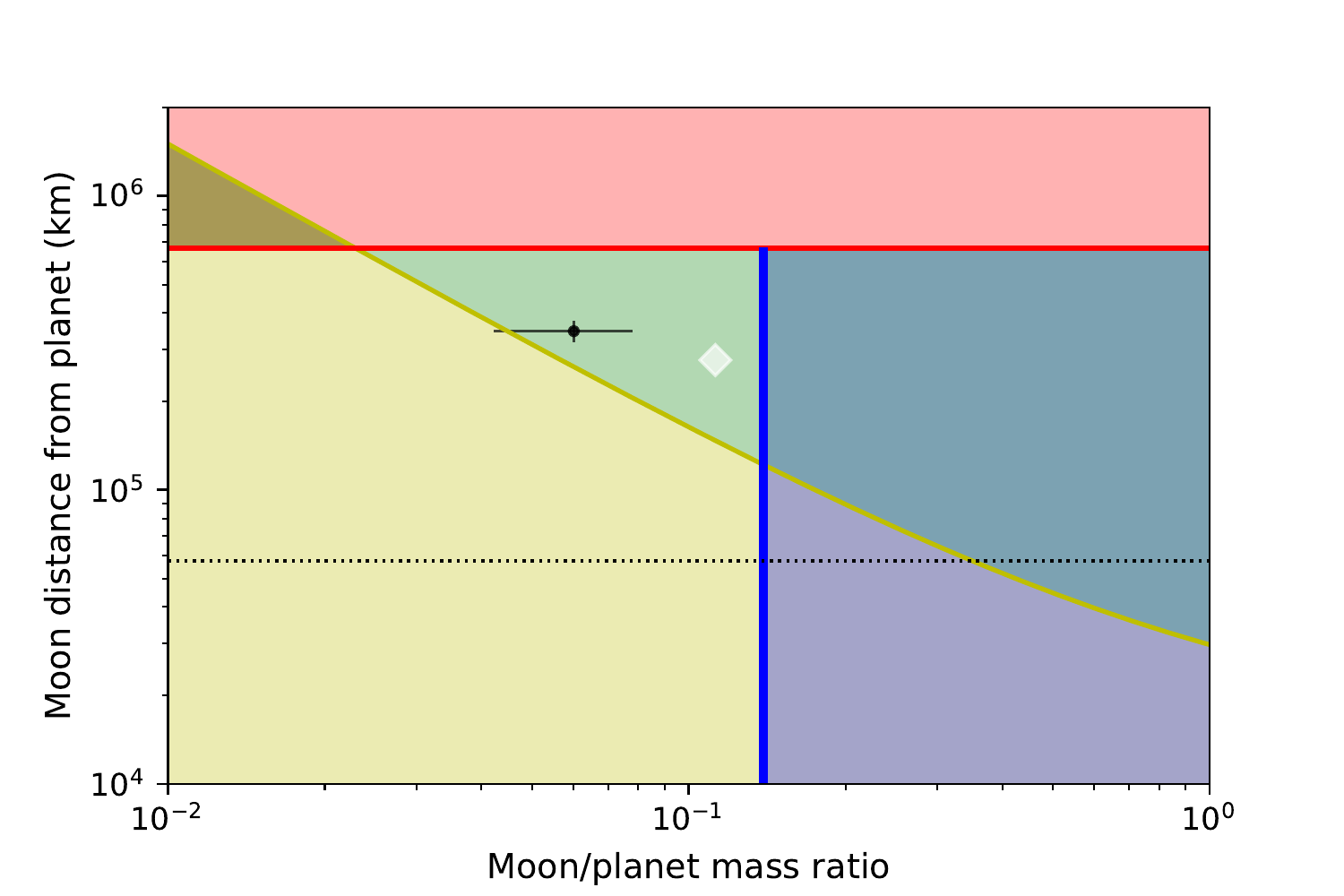} 
 \caption{\label{fig:koi3220bests}Quality of fit and sensitivity plot for KOI-3220.01\newline
  The symbols used are the same as in Figure \ref{fig:koi268bests}.}
\end{figure}

The reduced $\chi^2$ values for both hypotheses are well below unity, at 0.57 and 0.83 for the planet and moon hypotheses respectively.  The best-fit moon hypothesis requires a mass of just over 1.6 $M_{\earth}$, residing at a distance of 0.2 $R_{Hill}$.  Against a Sun-sized star, such a moon would produce a discernible transit, but this 1.4 $R_{\sun}$ moderately noisy star shifts the detection threshold significantly.  As a result, the moon is well inside the green zone of the sensitivity plot (Figure \ref{fig:koi3220bests}).  The best fit is near the threshold, but the posteriors indicate that a smaller moon farther out is possible, and it is would be well below the nominal detection limit.  We thus conclude that the TTVs of KOI-3220.01 could be caused by a large moon, but we cannot rule out a planet as the cause.  

\subsection{Poor exomoon candidate systems}

\subsubsection{KOI-1848}
\begin{tabular}{rlrl}
    Spectral Type & F6IV & Planet Period & 49.6 d \\
    Star Radius & 1.18 $R_{\sun}$ & Planet Radius & 2.7 $R_{\earth}$ \\
    Star Mass & 1.10 $M_{\sun}$ & Planet Mass & 8.1 $M_{\earth}$ \\
    CDPP (7.2 hr) & 61.9 ppm & Avg TTV Err & 9.4 min \\ 
\end{tabular}
\\ \\
KOI-1848.01 (Kepler-978b) is somewhat smaller than Uranus in both mass and radius.  It has the second highest value for TTV amplitude, but a relatively lower average error, resulting in one of the higher SNR values.

\begin{figure}
 \includegraphics[scale=0.6]{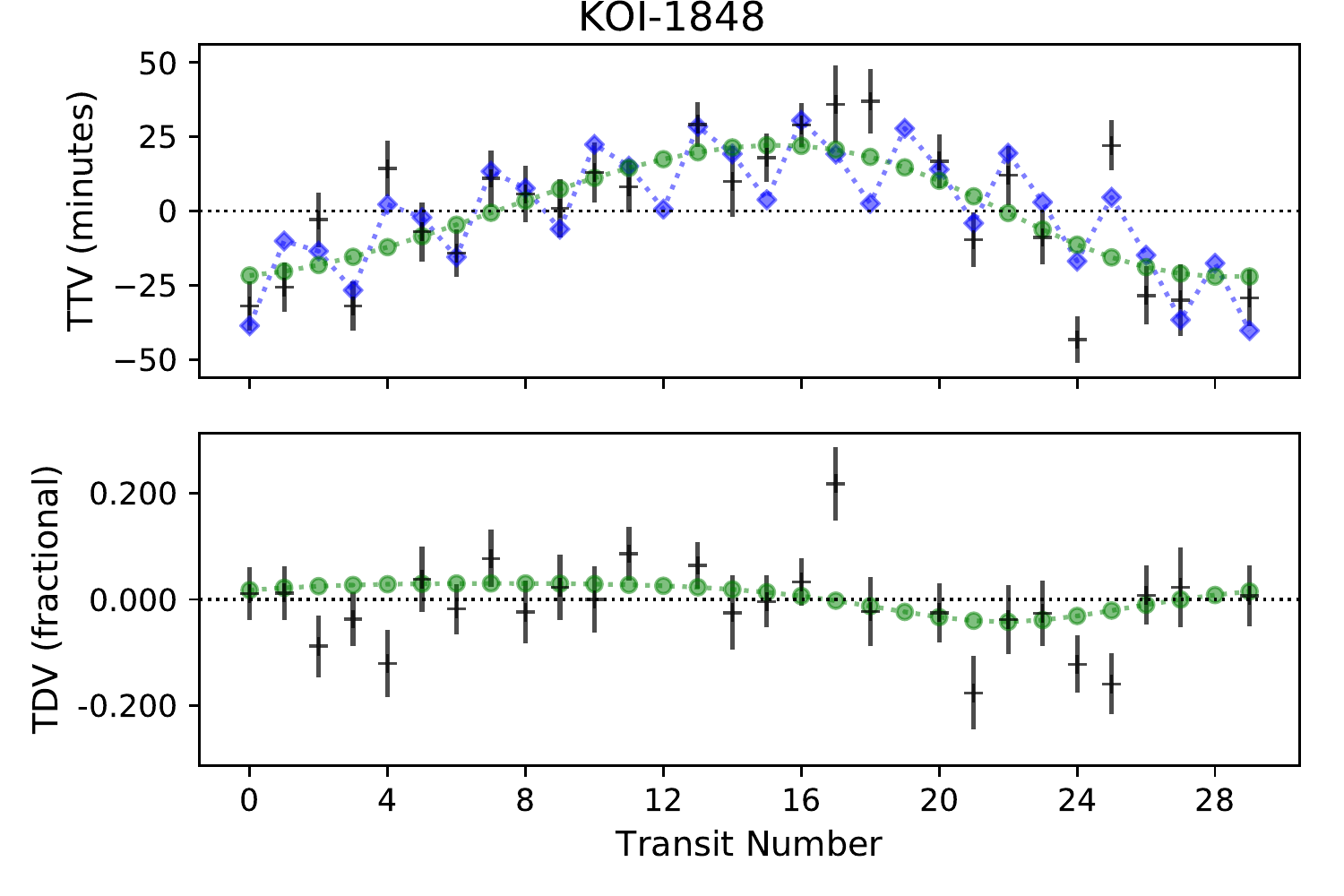}
 \includegraphics[scale=0.6]{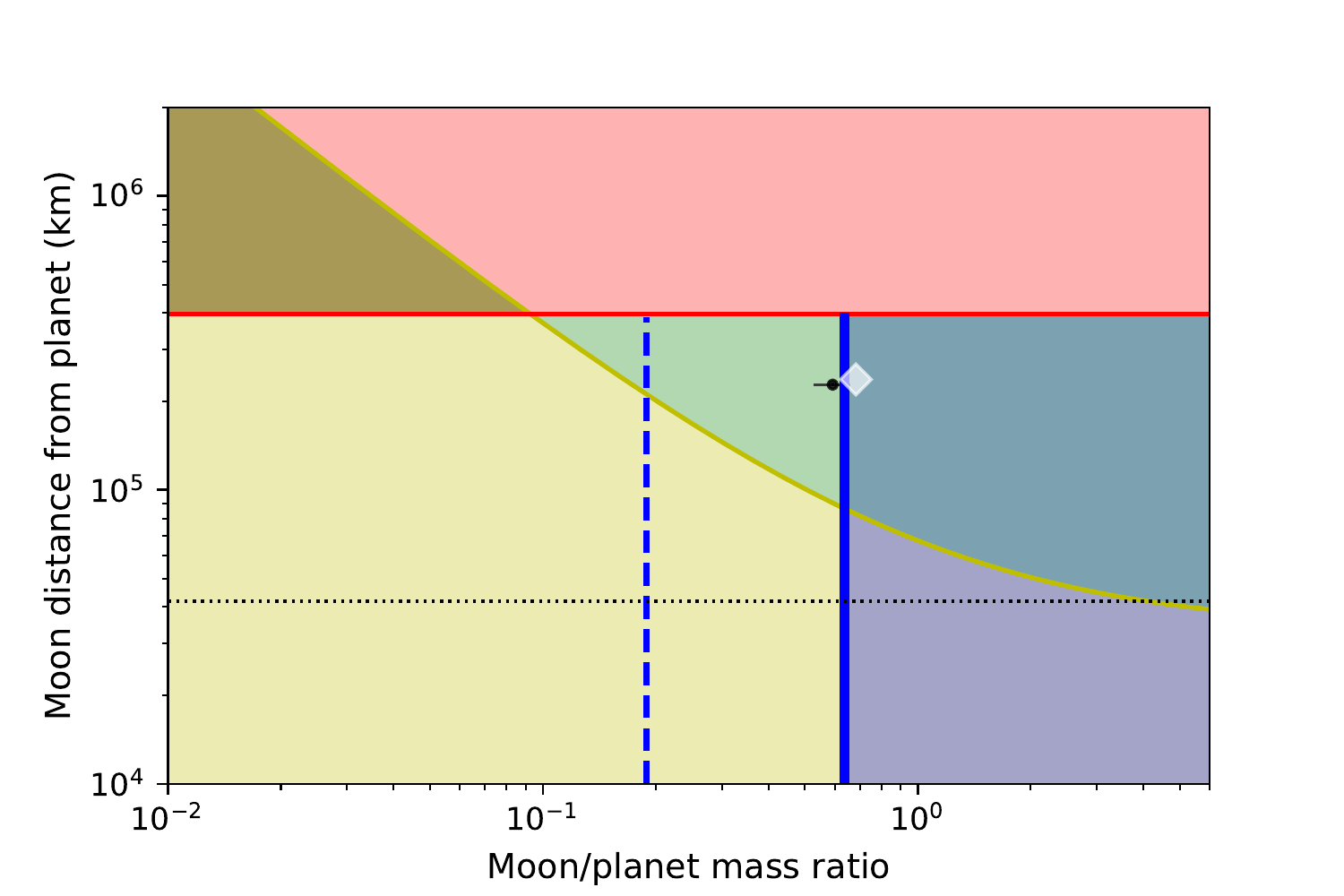}
 \caption{\label{fig:koi1848bests}Quality of fit and sensitivity plot for KOI-1848.01\newline
  The symbols used are the same as in Figure \ref{fig:koi268bests}.  The dashed blue line represents the detection threshold when assuming a Neptunian density rather than terrestrial. }
\end{figure}

The planet hypothesis results in a reduced $\chi^2$ value of 0.9, while the moon hypothesis is at 1.3.  However, the mass required for the exomoon model is approximately 5 $M_{\earth}$, more than two-thirds that of the planet, and suggesting a possible lower density makeup.  Figure \ref{fig:koi1848bests} shows an additional dashed blue line to indicate the detection threshold for a Neptune-like density moon, but regardless of its density, such a moon is likely large enough to detect photometrically. We conclude that the TTVS of KOI-1848.01 are unlikely to be due to an unseen moon.

\subsubsection{KOI-2469}
\begin{tabular}{rlrl}
    Spectral Type & K2 & Planet Period & 131.19 d \\
    Star Radius & 0.80 $R_{\sun}$ & Planet Radius & 2.4 $R_{\earth}$ \\
    Star Mass & 0.77 $M_{\sun}$ & Planet Mass & 6.6 $M_{\earth}$ \\
    CDPP (6.9 hr) & 143 ppm & Avg TTV Err & 15.5 min \\ 
\end{tabular}
\\ \\
KOI-2469.01 is another target with an "unconfirmed" status in NASA's Exoplanet Archive, but has a dispostion score of 1 \citep{nep2013,kep8cat2018}.  The star is in a near-tie for the noisiest of our targets with a CDPP of 143 ppm.  The known planet is a super-Earth \citep{ck2018} with the largest average TTV error of our sample, the largest TTV amplitude, and the bare minimum number of transits (10).

\begin{figure}
 \includegraphics[scale=0.6]{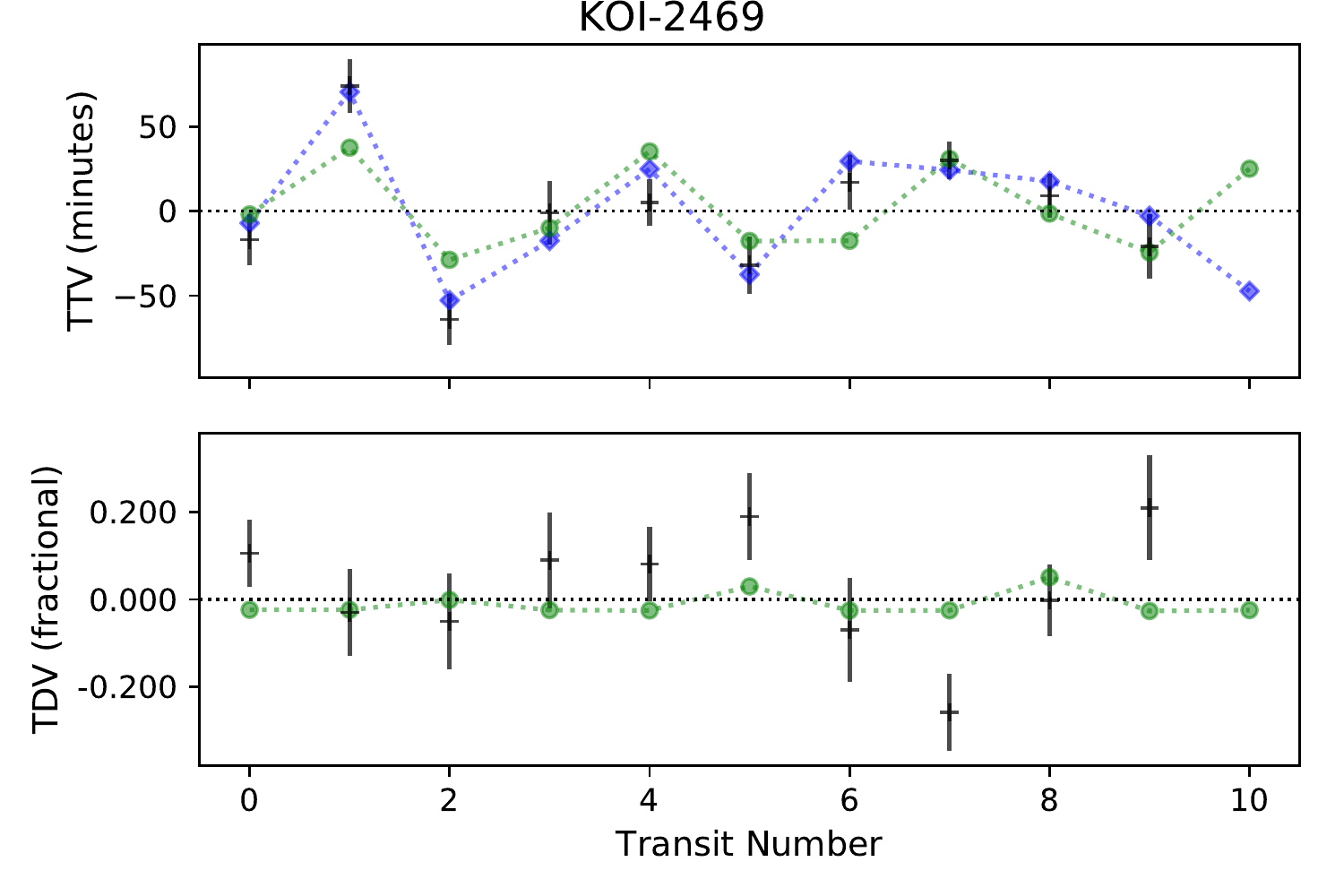}
 \includegraphics[scale=0.6]{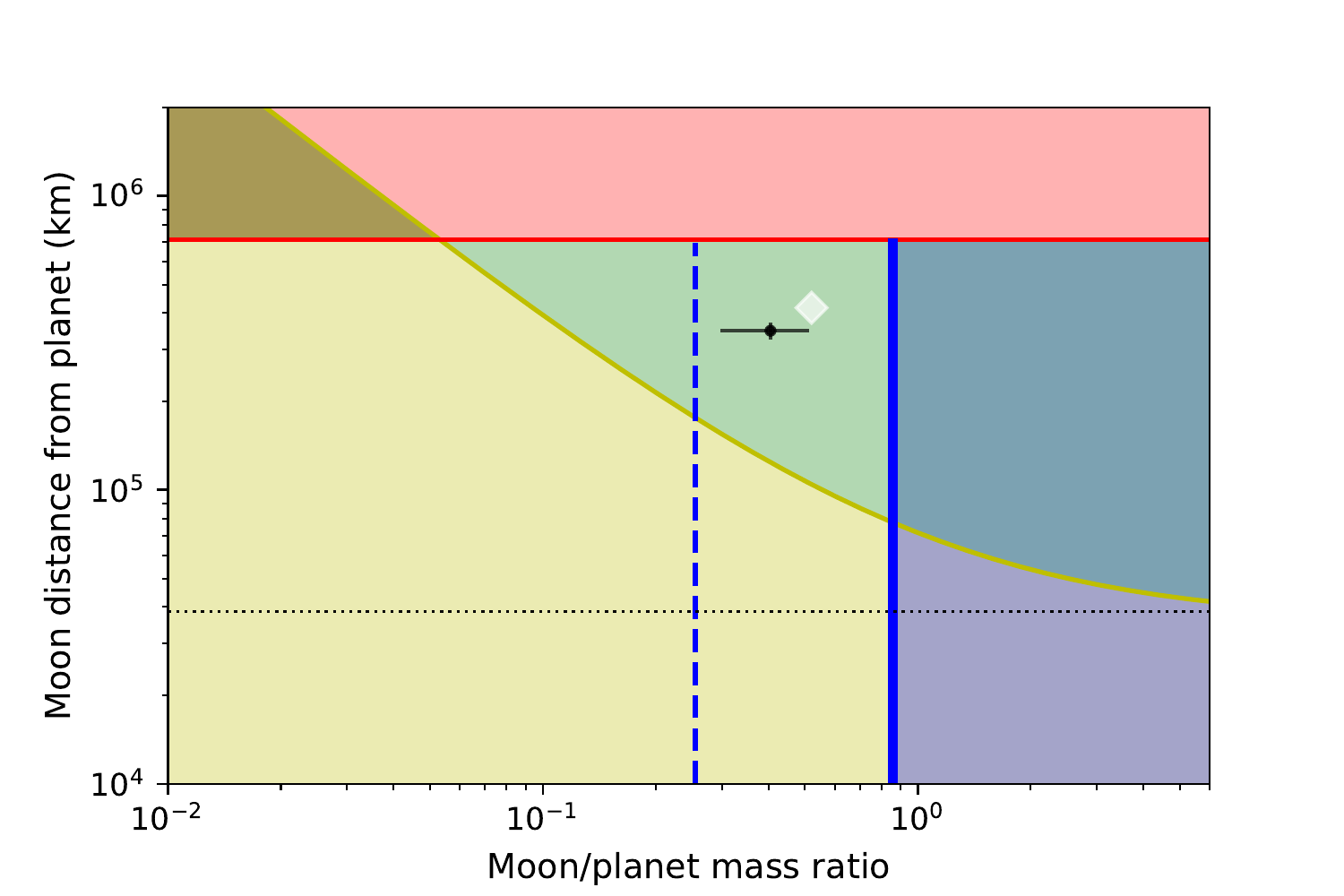} 
 \caption{\label{fig:koi2469bests}Quality of fit and sensitivity plot for KOI-2469.01\newline
  The symbols used are the same as in Figure \ref{fig:koi268bests}.  The dashed blue line represents the detection threshold when assuming a Neptunian density rather than terrestrial.}
\end{figure}

The high CDPP value pushes the blue line in the sensitivity plot to the right, and both the posterior and best-fit values are well inside the green zone.  The reduced $\chi^2$ value for planet hypothesis is 0.31, but 1.13 for the moon hypothesis.  The modelled mass of the moon could be anywhere from a quarter to half the mass of the planet, (approximately 2.5 to 3.5 $M_{\earth}$)  making this a binary planet, and suggesting a possible lower density.  Figure \ref{fig:koi2469bests} shows the detection threshold for a terrestrial world in blue, while the threshold for a Neptunian density is dashed.  An unseen moon could exist in the system but only if it were of terrestrial density.  However, given its much improved quality of fit, the planet hypothesis is more likely.

\subsubsection{Other systems with poor moon fits}
\begin{figure}
 \includegraphics[scale=0.6]{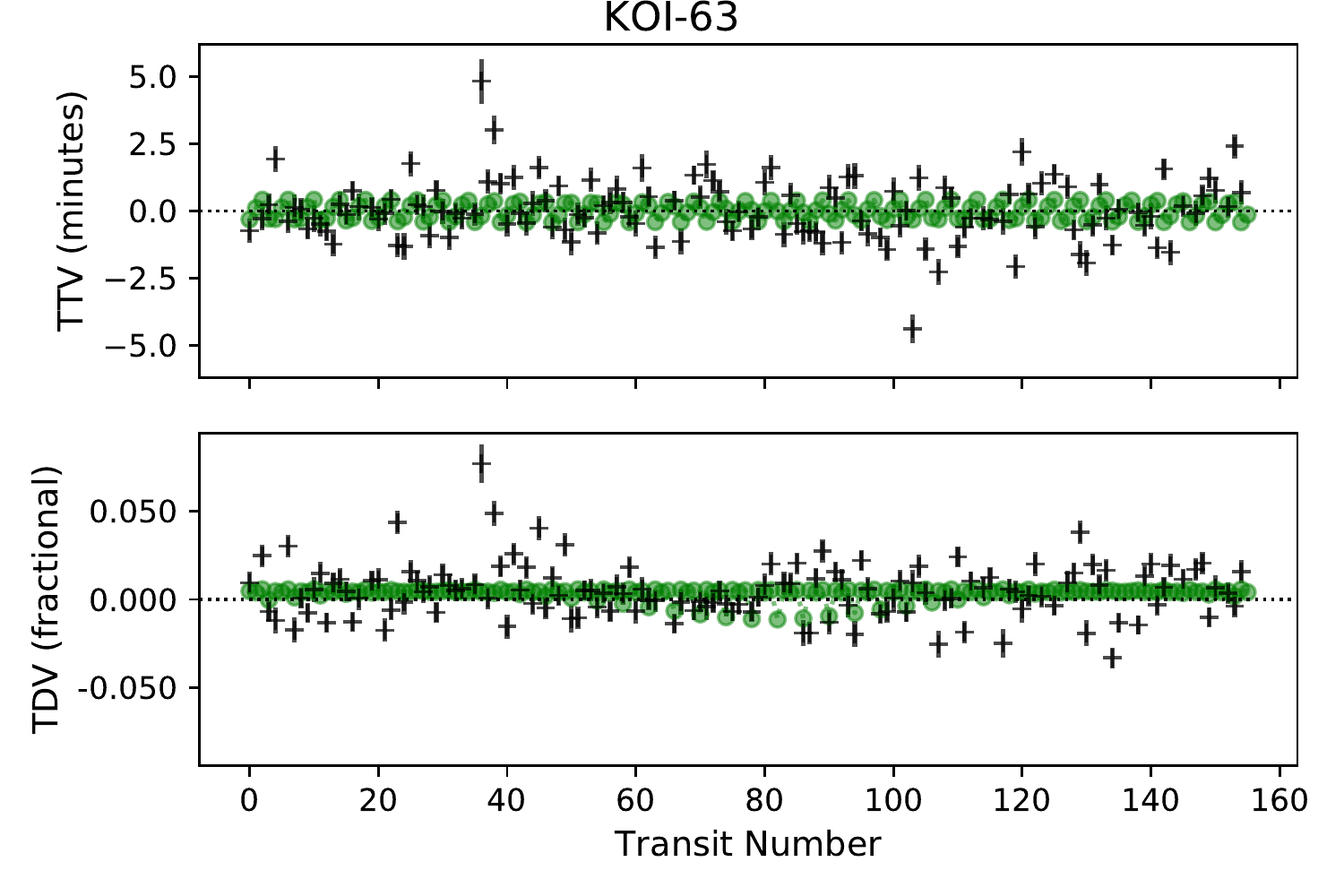}
 \includegraphics[scale=0.6]{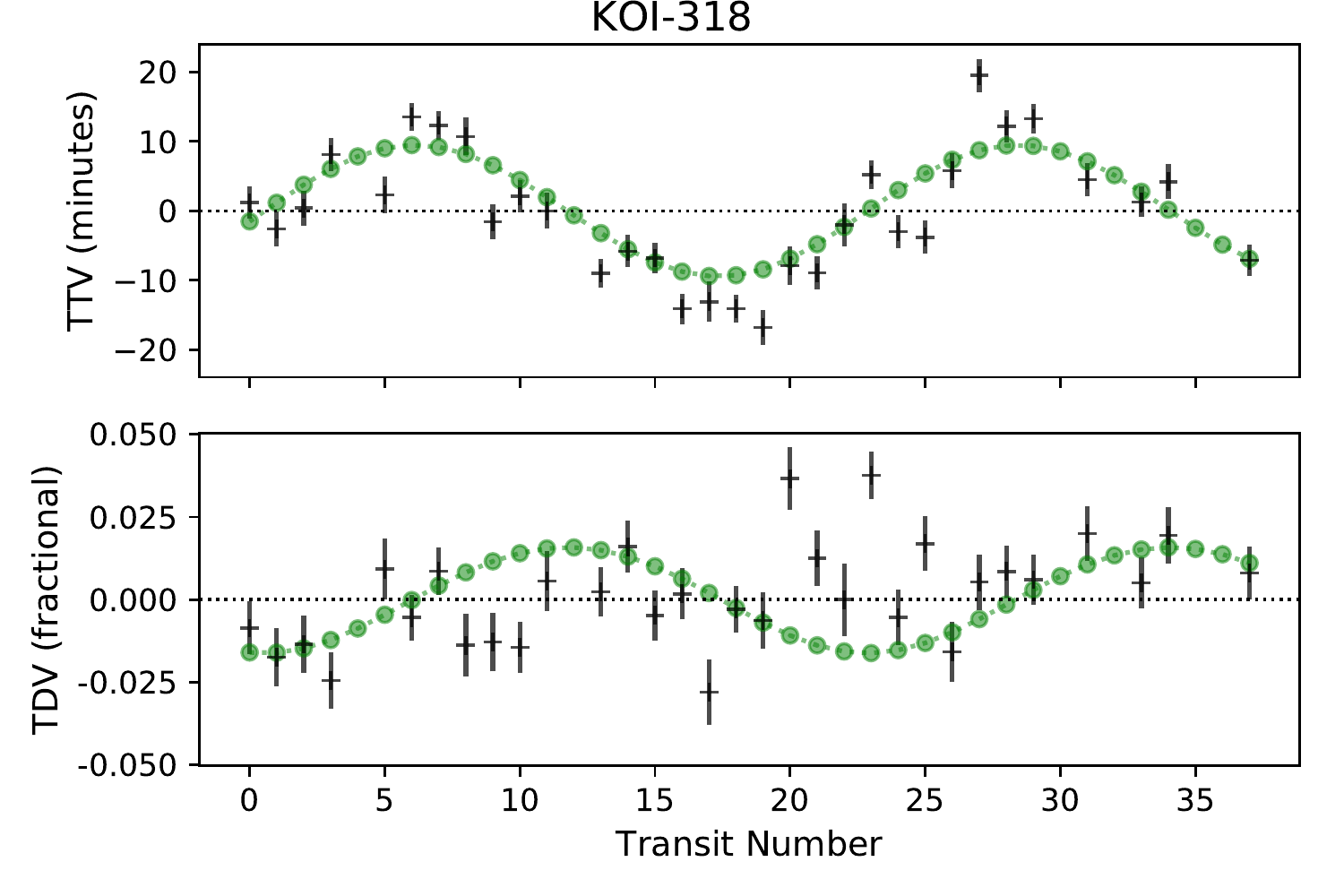} 
 \includegraphics[scale=0.6]{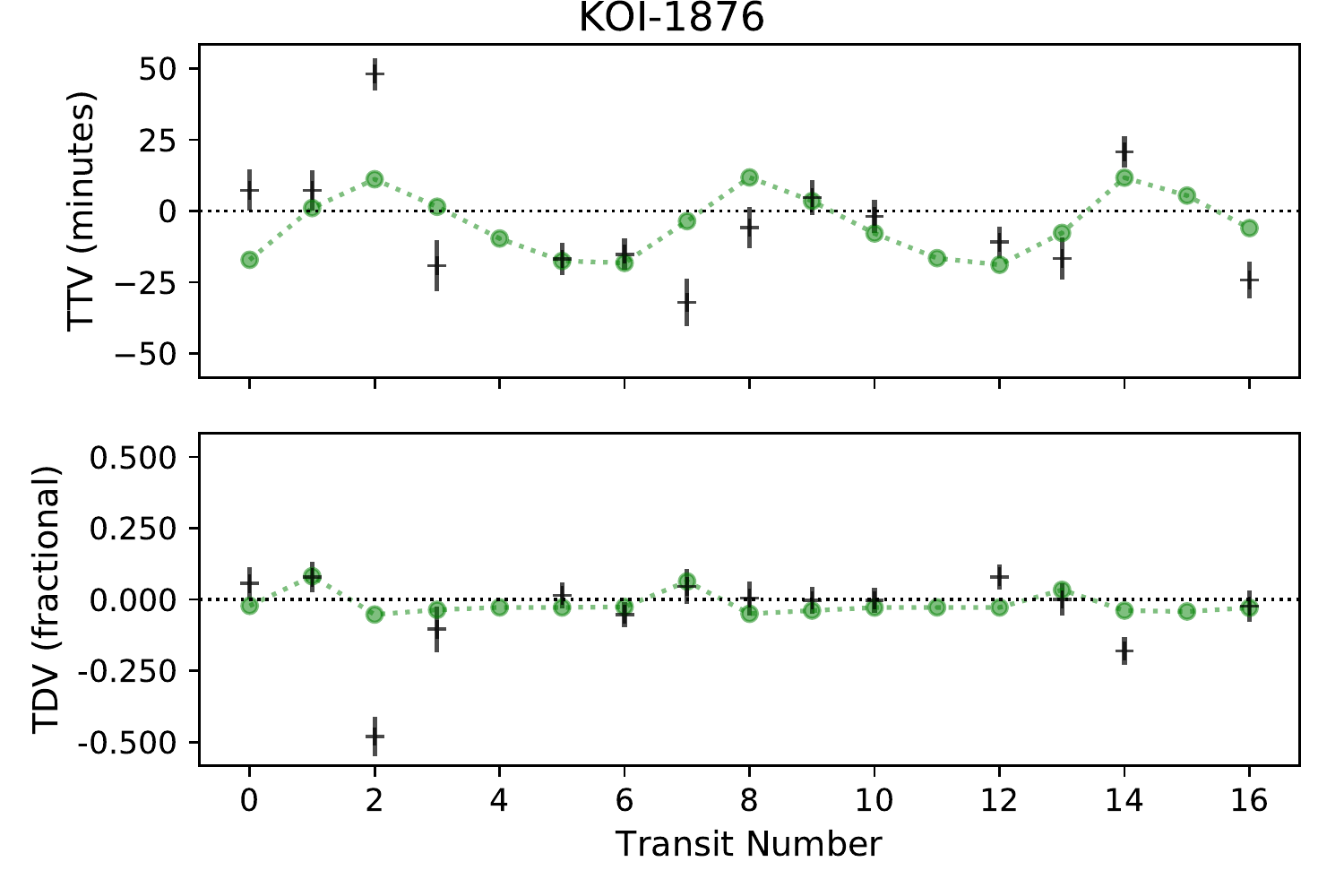} 
 \caption{\label{fig:badmoonsbests}Quality of fit for TTV and TDVs for KOIs 63.01, 318.01, and 1876.01.  The symbols used are the same as in Figure \ref{fig:koi268bests}, but no planet results are included.}
\end{figure}

The remaining targets KOI-63, KOI-318 and KOI-1876 all showed poor moon fits, with reduced $\chi^2$ values of 2.5, 1.9 and 3.3.  Of the three, KOI-318 shows the best TTV fit but the worst TDV fit of any target.  KOI-63 is our shortest period target, but the best moon fit misses the majority of the data points, as does the best fit for KOI-1876.  Combined, we thus consider it unlikely that a moon alone is the cause for the TTVs in these systems, and we did not proceed with the planet hypothesis for these targets or ascertain a position in a sensitivity plot.  The results for these systems are shown in Figure \ref{fig:badmoonsbests}.

\section{Discussion}
While we have shown that photometrically-undetected exomoons could create TTVs, our targets here are at best equally well-matched by the hypothesis of an additional unseen planet in the system. This is perhaps not surprising given that our exomoon model has only five parameters, while the additional planet model has ten, giving it additional flexibility. As a result it will be very difficult to establish the presence of an exomoon solely on the basis of TTVs produced, and future observational platforms with higher photometric sensitivity will be required to detect them with confidence. 

An important consideration throughout this investigation is the 
assumed density of the exomoon, which affects its cross-section and thereby the location of the blue photometric detection threshold line.  In all sensitivity plots shown, a solid blue line indicates the expected detection limit assuming the moon is of a terrestrial density, which results in a smaller cross-section for a given mass.  Where the modelled moon mass is particularly large, a dashed line corresponding to a Neptunian-density is shown, and this can appreciably decrease the size of the green zone.  

All the exomoon models presented here require significantly larger masses in proportion to their host planet than is seen in our solar system.  Not counting Charon around Pluto (mass ratio of 0.13), the largest moon proportional to its planet in our solar system is our own Moon, with a mass ratio of 0.0123.  The smallest hypothetical moon in our sample is in KOI-1472, at a factor of 0.043, and several of the targets have mass ratios greater than 0.1.  Multiple mechanisms for the formation of massive moons have been explored \citep{barrmoon2017, malamudmoon2020} and so these moon models cannot be excluded on that basis. In fact the most massive exomoons are likely to be the first discovered, much as was the case for exoplanets themselves.

The stellar radius is vital in establishing the transit detection threshold, and thus the size of the green zone in the sensitivity plots.  Here, we used stellar values from \citet{math2017} for both the mass and radius of the stars.  Just before submission of this work, an updated Gaia-Kepler catalog \citep{berger2020} was released. It provides stellar radii and mass values largely consistent with the ones we used (though KOI-2469 is absent).  Masses are within their mutual error bars and differ by a median of less than 5\%, and at most 10\%. Radii show greater differences, but are still mostly within their mutual error bars, differing by a median of 14\%.  The stars with more significant radii differences (those greater than 15\%) are usually larger in the Berger et al. catalog.  There is one notable exception, that of KOI-2728.  The value from Mathur et al. is 2.6 $R_{\sun}$ while Berger et al. estimates it at 1.3 $R_{\sun}$.  This updated stellar radius would reduce the transit detection threshold by a factor of 8 (shifting the blue line leftwards on the sensitivity diagram for KOI-2728).  In this case, a terrestrial moon would still lie inside the green zone, but a moon of Neptunian density would be on the edge of detectability.

\begin{table}
 \caption{Summary of Best-Fit Results: Planet vs Moon Hypothesis}
 \label{tab:qualresults}
 \begin{tabular}{llllll}
    & & TTV & & & \\
    & \# Data & SNR & Planet & Moon & Could Moon \\
    KOI & Points & (min) & $\chi^2$/N & $\chi^2$/N & Explain TTVs? \\
  \hline
    268.01 & 11 & 2.37 & 0.579 & 1.514 &  yes \\[1pt]
    303.01 & 21 & 1.56 & 0.581 & 0.793 &  yes \\[1pt]
    1302.01 & 24 & 1.79 & 0.457 & 0.804 &  yes \\[1pt]
    1472.01 & 17 & 1.66 & 0.329 & 0.865 &  yes \\[1pt]
    1848.01 & 27 & 2.46 & 0.873 & 1.343 &  no \\[1pt]    
    1888.01 & 12 & 1.84 & 0.883 & 0.682 &  yes \\[1pt]
    1925.01 & 11 & 1.57 & 0.656 & 0.622 &  yes \\[1pt]
    2469.01 & 10 & 2.42 & 0.307 & 1.133 &  no \\[1pt]
    2728.01 & 20 & 1.71 & 0.427 & 0.748 &  yes \\[1pt]
    3220.01 & 14 & 1.67 & 0.566 & 0.826 &  yes \\[1pt]
  \hline
 \end{tabular}
\end{table}

\begin{table}
 \caption{Physical Parameters of Potential Moons, Best Fit Results}
 \label{tab:moonphysparams}
 \begin{tabular}{lllll}
     &  & Moon /& Orbital & \\
     & Mass & Planet & Distance &  \\
     KOI & ($M_{\earth}$) & Mass Ratio & ($R_{Hill}$) & Eccentricity\\
  \hline
    268.01 & 0.817 & 0.088 & 0.217 & 0.281 \\[1pt]
    303.01 & 0.499 & 0.066 & 0.278 & 0.198 \\[1pt]
    1302.01 & 2.931 & 0.267 & 0.289 & 0.113 \\[1pt]
    1472.01 & 1.636 & 0.043 & 0.206 & 0.495 \\[1pt]
    1848.01 & 5.557 & 0.684 & 0.299 & 0.166 \\[1pt]    
    1888.01 & 1.551 & 0.078 & 0.235 & 0.027 \\[1pt]
    1925.01 & 0.300 & 0.300 & 0.222 & 0.024 \\[1pt]
    2469.01 & 3.441 & 0.521 & 0.294 & 0.368 \\[1pt]
    2728.01 & 6.057 & 0.247 & 0.295 & 0.130 \\[1pt]
    3220.01 & 1.586 & 0.111 & 0.208 & 0.269 \\[1pt]
  \hline
 \end{tabular}
\end{table}

\section{Conclusions}
We examined an unexplored portion of parameter space and evaluated whether exomoons could plausibly explain certain TTV signals seen in the Kepler data.  We rule out the existence of moons being the primary cause of the TTVs in five Kepler systems;  however, there are TTV signals consistent with exomoons in eight others.  We cannot definitively ascribe the observed TTVs in any particular system to an exomoon, as they all prove equally reproducible by a hypothetical additional planet.  The currently available TTV and TDV data do not appear sufficient as the sole method for detecting an exomoon.  While these TTVs $\it{could}$ be explained by an unseen moon, we lack sufficient information to claim any detections.

Of the Kepler data set, the systems examined here may be the best targets warranting further examination in the search for exomoon TTVs.  Followup studies using radial velocities (to potentially find or rule out large companion planets as TTV sources) and further transit studies with improved timing (especially for transit durations) would be required to definitively determine the existence of the potential exomoons examined here.  The upcoming PLATO mission \citep{platomission} with its higher cadence capability may shed light on these systems by obtaining higher precision timing observations.

\section{Acknowledgements} This work was funded in part by the Natural Sciences and Engineering Research Council of Canada Discovery Grants program (Grant no. RGPIN-2018-05659).  We thank the anonymous reviewer for comments that helped us greatly improve this work.

\section{Data Availability}
All data used in this paper comes from publicly available sources (NASA's Exoplanet Archive at https://exoplanetarchive.ipac.caltech.edu  or  SIMBAD at http://simbad.u-strasbg.fr/simbad/)  or from previously published papers with publicly available data (most notably \citet{hm2016} and \citet{ck2018}). 

\bibliographystyle{mnras}
\bibliography{CFBibtex}

\begin{thebibliography}{}
\makeatletter
\relax
\def\mn@urlcharsother{\let\do\@makeother \do\$\do\&\do\#\do\^\do\_\do\%\do\~}
\def\mn@doi{\begingroup\mn@urlcharsother \@ifnextchar [ {\mn@doi@}
  {\mn@doi@[]}}
\def\mn@doi@[#1]#2{\def\@tempa{#1}\ifx\@tempa\@empty \href
  {http://dx.doi.org/#2} {doi:#2}\else \href {http://dx.doi.org/#2} {#1}\fi
  \endgroup}
\def\mn@eprint#1#2{\mn@eprint@#1:#2::\@nil}
\def\mn@eprint@arXiv#1{\href {http://arxiv.org/abs/#1} {{\tt arXiv:#1}}}
\def\mn@eprint@dblp#1{\href {http://dblp.uni-trier.de/rec/bibtex/#1.xml}
  {dblp:#1}}
\def\mn@eprint@#1:#2:#3:#4\@nil{\def\@tempa {#1}\def\@tempb {#2}\def\@tempc
  {#3}\ifx \@tempc \@empty \let \@tempc \@tempb \let \@tempb \@tempa \fi \ifx
  \@tempb \@empty \def\@tempb {arXiv}\fi \@ifundefined
  {mn@eprint@\@tempb}{\@tempb:\@tempc}{\expandafter \expandafter \csname
  mn@eprint@\@tempb\endcsname \expandafter{\@tempc}}}

\bibitem[\protect\citeauthoryear{{Agol}, {Steffen}, {Sari}  \&
  {Clarkson}}{{Agol} et~al.}{2005}]{agol2005}
{Agol} E.,  {Steffen} J.,  {Sari} R.,   {Clarkson} W.,  2005, \mn@doi [\mnras]
  {10.1111/j.1365-2966.2005.08922.x}, \href
  {https://ui.adsabs.harvard.edu/abs/2005MNRAS.359..567A} {359, 567}

\bibitem[\protect\citeauthoryear{Akeson et~al.,}{Akeson et~al.}{2013}]{nep2013}
Akeson R.~L.,  et~al., 2013, \mn@doi [Publications of the Astronomical Society
  of the Pacific] {10.1086/672273}, 125, 989

\bibitem[\protect\citeauthoryear{Barr \& Bruck~Syal}{Barr \&
  Bruck~Syal}{2017}]{barrmoon2017}
Barr A.~C.,  Bruck~Syal M.,  2017, \mn@doi [Monthly Notices of the Royal
  Astronomical Society] {10.1093/mnras/stx078}, 466, 4868

\bibitem[\protect\citeauthoryear{{Berger}, {Huber}, {van Saders}, {Gaidos},
  {Tayar}  \& {Kraus}}{{Berger} et~al.}{2020}]{berger2020}
{Berger} T.~A.,  {Huber} D.,  {van Saders} J.~L.,  {Gaidos} E.,  {Tayar} J.,
  {Kraus} A.~L.,  2020, \mn@doi [\aj] {10.3847/1538-3881/159/6/280}, \href
  {https://ui.adsabs.harvard.edu/abs/2020AJ....159..280B} {159, 280}

\bibitem[\protect\citeauthoryear{Borucki et~al.,}{Borucki
  et~al.}{2010}]{bor2010}
Borucki W.~J.,  et~al., 2010, \mn@doi [Science] {10.1126/science.1185402}, 327,
  977

\bibitem[\protect\citeauthoryear{{Buchner} et~al.,}{{Buchner}
  et~al.}{2014}]{buch2014}
{Buchner} J.,  et~al., 2014, \mn@doi [\aap] {10.1051/0004-6361/201322971},
  \href {https://ui.adsabs.harvard.edu/abs/2014A&A...564A.125B} {564, A125}

\bibitem[\protect\citeauthoryear{{Chambers}, {Wetherill}  \& {Boss}}{{Chambers}
  et~al.}{1996}]{chawetbos96}
{Chambers} J.~E.,  {Wetherill} G.~W.,   {Boss} A.~P.,  1996, \mn@doi [\icarus]
  {10.1006/icar.1996.0019}, \href
  {https://ui.adsabs.harvard.edu/abs/1996Icar..119..261C} {119, 261}

\bibitem[\protect\citeauthoryear{{Chen} \& {Kipping}}{{Chen} \&
  {Kipping}}{2018}]{ck2018}
{Chen} J.,  {Kipping} D.~M.,  2018, \mn@doi [\mnras] {10.1093/mnras/stx2411},
  \href {https://ui.adsabs.harvard.edu/abs/2018MNRAS.473.2753C} {473, 2753}

\bibitem[\protect\citeauthoryear{{Deck}, {Agol}, {Holman}  \&
  {Nesvorn{\'y}}}{{Deck} et~al.}{2014}]{da2014}
{Deck} K.~M.,  {Agol} E.,  {Holman} M.~J.,   {Nesvorn{\'y}} D.,  2014, \mn@doi
  [\apj] {10.1088/0004-637X/787/2/132}, \href
  {https://ui.adsabs.harvard.edu/abs/2014ApJ...787..132D} {787, 132}

\bibitem[\protect\citeauthoryear{{ESA}}{{ESA}}{2017}]{platomission}
{ESA} 2017, {PLATO: Revealing habitable worlds around solar-like stars}, PLATO
  Definition Study Report (Red Book). European Space Agency ESA-SCI(2017)1

\bibitem[\protect\citeauthoryear{Everhart}{Everhart}{1985}]{eve85}
Everhart E.,  1985, in Carusi A.,  Valsecchi G.~B.,  eds, Dynamics of Comets:
  Their Origin and Evolution. Springer Netherlands, Dordrecht, pp 185--202

\bibitem[\protect\citeauthoryear{{Feroz}, {Hobson}  \& {Bridges}}{{Feroz}
  et~al.}{2009}]{feroz2009}
{Feroz} F.,  {Hobson} M.~P.,   {Bridges} M.,  2009, \mn@doi [\mnras]
  {10.1111/j.1365-2966.2009.14548.x}, \href
  {https://ui.adsabs.harvard.edu/abs/2009MNRAS.398.1601F} {398, 1601}

\bibitem[\protect\citeauthoryear{{Fox} \& {Wiegert}}{{Fox} \&
  {Wiegert}}{2019}]{foxwie19}
{Fox} C.,  {Wiegert} P.,  2019, \mn@doi [\mnras] {10.1093/mnras/sty2738}, \href
  {https://ui.adsabs.harvard.edu/abs/2019MNRAS.482..639F} {482, 639}

\bibitem[\protect\citeauthoryear{{Gilliland} et~al.,}{{Gilliland}
  et~al.}{2011}]{gill2011}
{Gilliland} R.~L.,  et~al., 2011, \mn@doi [\apjs] {10.1088/0067-0049/197/1/6},
  \href {https://ui.adsabs.harvard.edu/abs/2011ApJS..197....6G} {197, 6}

\bibitem[\protect\citeauthoryear{{Gladman}}{{Gladman}}{1993}]{gla93}
{Gladman} B.,  1993, \mn@doi [\icarus] {10.1006/icar.1993.1169}, \href
  {https://ui.adsabs.harvard.edu/abs/1993Icar..106..247G} {106, 247}

\bibitem[\protect\citeauthoryear{{Heller}, {Hippke}, {Placek}, {Angerhausen}
  \& {Agol}}{{Heller} et~al.}{2016}]{heller2016}
{Heller} R.,  {Hippke} M.,  {Placek} B.,  {Angerhausen} D.,   {Agol} E.,  2016,
  \mn@doi [\aap] {10.1051/0004-6361/201628573}, \href
  {https://ui.adsabs.harvard.edu/abs/2016A&A...591A..67H} {591, A67}

\bibitem[\protect\citeauthoryear{{Heller}, {Rodenbeck}  \& {Bruno}}{{Heller}
  et~al.}{2019}]{heller2019}
{Heller} R.,  {Rodenbeck} K.,   {Bruno} G.,  2019, \mn@doi [\aap]
  {10.1051/0004-6361/201834913}, \href
  {https://ui.adsabs.harvard.edu/abs/2019A&A...624A..95H} {624, A95}

\bibitem[\protect\citeauthoryear{{Hill}, {Kane}, {Seperuelo Duarte},
  {Kopparapu}, {Gelino}  \& {Wittenmyer}}{{Hill} et~al.}{2018}]{hillkane2018}
{Hill} M.~L.,  {Kane} S.~R.,  {Seperuelo Duarte} E.,  {Kopparapu} R.~K.,
  {Gelino} D.~M.,   {Wittenmyer} R.~A.,  2018, \mn@doi [\apj]
  {10.3847/1538-4357/aac384}, \href
  {https://ui.adsabs.harvard.edu/abs/2018ApJ...860...67H} {860, 67}

\bibitem[\protect\citeauthoryear{{Hinkel} \& {Kane}}{{Hinkel} \&
  {Kane}}{2013}]{hinkelkane2013}
{Hinkel} N.~R.,  {Kane} S.~R.,  2013, \mn@doi [\apj]
  {10.1088/0004-637X/774/1/27}, \href
  {https://ui.adsabs.harvard.edu/abs/2013ApJ...774...27H} {774, 27}

\bibitem[\protect\citeauthoryear{{Holczer} et~al.,}{{Holczer}
  et~al.}{2016}]{hm2016}
{Holczer} T.,  et~al., 2016, \mn@doi [\apjs] {10.3847/0067-0049/225/1/9}, \href
  {https://ui.adsabs.harvard.edu/abs/2016ApJS..225....9H} {225, 9}

\bibitem[\protect\citeauthoryear{{Holman} \& {Murray}}{{Holman} \&
  {Murray}}{2005}]{holmur2005}
{Holman} M.~J.,  {Murray} N.~W.,  2005, \mn@doi [Science]
  {10.1126/science.1107822}, \href
  {https://ui.adsabs.harvard.edu/abs/2005Sci...307.1288H} {307, 1288}

\bibitem[\protect\citeauthoryear{Holman \& Wiegert}{Holman \&
  Wiegert}{1999}]{holwie99}
Holman M.,  Wiegert P.,  1999, AJ, 117, 621

\bibitem[\protect\citeauthoryear{{Kipping}}{{Kipping}}{2009}]{kipp2009}
{Kipping} D.~M.,  2009, \mn@doi [\mnras] {10.1111/j.1365-2966.2008.13999.x},
  \href {https://ui.adsabs.harvard.edu/abs/2009MNRAS.392..181K} {392, 181}

\bibitem[\protect\citeauthoryear{{Kipping}, {Forgan}, {Hartman},
  {Nesvorn{\'y}}, {Bakos}, {Schmitt}  \& {Buchhave}}{{Kipping}
  et~al.}{2013}]{hek3}
{Kipping} D.~M.,  {Forgan} D.,  {Hartman} J.,  {Nesvorn{\'y}} D.,  {Bakos}
  G.~{\'A}.,  {Schmitt} A.,   {Buchhave} L.,  2013, \mn@doi [\apj]
  {10.1088/0004-637X/777/2/134}, \href
  {https://ui.adsabs.harvard.edu/abs/2013ApJ...777..134K} {777, 134}

\bibitem[\protect\citeauthoryear{{Kipping}, {Nesvorn{\'y}}, {Buchhave},
  {Hartman}, {Bakos}  \& {Schmitt}}{{Kipping} et~al.}{2014}]{hek4}
{Kipping} D.~M.,  {Nesvorn{\'y}} D.,  {Buchhave} L.~A.,  {Hartman} J.,  {Bakos}
  G.~{\'A}.,   {Schmitt} A.~R.,  2014, \mn@doi [\apj]
  {10.1088/0004-637X/784/1/28}, \href
  {https://ui.adsabs.harvard.edu/abs/2014ApJ...784...28K} {784, 28}

\bibitem[\protect\citeauthoryear{{Kipping}, {Schmitt}, {Huang}, {Torres},
  {Nesvorn{\'y}}, {Buchhave}, {Hartman}  \& {Bakos}}{{Kipping}
  et~al.}{2015}]{hek5}
{Kipping} D.~M.,  {Schmitt} A.~R.,  {Huang} X.,  {Torres} G.,  {Nesvorn{\'y}}
  D.,  {Buchhave} L.~A.,  {Hartman} J.,   {Bakos} G.~{\'A}.,  2015, \mn@doi
  [\apj] {10.1088/0004-637X/813/1/14}, \href
  {https://ui.adsabs.harvard.edu/abs/2015ApJ...813...14K} {813, 14}

\bibitem[\protect\citeauthoryear{{Kreidberg}, {Luger}  \& {Bedell}}{{Kreidberg}
  et~al.}{2019}]{kreid2019}
{Kreidberg} L.,  {Luger} R.,   {Bedell} M.,  2019, \mn@doi [\apjl]
  {10.3847/2041-8213/ab20c8}, \href
  {https://ui.adsabs.harvard.edu/abs/2019ApJ...877L..15K} {877, L15}

\bibitem[\protect\citeauthoryear{Malamud, Perets, Schäfer  \& Burger}{Malamud
  et~al.}{2020}]{malamudmoon2020}
Malamud U.,  Perets H.~B.,  Schäfer C.,   Burger C.,  2020, \mn@doi [Monthly
  Notices of the Royal Astronomical Society] {10.1093/mnras/staa211}, 492, 5089

\bibitem[\protect\citeauthoryear{{Mart{\'\i}nez-Rodr{\'\i}guez}, {Caballero},
  {Cifuentes}, {Piro}  \& {Barnes}}{{Mart{\'\i}nez-Rodr{\'\i}guez}
  et~al.}{2019}]{martinez2019}
{Mart{\'\i}nez-Rodr{\'\i}guez} H.,  {Caballero} J.~A.,  {Cifuentes} C.,  {Piro}
  A.~L.,   {Barnes} R.,  2019, \mn@doi [\apj] {10.3847/1538-4357/ab5640}, \href
  {https://ui.adsabs.harvard.edu/abs/2019ApJ...887..261M} {887, 261}

\bibitem[\protect\citeauthoryear{{Mathur} et~al.,}{{Mathur}
  et~al.}{2017}]{math2017}
{Mathur} S.,  et~al., 2017, \mn@doi [\apjs] {10.3847/1538-4365/229/2/30}, \href
  {https://ui.adsabs.harvard.edu/abs/2017ApJS..229...30M} {229, 30}

\bibitem[\protect\citeauthoryear{{Nicholson}, {Cuk}, {Sheppard}, {Nesvorny}  \&
  {Johnson}}{{Nicholson} et~al.}{2008}]{niccukshe08}
{Nicholson} P.~D.,  {Cuk} M.,  {Sheppard} S.~S.,  {Nesvorny} D.,   {Johnson}
  T.~V.,  2008, {Irregular Satellites of the Giant Planets}.
p.~411

\bibitem[\protect\citeauthoryear{{Rodenbeck}, {Heller}  \& {Gizon}}{{Rodenbeck}
  et~al.}{2020}]{rodenbeck2020}
{Rodenbeck} K.,  {Heller} R.,   {Gizon} L.,  2020, \mn@doi [\aap]
  {10.1051/0004-6361/202037550}, \href
  {https://ui.adsabs.harvard.edu/abs/2020A&A...638A..43R} {638, A43}

\bibitem[\protect\citeauthoryear{{Sartoretti, P.} \& {Schneider,
  J.}}{{Sartoretti, P.} \& {Schneider, J.}}{1999}]{sarsch1999}
{Sartoretti, P.} {Schneider, J.} 1999, \mn@doi [Astron. Astrophys. Suppl. Ser.]
  {10.1051/aas:1999148}, 134, 553

\bibitem[\protect\citeauthoryear{{Szab{\'o}}, {Szab{\'o}}, {D{\'a}lya},
  {Simon}, {Hodos{\'a}n}  \& {Kiss}}{{Szab{\'o}} et~al.}{2013}]{szabo2013}
{Szab{\'o}} R.,  {Szab{\'o}} G.~M.,  {D{\'a}lya} G.,  {Simon} A.~E.,
  {Hodos{\'a}n} G.,   {Kiss} L.~L.,  2013, \mn@doi [\aap]
  {10.1051/0004-6361/201220132}, \href
  {https://ui.adsabs.harvard.edu/abs/2013A&A...553A..17S} {553, A17}

\bibitem[\protect\citeauthoryear{Teachey \& Kipping}{Teachey \&
  Kipping}{2018}]{hek1625}
Teachey A.,  Kipping D.~M.,  2018, \mn@doi [Science Advances]
  {10.1126/sciadv.aav1784}, 4

\bibitem[\protect\citeauthoryear{Teachey, Kipping  \& Schmitt}{Teachey
  et~al.}{2017}]{hek6}
Teachey A.,  Kipping D.~M.,   Schmitt A.~R.,  2017, \mn@doi [The Astronomical
  Journal] {10.3847/1538-3881/aa93f2}, 155, 36

\bibitem[\protect\citeauthoryear{{Thompson} et~al.,}{{Thompson}
  et~al.}{2018}]{kep8cat2018}
{Thompson} S.~E.,  et~al., 2018, \mn@doi [\apjs] {10.3847/1538-4365/aab4f9},
  \href {https://ui.adsabs.harvard.edu/abs/2018ApJS..235...38T} {235, 38}

\bibitem[\protect\citeauthoryear{{Wenger} et~al.,}{{Wenger}
  et~al.}{2000}]{simbad2000}
{Wenger} M.,  et~al., 2000, \mn@doi [\aaps] {10.1051/aas:2000332}, \href
  {https://ui.adsabs.harvard.edu/abs/2000A&AS..143....9W} {143, 9}

\bibitem[\protect\citeauthoryear{{Wisdom} \& {Holman}}{{Wisdom} \&
  {Holman}}{1991}]{wishol1999}
{Wisdom} J.,  {Holman} M.,  1991, \mn@doi [\aj] {10.1086/115978}, \href
  {https://ui.adsabs.harvard.edu/abs/1991AJ....102.1528W} {102, 1528}

\makeatother
\end{thebibliography}

\clearpage
\onecolumn
%\section{Appendix}
\appendix
\begin{landscape}
\begin{table*}
 \caption{Planet Hypothesis Best Fit Parameters and Posteriors}
 \label{tab:planetresults}
 \begin{tabular}{llllllllllll}
    & KOI-268 & KOI-303 & KOI-1302 & KOI-1472 & KOI-1848 & KOI-1888 & KOI-1925 & KOI-2469 & KOI-2728 & KOI-3220 &  \\
    \hline
    $M_{b}$ Best Fit & 0.244 & 0.628 & 0.774 & 0.313 & 0.654 & 0.148 & 0.606 & 2.28 & 0.165 & 0.427 & $M_{J}$\\ 
    $M_{b}$ Posterior & $0.120\substack{+0.037\\-0.038}$ & $0.128\substack{+0.128\\-0.056}$ & $0.714\substack{+0.356\\-0.262}$ & $0.142\substack{+0.054\\-0.044}$ & $0.645\substack{+0.136\\-0.126}$ & $0.084\substack{+0.438\\-0.065}$ & $0.078\substack{+0.033\\-0.027}$ & $1.124\substack{+0.49\\-0.33}$ & $0.179\substack{+0.046\\-0.043}$ & $0.286\substack{+0.097\\-0.189}$ &  \\ \\[1pt]
    $P_{b}$ Best Fit & 315.238 & 275.276 & 235.897 & 328.088 & 149.387 & 336.219 & 313.656 & 419.721 & 100.319 & 285.853 & days\\ 
    $P_{b}$ Posterior & $322.7\substack{+5.6\\-6.1}$ & $274.9\substack{+33\\-18.8}$ & $235.4\substack{+0.8\\-1}$ & $328.2\substack{+2.9\\-2.4}$ & $149.1\substack{+0.9\\-0.8}$ & $397.1\substack{+39\\-43.2}$ & $316.7\substack{+8.3\\-4.8}$ & $422.5\substack{+5.9\\-5.9}$ & $100.2\substack{+0.2\\-0.3}$ & $289.6\substack{+4.2\\-3.6}$ &  \\ \\[1pt]
    $e_{b}$ Best Fit & 0.106 & 0.176 & 0.205 & 0.057 & 0.006 & 0.072 & 0.038 & 0.096 & 0.127 & 0.08 & \\ 
    $e_{b}$ Posterior & $0.042\substack{+0.035\\-0.025}$ & $0.215\substack{+0.037\\-0.073}$ & $0.201\substack{+0.047\\-0.043}$ & $0.05\substack{+0.024\\-0.022}$ & $0.012\substack{+0.016\\-0.008}$ & $0.128\substack{+0.091\\-0.065}$ & $0.055\substack{+0.048\\-0.034}$ & $0.088\substack{+0.024\\-0.021}$ & $0.094\substack{+0.043\\-0.052}$ & $0.036\substack{+0.04\\-0.022}$ &  \\ \\[1pt]
    $i_{b}$ Best Fit & 73.017 & 92.891 & 93.455 & 109.814 & 94.867 & 96.091 & 100.355 & 73.294 & 80.521 & 120.846 & deg \\ 
    $i_{b}$ Posterior & $84.9\substack{+9.4\\-9.1}$ & $92.3\substack{+8.2\\-11.6}$ & $59.3\substack{+8.2\\-8.4}$ & $93.8\substack{+11.8\\-18.7}$ & $85.9\substack{+6.2\\-3.2}$ & $96.7\substack{+24.8\\-32.9}$ & $94.3\substack{+12.2\\-19.4}$ & $89.6\substack{+14.9\\-14}$ & $82.2\substack{+12.9\\-7.8}$ & $91.1\substack{+20.8\\-23.1}$ & \\ \\[1pt]
    $\Omega_{b}$ Best Fit & -19.055 & -7.919 & 0.103 & 5.3 & -4.537 & -16.791 & 4.995 & 2.197 & 8.276 & -42.709 & deg \\
    $\Omega_{b}$ Posterior & $-7.6\substack{+13.5\\-9.0}$ & $-0.5\substack{+7.9\\-8.1}$ & $-0.1\substack{+10.9\\-11.6}$ & $2.0\substack{+8.1\\-7.6}$ & $3.4\substack{+3.9\\-5.0}$ & $-12\substack{+28.1\\-18.7}$ & $2.6\substack{+10.4\\-11.1}$ & $0.4\substack{+7.0\\-7.2}$ & $11.7\substack{+5.8\\-11.8}$ & $-26\substack{+64.7\\-14.9}$ & \\ \\[1pt]
    $\omega_{b}$ Best Fit & 143.035 & 101.62 & 214.62 & 249.454 & 227.684 & 313.707 & 67.048 & 154.021 & 4.665 & 219.65 & deg \\ 
    $\omega_{b}$ Posterior & $67\substack{+48\\-38}$ & $106\substack{+83\\-24}$ & $218\substack{+17\\-24}$ & $184\substack{+35\\-39}$ & $180\substack{+30\\-57}$ & $123\substack{+116\\-76}$ & $159\substack{+45\\-52}$ & $161\substack{+13\\-13}$ & $19\substack{+29\\-14}$ & $56\substack{+150\\-37}$ &\\ \\[1pt]
    $Mn_{b}$ Best Fit & 127.644 & 147.105 & 65.011 & 44.055 & 10.528 & 73.61 & 166.682 & 64.295 & 197.139 & 58.094 & deg\\
    $Mn_{b}$ Posterior & $218\substack{+45\\-56}$ & $151\substack{+71\\-26}$ & $65\substack{+16\\-11}$ & $95\substack{+32\\-33}$ & $38\substack{+54\\-27}$ & $176\substack{+60\\-72}$ & $175\substack{+51\\-33}$ & $71\substack{+11\\-11}$ & $152\substack{+36\\-38}$ & $237\substack{+36\\-168}$ & \\ \\[1pt]
    \hline
    $e_{c}$ Best Fit & 0.021 & 0.003 & 0.008 & 0.043 & 0.061 & 0.104 & 0.149 & 0.056 & 0.09 & 0.05 & \\ 
    $e_{c}$ Posterior & $0.04\substack{+0.024\\-0.018}$ & $0.053\substack{+0.05\\-0.036}$ & $0.013\substack{+0.019\\-0.009}$ & $0.13\substack{+0.043\\-0.044}$ & $0.041\substack{+0.023\\-0.024}$ & $0.065\substack{+0.062\\-0.041}$ & $0.101\substack{+0.024\\-0.023}$ & $0.092\substack{+0.045\\-0.038}$ & $0.113\substack{+0.034\\-0.024}$ & $0.118\substack{+0.059\\-0.058}$ & \\ \\[1pt]
    $\omega_{c}$ Best Fit & 68.07 & 1.756 & 202.692 & 34.727 & 140.916 & 33.983 & 180.321 & 198.221 & 69.78 & 57.83 & deg \\ 
    $\omega_{c}$ Posterior & $59\substack{+12\\-11}$ & $349\substack{+5\\-6}$ & $179\substack{+18\\-16}$ & $31\substack{+8\\-11}$ & $151\substack{+21\\-13}$ & $175\substack{+12\\-15}$ & $39\substack{+15\\-15}$ & $188\substack{+9\\-12}$ & $48\substack{+17\\-15}$ & $46\substack{+9\\-10}$ & \\ \\[1pt]
    $Mn_{c}$ Best Fit & 19.415 & 87.634 & 244.836 & 49.17 & 310.935 & 46.487 & 286.558 & 256.632 & 16.542 & 29.035 & deg \\
    $Mn_{c}$ Posterior & $27\substack{+10\\-10}$ & $94\substack{+6\\-6}$ & $270\substack{+15\\-19}$ & $45\substack{+7\\-7}$ & $300\substack{+13\\-22}$ & $284\substack{+12\\-13}$ & $42\substack{+13\\-12}$ & $273\substack{+10\\-9}$ & $33\substack{+10\\-13}$ & $35\substack{+8\\-8}$ & \\ \\[1pt]
    \hline
    Reduced $\chi^2$\\of Best Fit & 0.579 & 0.581 & 0.457 & 0.329 & 0.873 & 0.883 & 0.656 & 0.307 & 0.427 & 0.566 & \\[1pt]
 \end{tabular}
\end{table*}
\end{landscape}

\begin{landscape}
\begin{table*}
 \caption{Moon Hypothesis Best Fit Parameters and Posteriors}
 \label{tab:moonresults}
 \begin{tabular}{llllllllllll}
    & KOI-268 & KOI-303 & KOI-1302 & KOI-1472 & KOI-1848 & KOI-1888 & KOI-1925 & KOI-2469 & KOI-2728 & KOI-3220 & \\
    \hline
    $M_{m}$ Best Fit & 0.817 & 0.499 & 2.931 & 1.636 & 5.557 & 1.551 & 0.300 & 3.441 & 6.057 & 1.586 & $M_{\earth}$\\ 
    $M_{m}$ Posterior & $0.44\substack{+0.09\\-0.09}$ & $0.36\substack{+0.1\\-0.09}$ & $2.91\substack{+0.57\\-0.52}$ & $0.70\substack{+0.36\\-0.27}$ & $4.82\substack{+0.36\\-0.54}$ & $1.20\substack{+0.32\\-0.31}$ & $0.18\substack{+0.05\\-0.05}$ & $2.67\substack{+0.68\\-0.71}$ & $4.88\substack{+1.02\\-1.05}$ & $0.85\substack{+0.27\\-0.25}$ & \\ \\[1pt]
    $a_{m}$ Best Fit & 0.217 & 0.278 & 0.289 & 0.206 & 0.299 & 0.235 & 0.222 & 0.294 & 0.295 & 0.208 & $R_{Hill}$ \\ 
    $a_{m}$ Posterior & $0.257\substack{+0.028\\-0.044}$ & $0.277\substack{+0.014\\-0.050}$ & $0.290\substack{+0.007\\-0.012}$ & $0.244\substack{+0.031\\-0.03}$ & $0.287\substack{+0.008\\-0.009}$ & $0.285\substack{+0.009\\-0.017}$ & $0.276\substack{+0.017\\-0.056}$ & $0.245\substack{+0.033\\-0.032}$ & $0.289\substack{+0.007\\-0.008}$ & $0.261\substack{+0.027\\-0.043}$ & \\ \\[1pt]
    $Mean_{m}$ Best Fit & 89.844 & 321.701 & 270.872 & 139.233 & 92.752 & 257.304 & 238.324 & 194.074 & 35.963 & 17.501 & deg\\ 
    $Mean_{m}$ Posterior & $127\substack{+107\\-52}$ & $219\substack{+106\\-124}$ & $74\substack{+72\\-51}$ & $175\substack{+73\\-99}$ & $121\substack{+108\\-62}$ & $135\substack{+119\\-93}$ & $139\substack{+148\\-104}$ & $167\substack{+27\\-28}$ & $76\substack{+242\\-53}$ & $89\substack{+166\\-58}$ & \\ \\[1pt]
    $e_{m}$ Best Fit & 0.281 & 0.198 & 0.113 & 0.495 & 0.166 & 0.027 & 0.024 & 0.368 & 0.13 & 0.269 & \\ 
    $e_{m}$ Posterior & $0.122\substack{+0.108\\-0.051}$ & $0.113\substack{+0.113\\-0.089}$ & $0.036\substack{+0.046\\-0.023}$ & $0.231\substack{+0.216\\-0.156}$ & $0.051\substack{+0.031\\-0.029}$ & $0.053\substack{+0.041\\-0.030}$ & $0.07\substack{+0.119\\-0.047}$ & $0.301\substack{+0.096\\-0.161}$ & $0.051\substack{+0.055\\-0.030}$ & $0.136\substack{+0.085\\-0.090}$ & \\ \\[1pt]
    $\omega_{m}$ Best Fit & 101.759 & 82.578 & 229.435 & 310.739 & 269.987 & 48.75 & 207.235 & 85.108 & 246.46 & 210.756 & deg \\
    $\omega_{m}$ Posterior & $90\substack{+161\\-37}$ & $156\substack{+117\\-102}$ & $127\substack{+81\\-60}$ & $278\substack{+30\\-195}$ & $242\substack{+52\\-109}$ & $178\substack{+89\\-113}$ & $117\substack{+139\\-73}$ & $275\substack{+19\\-31}$ & $249\substack{+55\\-62}$ & $189\substack{+78\\-105}$ & \\ \\[1pt]
    \hline
    Reduced $\chi^2$ Best Fit& 1.514 & 0.793 & 0.804 & 0.865 & 1.343 & 0.682 & 0.622 & 1.133 & 0.748 & 0.826 & \\[1pt]
 \end{tabular}
\end{table*}
\end{landscape}
%%%%%%%%%%%%%%%%%%%%%%%%%%%%%%%%%%%%%%%%%%%%%%%%%%
% Don't change these lines
\bsp	% typesetting comment
\label{lastpage}
\end{document}